\documentclass{aa}  
\usepackage{graphicx}
\usepackage{xcolor}
\usepackage{rotating}
\usepackage{tikz}
\usepackage{txfonts}
\usepackage{supertabular}
\usepackage{longtable}
\usepackage{tabularx}
\usepackage{lscape}
\usepackage{soul}
\usepackage{appendix}
\usepackage{float}
\bibpunct{(}{)}{;}{a}{}{,} % to follow the A&A style
\newcommand{\jklang}[1]{\textbf{\textcolor{blue}{[check language]}}}

\makeatletter
\newcommand\footnoteref[1]{\protected@xdef\@thefnmark{\ref{#1}}\@footnotemark}
\makeatother
\usepackage[para]{footmisc}

\begin{document}

   \title{Bird's eye view of molecular clouds in the Milky Way}
   %Bird's-eye view of the local Galactic environment: \\ Column Density Statistics}
   \subtitle{II. Cloud kinematics from subparsec to kiloparsec scales}
   %The Larson relations from sub-pc to kpc scales}
   %Column Density Statistics}
    %\shorttitle{Birds-eye view: Column Density}
    %\shortauthors{Spilker, Kainulainen \& Zhang}
   \author{Andri Spilker
          \inst{1}
          \and
          Jouni Kainulainen
          \inst{1}
          \and
          Jan Orkisz
          \inst{1}
          %\and
          %++?
          %\fnmsep%\thanks{Just to show the usage
          %of the elements in the author field}
          }

   \institute{Chalmers University of Technology, Department of Space, Earth and Environment, SE-412 93 Gothenburg, Sweden\\
              \email{andri.spilker@chalmers.se}
             }
   \date{Received -----; accepted -----}

% \abstract{}{}{}{}{} 
% 5 {} token are mandatory
  \abstract
  {The kinematics of molecular gas are crucial for setting the stage for star formation. One key question related to the kinematic properties of gas is how they depend on the spatial scale.}%. 
  {We aim to describe the CO spectra, velocity dispersions, and especially the linewidth-size relation of molecular gas from cloud (parsec-) scales to kiloparsec scales in a complete region within the Milky Way disk.}
  {We used the census of molecular clouds within 2 kpc from our earlier work, together with CO emission data for them from the literature. We studied the kinematics and the Larson relations for the sample of individual clouds. We also mimicked a face-on view of the Milky Way and analysed the kinematics of the clouds within apertures of 0.25-2 kpc in size. In this way, we describe the scale-dependence of the CO gas kinematics and Larson's relations.} 
  {We describe the spectra of CO gas at cloud scales and in apertures between 0.25-2 kpc in our survey area. The spectra within the apertures are relatively symmetric, but show non-Gaussian high-velocity wings. At cloud scales, our sample shows a linewidth-size relation $\sigma_v=1.5\cdot R^{0.3\pm0.1}$ with a large scatter. The mass-size relation in the sample of clouds is $\mathrm{M_{CO}}= 794 \cdot R^{1.5\pm0.5}$. The relations are also present for the apertures at kiloparsec-scales. The best-fit linewidth-size relation for the apertures is $\sigma_v=0.5 \cdot R^{0.35\pm0.01}$, and the best-fit mass-size relation is $\mathrm{M_{CO}}= 229 \cdot R^{1.4\pm0.1}$. A suggestive dependence on Galactic environment is seen. Apertures closer to the Galactic centre and the Sagittarius spiral arm have slightly higher velocity dispersions. We explore the possible effect of a diffuse component in the survey area and find that such a component would widen the CO spectra and could flatten the linewidth-size relation. Understanding the nature of the possible diffuse CO component and its effects on observations is crucial for connecting Galactic and extragalactic data.}
  {}

    \maketitle
%
%________________________________________________________________

\section{Introduction}
%MOVE 1: Establishing a Territory
%   STEP 1: Claiming Centrality
%   STEP 2: Making Topic Generalisations
%   STEP 3: Reviewing Previous Research

%MOVE 2: Establishing a Niche
%   STEP 1A: Counter-Claiming
%   STEP 1B: Indicating a Gap
%   STEP 1C: Question-Raising
%   STEP 1D: Continuing a Tradition

%MOVE 3: Occupying the Niche
%   STEP 1A: Outlining Purposes
%   STEP 1B: Announcing Present Research
%   STEP 2: Announcing Main Findings
%   STEP 3: Indicating Structure of the Article
%   STEP 4: Evaluation of Findings*
%* not originally included in Swale's analysis

%   OR:
%   WHAT IS THE ISSUE?
%   WHAT DO WE KNOW ABOUT IT?
%   WHAT IS MISSING?
%   THIS WORK

Stars in galaxies form from gas in molecular clouds, and the movement of this gas can give us insight into the physical conditions within the clouds and within the galaxies. An open question is which physical processes set the properties of stars and their parent clouds, and on which scales. Important insight has been gained by studying cloud kinematics and other cloud properties in the past: Molecular clouds have been found to be turbulent, gravitationally bound, and to have approximately constant surface densities \citep{larson1981turbulence,heyer2015molecular}. The internal (subparsec, subpc) velocity structure of molecular clouds can only be studied in our Milky Way, but extragalactic studies indicate that the velocity structure on galaxy disk scales (kiloparsec, kpc) also affects the molecular clouds and possibly their star formation activity \citep{schinnerer2013pdbi,hughes2013probability,sun2020molecular,leroy2021phangs,rosolowsky2021giant}. Bridging the gap between the scales studied in the Milky Way and external galaxies could lead us to an improved understanding of which scales and physical processes are important in setting the properties of molecular clouds and the stage for star formation.

The kinematics of Milky Way molecular clouds have previously been studied in small samples of nearby clouds \citep[e.g.][]{larson1981turbulence,falgarone1992small,heyer2004universality}, in larger samples \citep[e.g.,][]{solomon1987mass,roman2009kinematic}, and even in the entire Milky Way \citep{miville2017physical}. However, studies with significant numbers of clouds mainly rely on kinematic distances and on various cloud-finding algorithms for cloud definitions. The lack of accurate distances makes it difficult to pinpoint the location of the clouds in the Milky Way disk and to relate them to their larger-scale Galactic environment. The varying cloud definitions also make it difficult to reconcile the data with more detailed observations of the internal structure and star formation properties of the clouds. In order to study the full range of scales involved, it is necessary to relate molecular clouds both to their larger-scale galactic environment and to their intricate internal structure and star formation properties.

In extragalactic studies, the scales from entire galaxy disks (tens of kpc) to clouds \citep[few tens of pc; i.e. PHANGS;][]{leroy2021phangs}  can be observed. Within these scales, the variations of the properties of the interstellar medium (ISM) as a function of scale have been studied. For example, \cite{leroy2016portrait} have shown that the ISM of M51 has higher surface densities, lower line widths, and more self-gravity at smaller scales, while \cite{schinnerer2019} demonstrated that the overlap between CO and H$\alpha$ emission in eight nearby galaxies varies as a function of scale and resolution. However, extragalactic studies cannot yet go beyond the cloud scale and down to the filaments and cores that are more directly linked to star formation. In the Milky Way, these substructures in clouds can be resolved by observations, but the task of linking the small-scale star formation properties with the Galactic scales is rarely undertaken.

In \cite{spilker2021bird} (hereafter Paper I), we developed the bird's eye Milky Way experiment to study molecular cloud properties as a function of scale and to create a new interface between Galactic and extragalactic works. In this experiment, we compiled a sample of molecular clouds within a distance of 2 kpc that is as complete as possible and studied them individually and in a bird's-eye view, that is, in a face-on perspective of the Galactic disk. This enabled us to describe the column density statistics and star formation activity as a function of scale from cloud scales to kiloparsec scales. We further showed how such analysis can lead to new constraints for important parameters such as column density probability density functions (PDFs) and provide insight into the build-up of relations such as the Kennicutt-Schmidt relation. 

In this paper, we take the next logical step in developing the bird's eye Milky Way experiment and study the gas kinematics in the same manner. We use as a main diagnostic tool the velocity standard deviations of the clouds, measured from CO$(J=1-0)$ emission data. This enables us to describe the gas kinematics in the solar neighbourhood gas as a function of spatial scale and to revisit the \cite{larson1981turbulence} linewidth-size and mass-size relations. Furthermore, it enables us to describe the dependence of the velocity dispersion of molecular gas on Galactic environment. In this way, we study molecular cloud kinematics from subpc to kpc scales, taking an important step in understanding the observational origin and interpretation of the \cite{larson1981turbulence} relations and towards connecting studies of molecular gas kinematics in the Milky Way with external galaxies.

%%%%%%%%%%%%%%%%%%%%%%%%%%%%%%%%%
%%%%%%%%%%%%%%%%%%%%%%%%%%%%%%%%%
\section{Data and methods}
\label{sec:method}
%%%%%%%%%%%%%%%%%%%%%%%%%%%%%%%%%
%%%%%%%%%%%%%%%%%%%%%%%%%%%%%%%%%

In the following, we first describe the cloud sample (Sect. \ref{sec:m_sample}) and the CO data (Sect. \ref{sec:m_codata}) used in the paper. Then, we describe the methods related to the analysis of individual clouds (Sect. \ref{sec:m_ind}) and to the analysis of clouds from the Bird's Eye perspective (Sect. \ref{sec:m_aps}).

%%%%%%%%%%%%%%%%%%%%%%%%%%%%%%%%%
\subsection{Cloud sample}
\label{sec:m_sample}
%%%%%%%%%%%%%%%%%%%%%%%%%%%%%%%%%

The cloud sample considered in this paper is the same as in Paper I. This sample of molecular clouds within a distance of 2 kpc is as complete as possible. It was established by searching cloud-like objects from various CO and dust-based data and catalogues in the literature (see Paper I for details). The sample contains 72 clouds that are discernible in CO or dust extinction or emission. Eight of the clouds are not covered by the \cite{dame2001milky} survey (cf. Section \ref{sec:m_codata}) and are therefore not included in this paper. Our final sample therefore consist of 64 clouds. The cloud sample and its properties are listed in Table \ref{tab:veltable}.

%%%%%%%%%%%%%%%%%%%%%%%%%%%%%%%%%
\subsection{CO data}
\label{sec:m_codata}
%%%%%%%%%%%%%%%%%%%%%%%%%%%%%%%%%

We used the CO$(J=1-0)$ data published by \cite{dame2001milky} to study the kinematics of our cloud sample. The \cite{dame2001milky} survey covers $\sim45\%$ of the sky within $30^\circ$ of the equator and almost all molecular clouds larger than a few degrees. The angular resolution is $\sim8.5'$, which corresponds to a physical resolution of $\sim0.2$ pc at a distance of 100 pc and $\sim5$ pc at a distance of 2 kpc. The spectral resolution of the data for the most part is 0.65 km/s (63\% of the clouds), but for some areas of the map, it is 1.3 or 0.26 km/s \citep[see Table 1 in][]{dame2001milky}. For 84\% of the clouds, the total spectrum is clearly resolved. The clouds with unresolved spectra are generally very small. 

The CO data were used to derive physical properties of the clouds and apertures. The most fundamental parameter for studying kinematics is the standard deviation of the total spectrum, $\sigma_v$ (velocity dispersion). We calculated $\sigma_v$ as the square root of the second moment of the spectra. $\sigma_v$ can be related to the turbulent properties of the clouds and apertures through the Mach number. The Mach number is evaluated as $\mathcal{M} = \sigma_{nth,1D}/c_s$, where $c_{\mathrm{s}}=\sqrt{\frac{k_b T}{\mu m_p}}$ is the isothermal sound speed, $\mu$ is the mean molecular mass
($\mu$ = 2.33 amu), $m_p$ is the proton mass, $k_b$ is the Boltzmann constant, and $T$ is the temperature. We calculated rough Mach numbers adopting an arbitrary constant temperature of 18 K \citep[as in][]{dame2001milky}. When we assume that all non-thermal contributions to the velocity dispersion are due to turbulence, the Mach number is
\begin{equation}
    \mathcal{M}_\mathrm{s, 3D} \approx \sqrt{3} \frac{\sigma_{v_{\text {turb}, \mathrm{1D}}}}{c_{\mathrm{s}}}=\sqrt{3}\left[\left(\frac{\sigma_{v}}{c_{\mathrm{s}}}\right)^{2}-\left(\frac{\mu}{\mu_{o b s}}\right)\right]^{1 / 2}.
\end{equation}
Here $\sigma_v$ is the velocity dispersion along the line of sight, and $\mu_{obs}$ is the molecular mass of the observed molecule (28 amu for $^{12}$CO). 
The integrated intensity of a CO spectrum $\int T_B(v) dv$ is proportional to the column density $N$ by the $X_{CO}$ factor $N=X_{CO}\int T_B(v) dv$ \citep{bohlin1978survey}. We used 
\begin{equation}
    X_{CO}=2\times10^{20}\mathrm{cm}^{-2}(\mathrm{K\:km\:s}^{-1})^{-1},
\end{equation}
as recommended by \cite{bolatto2013co}. It is then possible to use the column density to compute the mass, $M=\mu m_p N d^2$, where $d$ is the distance to the cloud. The cloud and aperture mass can then be estimated from the integrated intensity of the CO spectra as
\begin{equation}
    \mathrm{M_{CO}} = X_{CO} \mu m_p \cdot \int T_B(v) dv \cdot d^2. 
    \label{eq:mass}
\end{equation}

The total CO mass of the clouds in our sample is $4.9 \cdot 10^6$ M$_\odot$. This is 56\% of the mass from extinction (Paper I). Several factors can contribute to this difference. Dust traces a wider range of column densities than CO. On the one hand, a significant amount of molecular gas is CO dark \citep[e.g.][]{goodman2009true}, but traced by dust. On the other hand, CO becomes optically thick, or depletes, at high column densities where dust still traces the total column. It is also unclear how accurate the adopted X-factor is exactly in the solar neighbourhood, as it may have some dependence on galactic radius \citep[i.e.][]{bolatto2013co}. 

The CO-derived mass of the clouds in our sample is 58\% of the mass of CO clouds in this area identified by \cite{miville2017physical}. The cloud catalogue by \cite{miville2017physical} assigns virtually all the CO gas to clouds (it includes 98\% of the CO emission within $\pm5^{\circ}$ of the Milky Way disk), so it is not surprising that we recover significantly less mass in our hand-picked sample. As established in Paper I, we are unlikely to miss clouds larger than $\sim10^4$ M$_\odot$ that contain a significant fraction of molecular gas, but we may miss smaller and more diffuse clouds.

%%%%%%%%%%%%%%%%%%%%%%%%%%%%%%%%%
\subsection{Kinematic analysis of the individual clouds}
\label{sec:m_ind}
%%%%%%%%%%%%%%%%%%%%%%%%%%%%%%%%%

We first used the CO data to study the sample as individual clouds. The cloud sample with its properties is listed in Table \ref{tab:veltable}, and an overview of the clouds in the position-position-velocity ($p$-$p$-$v$) space is shown in Appendix \ref{app:overview}. To limit contamination from foreground and background gas, the clouds were cropped of parts that were not coherent in velocity. This was done by examining the CO $p$-$p$-$v$ cubes of the clouds. When a cloud had more than one peak in velocity, we chose the component closest to the velocity corresponding to the distance of the cloud. %\st{The cropped part of the spectra are shown in grey in in appendix \ref{app:overview}}. 
With the cropped cloud spectra in hand, we derived the velocity standard deviations and masses of the clouds. The masses and linewidths were compared to the sizes from the area above 3 mag of extinction (from Paper I). We estimated the uncertainty on $\sigma_v$ to be $\delta \sigma_v = \frac{\sigma_v}{\sqrt{N}}$ (where $N$ is the number of spectral channels), and the uncertainty on the sizes of the clouds was estimated to be the change in size when we adjusted the magnitude limit by $\pm 20\%$. The uncertainty in mass was taken to be $\pm 20\%$. To fit the linewidth-size and mass-size relations we performed an orthogonal distance regression fit with the Python Scipy package \texttt{ODR}, that takes the uncertainties on all quantities into account \citep{2020SciPy-NMeth}.

%%%%%%%%%%%%%%%%%%%%%%%%%%%%%%%%%
\subsection{Kinematic analysis from the Bird's Eye perspective}
\label{sec:m_aps}
%%%%%%%%%%%%%%%%%%%%%%%%%%%%%%%%%

We then used the CO data to analyse the survey area from a bird's eye perspective, following the approach of Paper I. This was possible due to recently derived accurate distances to the molecular clouds based on data from Gaia by \cite{zucker2019large,zucker2020compendium}. In this analysis, we constructed the total CO emission spectra of clouds within apertures of different sizes as they would appear if the Milky Way were viewed from a face-on angle. To facilitate the experiment, we conjectured that the CO velocity distributions of individual clouds from the face-on perspective are the same as in the plane of the sky, except for the contribution from the rotation of the galaxy. The rotation of the galaxy was taken into account by subtracting from the spectra the systematic velocity of the clouds that is due to the rotation of the galaxy disk. The disk motion was calculated using the Galactic rotation curve from \cite{reid2019trigonometric} and the accurate distances from \cite{zucker2019large,zucker2020compendium}. 

\begin{figure}%[h!]
    \centering
    \includegraphics[width=0.49\textwidth]{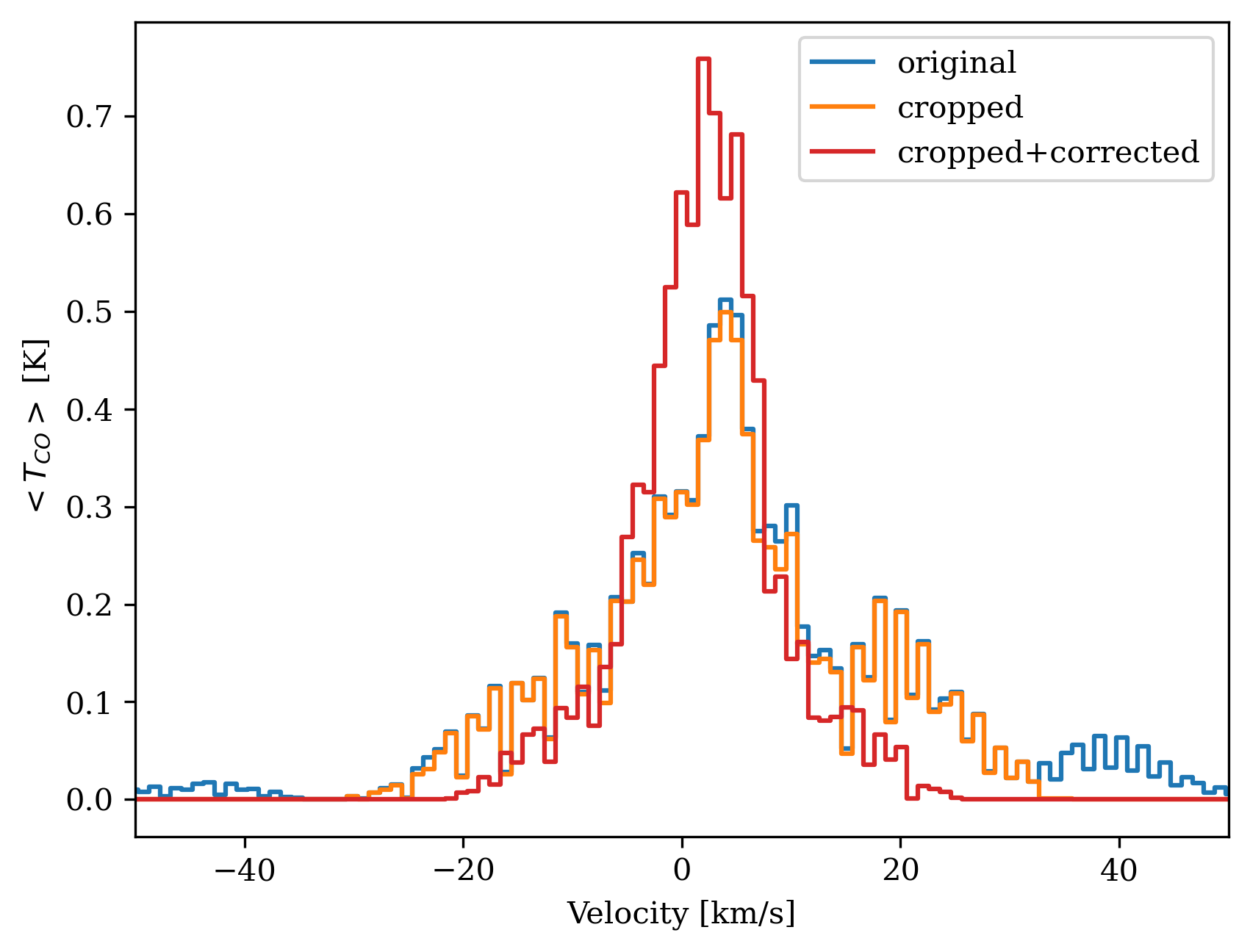}
    \caption{Total CO spectrum of the molecular clouds within a distance of 2 kpc. The original spectrum before any corrections is shown in blue, the spectrum after the cropping of the clouds is shown in orange, and the final spectrum with cropped spectra and subtracted disk velocity is shown in red. The same spectra in mass units are shown in Fig. \ref{fig:MT_tot}.
    }
    \label{fig:spect_tot}
\end{figure}

After this correction, the cloud spectra represent the motions of the clouds independent of the galaxy rotation. These motions include the internal movement of gas within the clouds and any peculiar motion the cloud might have in the disk, that is, motion that does not perfectly follow the disk rotation. When the spectra of all the clouds in our $R=2$ kpc survey area were combined, we obtained an overview of the total CO spectrum of the molecular clouds in our local Galactic environment. This total spectrum is shown in Fig. \ref{fig:spect_tot} before and after the corrections (velocity cropping and subtraction of the Galactic rotation). 

In this Bird's Eye Experiment, we obtained the total CO spectra within the apertures of different sizes by summing over the spectra of the clouds within each aperture. We summed the spectra using the cloud mass, $\mathrm{M_{CO}}$, as the weight so that the  contribution of a cloud in the aperture spectrum was proportional to its mass, just as it would be if all the clouds were at the same distance (e.g. when viewed face-on in another galaxy). In this calculation, the clouds are considered point sources, that is, the cloud coordinates alone determine whether the entire spectrum of a cloud is included in the aperture. 

The above procedure gives us access to the velocity dispersions (again as the square root of the second moment of the spectra), to the full width at half maxima (FHWM, different from $2\sqrt{2\ln2}\sigma$ because the spectra are not exactly Gaussians), and to the masses of the CO gas within the apertures. We constructed maps of the aperture properties, sampling the entire 2 kpc radius survey area. This was performed for apertures with radii of 1, 0.5, and 0.25 kpc to study the scale and environment dependence of the aperture properties. The apertures were placed at Nyquist sampling intervals, as in Paper I.

%%%%%%%%%%%%%%%%%%%%%%%%%%%%%%%%%
%%%%%%%%%%%%%%%%%%%%%%%%%%%%%%%%%
\section{Results}
\label{sec:results}
%%%%%%%%%%%%%%%%%%%%%%%%%%%%%%%%%
%%%%%%%%%%%%%%%%%%%%%%%%%%%%%%%%%

This section is divided into two parts: We first describe the kinematics for the most complete sample of molecular clouds within 2 kpc (Sect. \ref{sec:r_ind}) and then describe how the data appear when viewed from a bird's eye perspective (Sect. \ref{sec:r_aps}). In both cases, we study the velocity dispersions and masses, and their scale dependences (the linewidth-size and mass-size relations).

%%%%%%%%%%%%%%%%%%%%%%%%%%%%%%%%%%%%%%%%%%%
\subsection{Kinematics of the individual clouds}
\label{sec:r_ind}
%%%%%%%%%%%%%%%%%%%%%%%%%%%%%%%%%%%%%%%%%%%

\begin{figure}%[h!]
    \centering
    \includegraphics[width=0.45\textwidth]{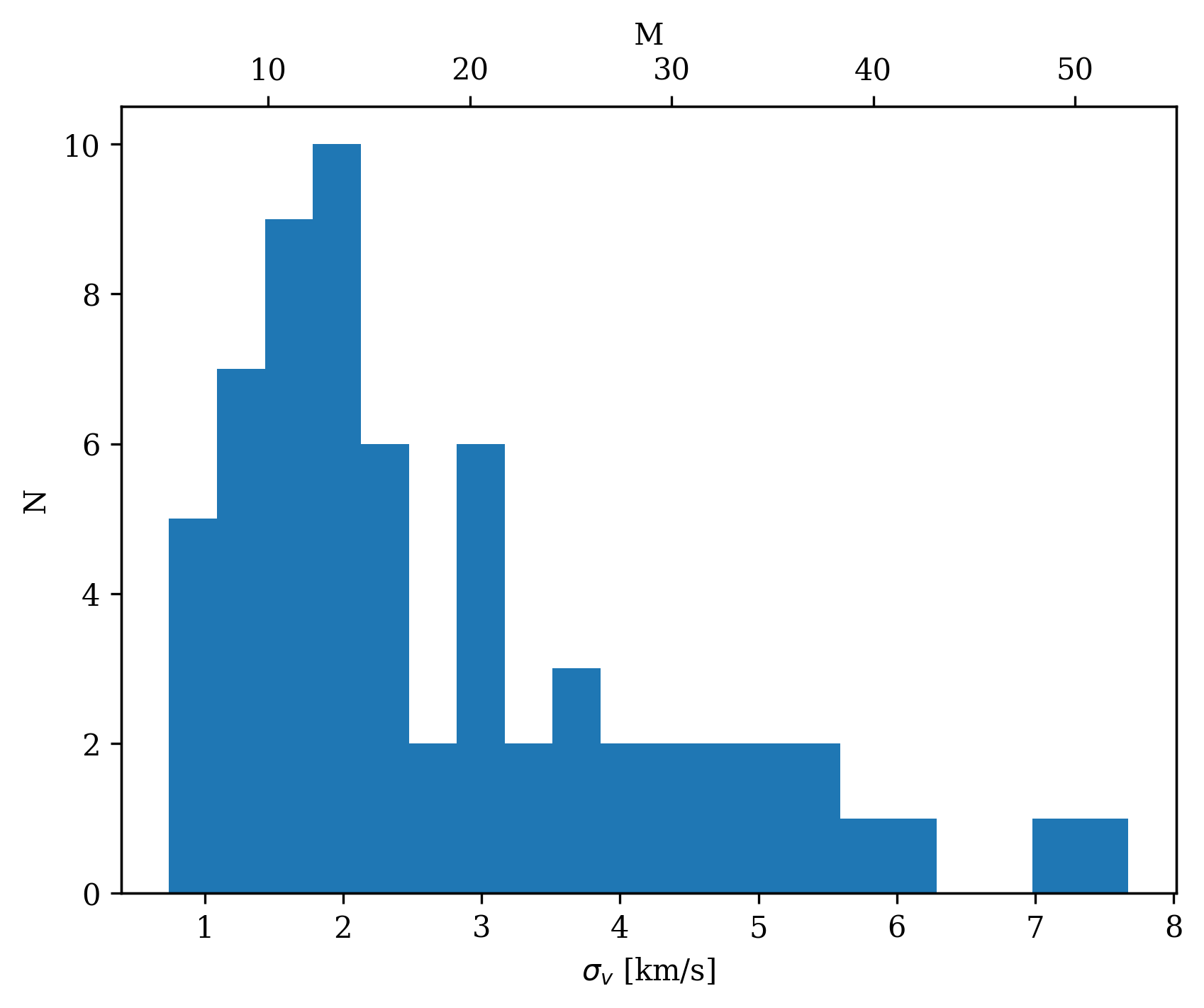}
    \caption{Histogram of the velocity dispersions of individual clouds. The approximate Mach number is shown on the top x-axis. 
    }
    \label{fig:sigmahist}
\end{figure}

We first study the kinematics of the cloud sample. The distribution of the velocity dispersions of the clouds and the corresponding Mach numbers are shown in Fig. \ref{fig:sigmahist}. The mean of the distribution is 2.75 km/s ($\mathcal{M}=18.9$) and the median is 2.22 km/s ($\mathcal{M}=15.3$). The interquartile range is 1.4-3.5 km/s, and the distribution has a tail to high velocity dispersions. The distribution is similar to that for the whole Milky Way \citep{miville2017physical}. The clouds with $\sigma_\mathrm{v} > 5$ km s$^{-1}$ are L1340-55, Ara, Lagoon, M20, M17, and w3-w4-w5. Lagoon, M20, and M17 are also outliers in terms of their density distributions (Paper I), and are massive clouds towards the Galactic centre. 

\iffalse
\begin{figure}%[h!]
    \centering
    \includegraphics[width=0.45\textwidth]{figs/singleclouds/size-linewidth.png}
    \caption{Linewidth-size relation for the clouds in the sample. Symbol size is proportional to the cloud mass, and the blue solid line shows a fit to the data. The dotted grey line shows the fit from \cite{larson1981turbulence}.}
    \label{fig:sizeline}
\end{figure}

\begin{figure}%[h!]
    \centering
    \includegraphics[width=0.45\textwidth]{figs/singleclouds/mass-radius.png}
    \caption{Mass-size relation for the clouds in the sample. Symbol size is proportional to the cloud mass, and the blue solid line shows a fit to the data. The dotted grey line shows the relation from \cite{larson1981turbulence}.}
    \label{fig:massrad}
\end{figure}
\fi

\begin{figure}%[h!]
    \centering
    \includegraphics[width=0.49\textwidth]{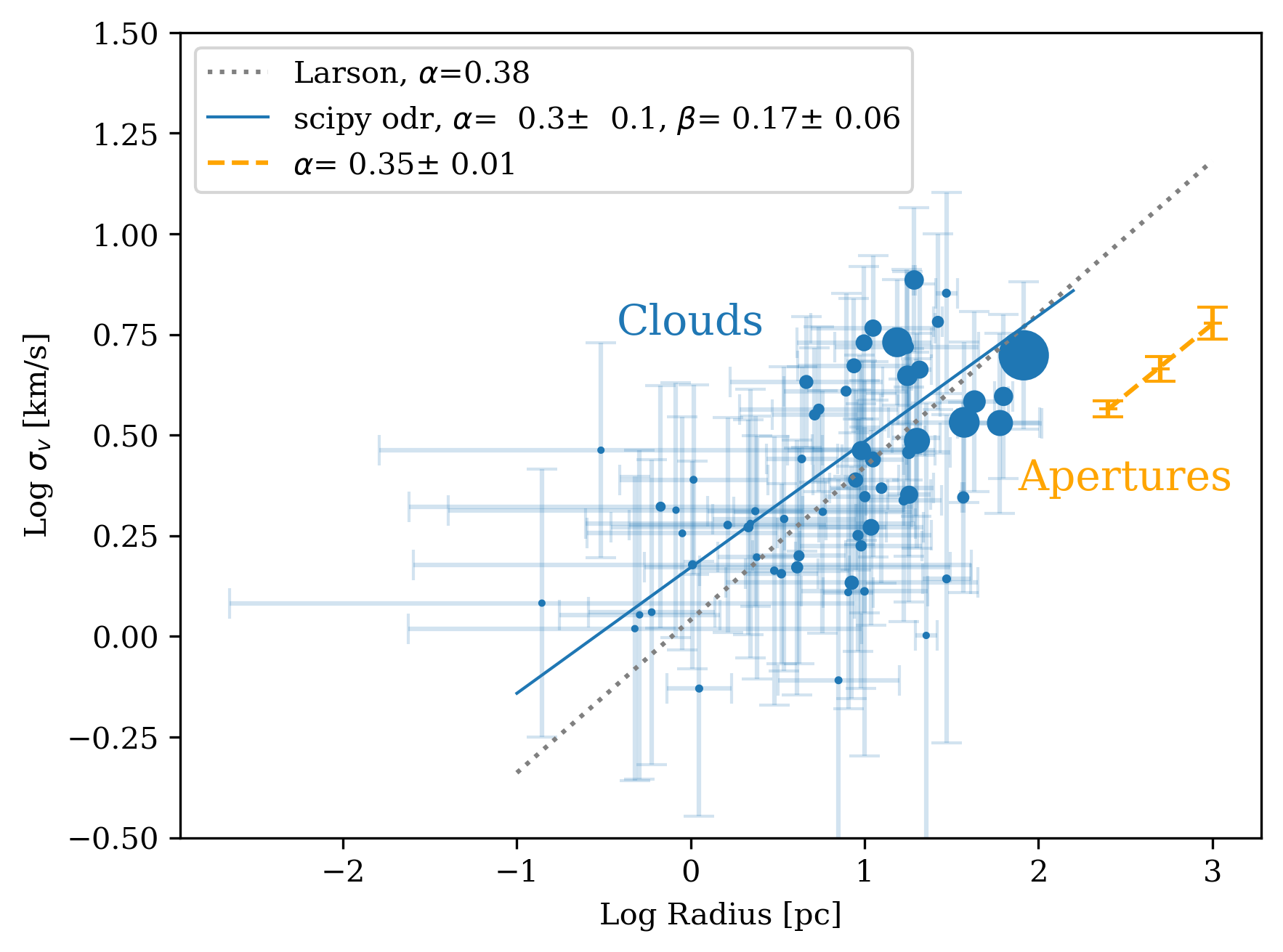}
    \includegraphics[width=0.46\textwidth]{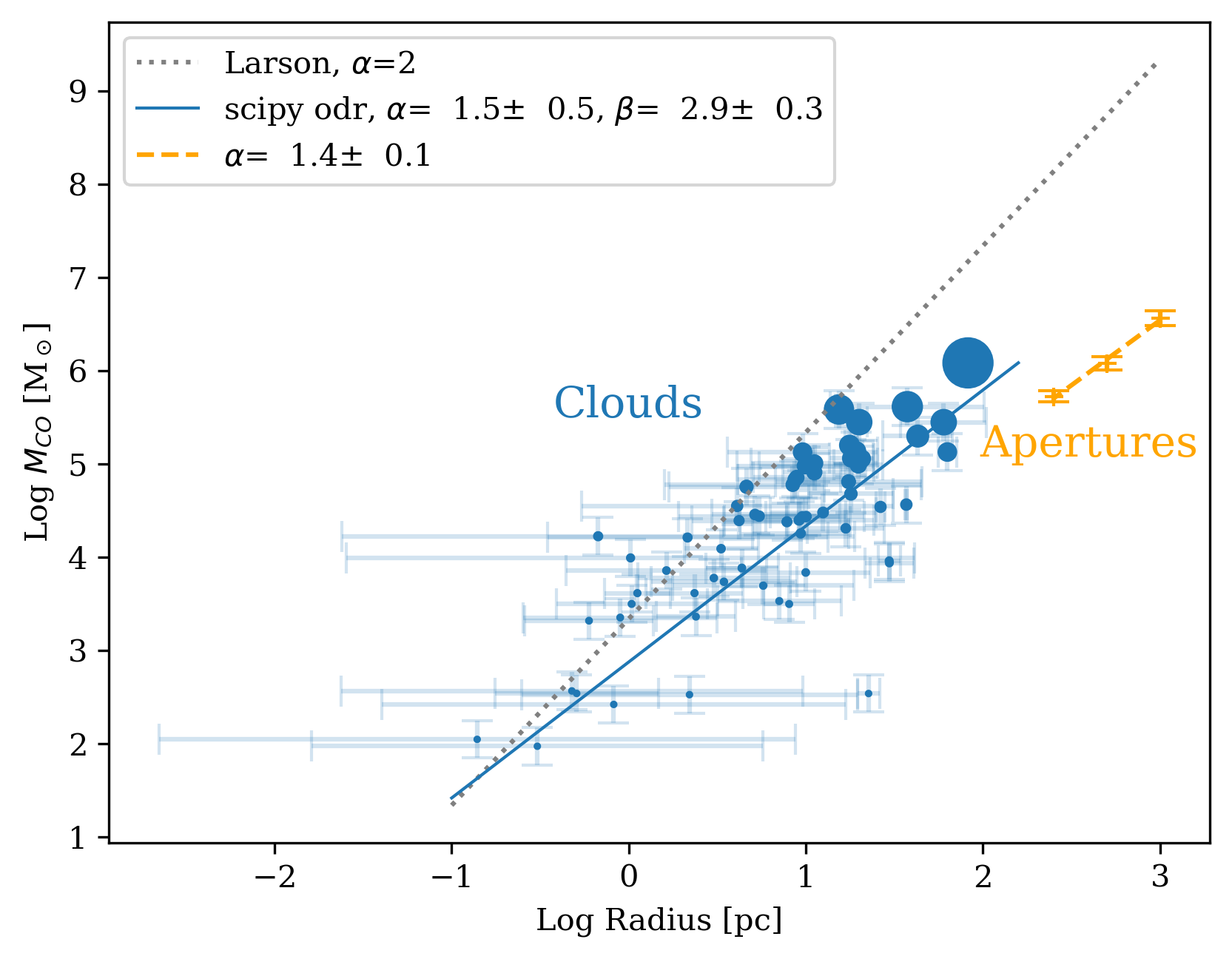}
    \caption{Linewidth-size (top) and mass-size (bottom) relations for the clouds and apertures. Clouds are shown in blue, and the symbol size is proportional to the cloud mass. The error bars are as described in Sect. \ref{sec:m_ind}. Apertures are shown in orange, with error bars indicating the standard deviation between the apertures. The solid blue line shows a fit to the cloud data, the dashed orange line shows a fit to the aperture data, and the dotted grey line shows the fit from \cite{larson1981turbulence}.}
    \label{fig:larson}
\end{figure}

\begin{figure}%[h!]
\centering
\includegraphics[width=0.45\textwidth]{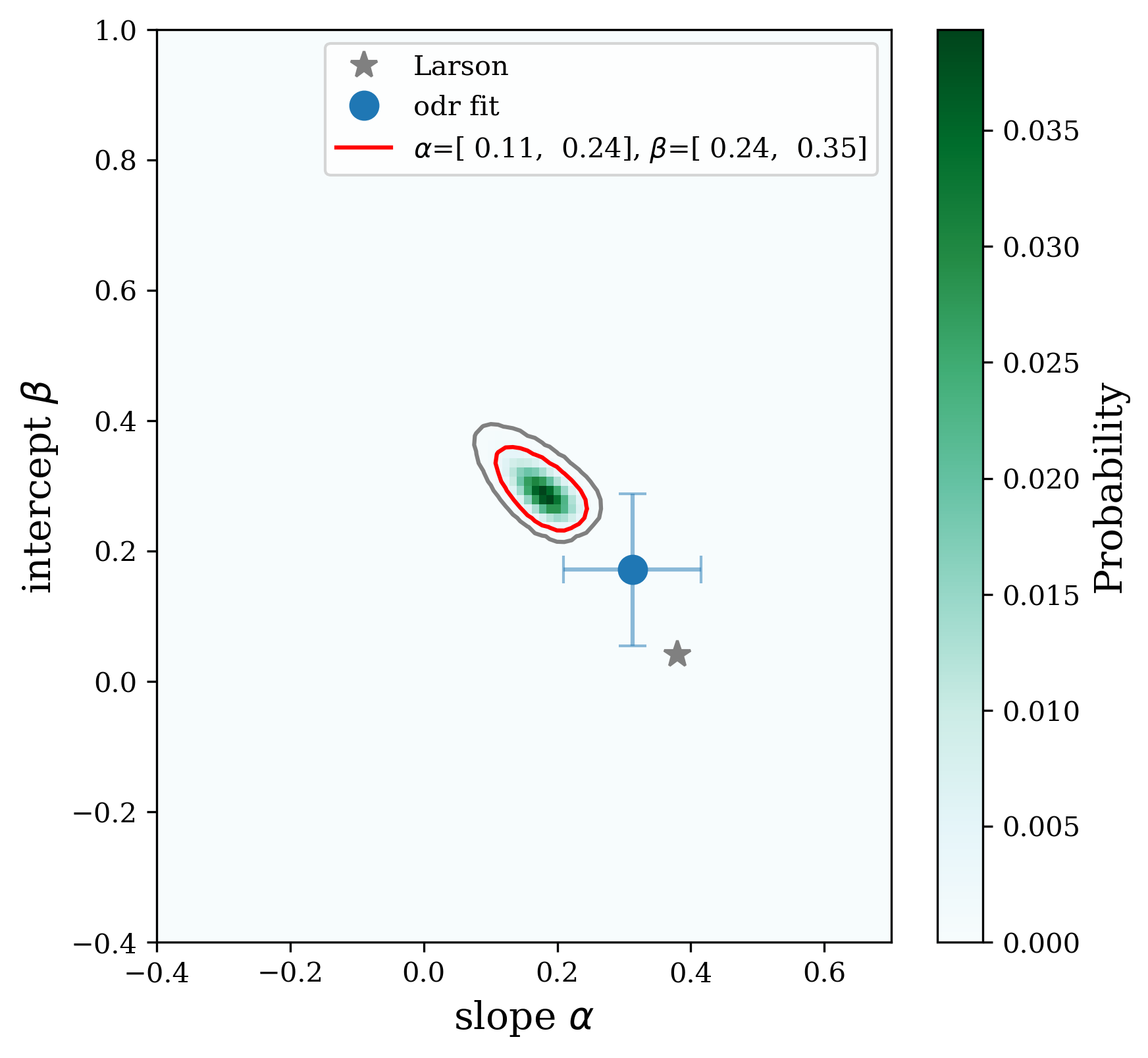}
\includegraphics[width=0.44\textwidth]{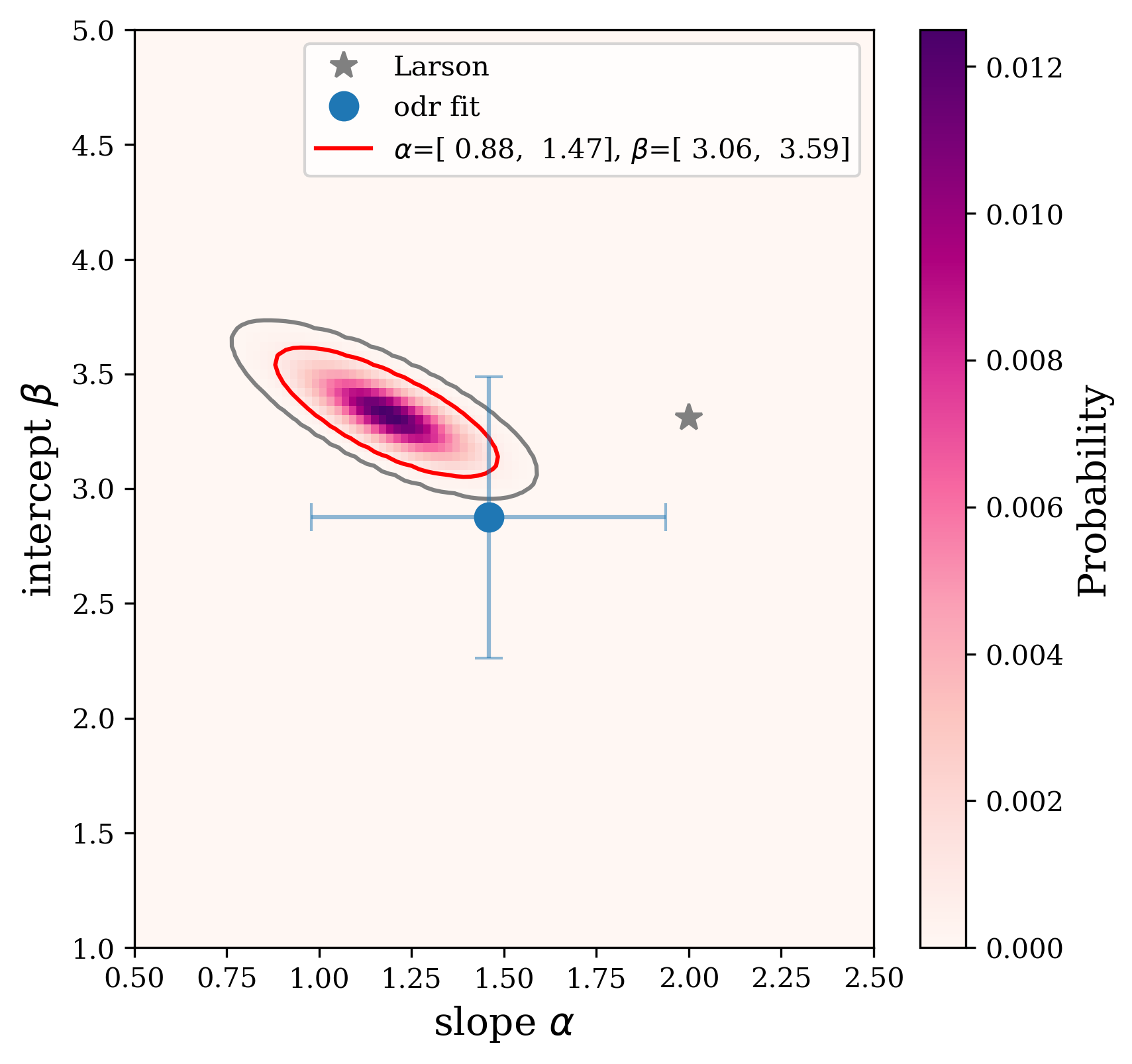}
    \caption{Probability maps in gradient-intercept space for linewidth-size relation (top) and mass-size relation (bottom), assuming a probability of one that the parameters are within the ranges of the plot. The original \cite{larson1981turbulence} fits are denoted by a grey star, and the fits from Fig. \ref{fig:larson} are denoted by a blue circle. The 95\% confidence intervals are encircled in red, and the  99.7\% confidence intervals are presented in grey.}
    \label{fig:chi2_larson}
\end{figure}

The linewidth-size relation for the sample is shown in Fig. \ref{fig:larson} (top panel). There is a correlation in the data (Spearman's rank coefficient $R_s=0.53$ and Pearson coefficient $p=9\cdot10^{-6}$), but the scatter is large. The fitted linewidth-size relation is $\sigma_v=1.5\cdot R^{0.3\pm0.1}$. This is consistent with the original \cite{larson1981turbulence} relation and other works \citep[e.g.][]{miville2017physical}. Figure \ref{fig:larson} (bottom panel) shows the mass-size relation for the cloud sample. The correlation is stronger ($R_s=0.77$ and $p=2\cdot10^{-13}$), yielding the relation $\mathrm{M_{CO}}= 794 \cdot R^{1.5\pm0.5}$. The exponent of $\sim$2 from \cite{larson1981turbulence} implies a constant column density of molecular clouds, and within 2$\sigma$ error, our data are consistent with this result. 

Figure \ref{fig:chi2_larson} shows probability maps for different values of the slope and intercept for the linewidth-size and mass-size relationships. These probabilities were calculated from $P\propto e^{-\chi^2/2}$, where $\chi^2$ is the goodness-of-fit parameter calculated without uncertainties. The 95\% confidence intervals given here do not quite coincide with the ODR fits from Fig. \ref{fig:larson}, and we interpret this as being due to the relatively large uncertainties on the measurements. This then results in a wide range of possible values for the intercepts and slopes of the linewidth-size and mass-size relations. 

%%%%%%%%%%%%%%%%%%%%%%%%%%%%%%%%%%%%%%%%%%%%%%%%
\subsection{Kinematics from the bird's eye perspective}
\label{sec:r_aps}
%%%%%%%%%%%%%%%%%%%%%%%%%%%%%%%%%%%%%%%%%%%%%%%%

\setcounter{table}{1}
\begin{table*}
\caption{Properties of the aperture spectra.} % title of Table
\label{tab:aps}      % is used to refer this table in the text
\centering                          % used for centering table
\begin{tabular}{c c c c c c c c}        % centered columns (4 columns)
\hline\hline                 % inserts double horizontal lines
R  & N\tablefootmark{a} & $\mathrm{M_{CO}}$ & FWHM & $\sigma_\mathrm{v}$ & $<v>$ & $\sigma_{\mathrm{v}sun}$\tablefootmark{b} & $<v>_{sun}$\tablefootmark{b} \\ %& $\alpha_{vir}$ \\   
\hline
kpc & & [log M$_\odot$] & [km/s] & [km/s] & [km/s] & [km/s] & [km/s] \\
\hline      
2.00\tablefootmark{c} & 1 & 6.7 & 10.1 & 6.8 & 1.2 & 6.8 & 1.2 \\ 
1.00    &       12      &       6.6     $\pm$   0.1     &   10.6$\pm$   1.5 &   6.0 $\pm$   0.5     &       0.5     $\pm$   0.9     &       4.2     &       1.8  \\ 
0.50    &       39      &       6.1     $\pm$   0.1 &   9.5     $\pm$   0.8 &   4.6 $\pm$   0.3     &       1.2     $\pm$   0.9     &       3.8     &       2.6      \\ 
0.25    &       87      &       5.7     $\pm$   0.1 &   7.8     $\pm$   0.6 &   3.7 $\pm$   0.2     &       1.3     $\pm$   0.6     &       2.8     &       4.2      \\ 
\hline                                   %inserts single line
\end{tabular}
\tablefoot{\\
\tablefoottext{a}{Number of apertures considered.}
\tablefoottext{b}{Values for Sun-centred apertures.}
\tablefoottext{c}{This is the full survey area.}
}
\end{table*}

We now study the same cloud sample as it would look if seen from outside the Milky Way in a bird's eye view through apertures with radii ranging from 0.25 to 2 kpc. We study the properties of the aperture spectra and their dependence on scale and positioning within the Galactic environment. 

% 1: Total spectra

\begin{figure}%[h!]
    \centering
    \includegraphics[width=0.49\textwidth]{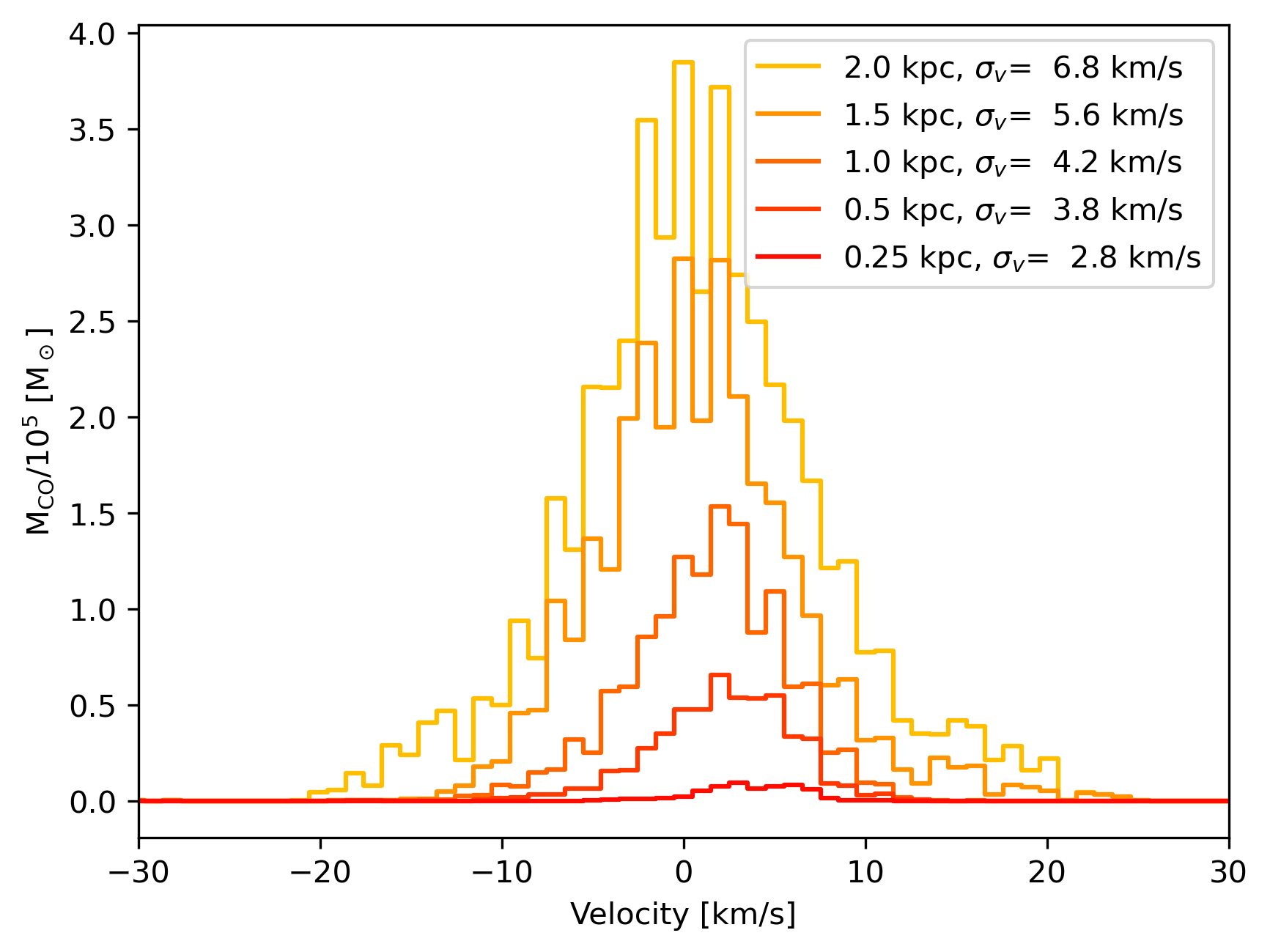} %apertures/Vel_in_r2000_newcuts_d2.png}
    \caption{Total CO spectra, transformed into units of solar masses, for Sun-centred apertures of various sizes. The radius of the apertures and the velocity dispersion of the spectra are given in the legend.}
    \label{fig:aps_sun}
\end{figure}

\begin{figure}%[h!]
    \centering
    \includegraphics[width=0.49\textwidth]{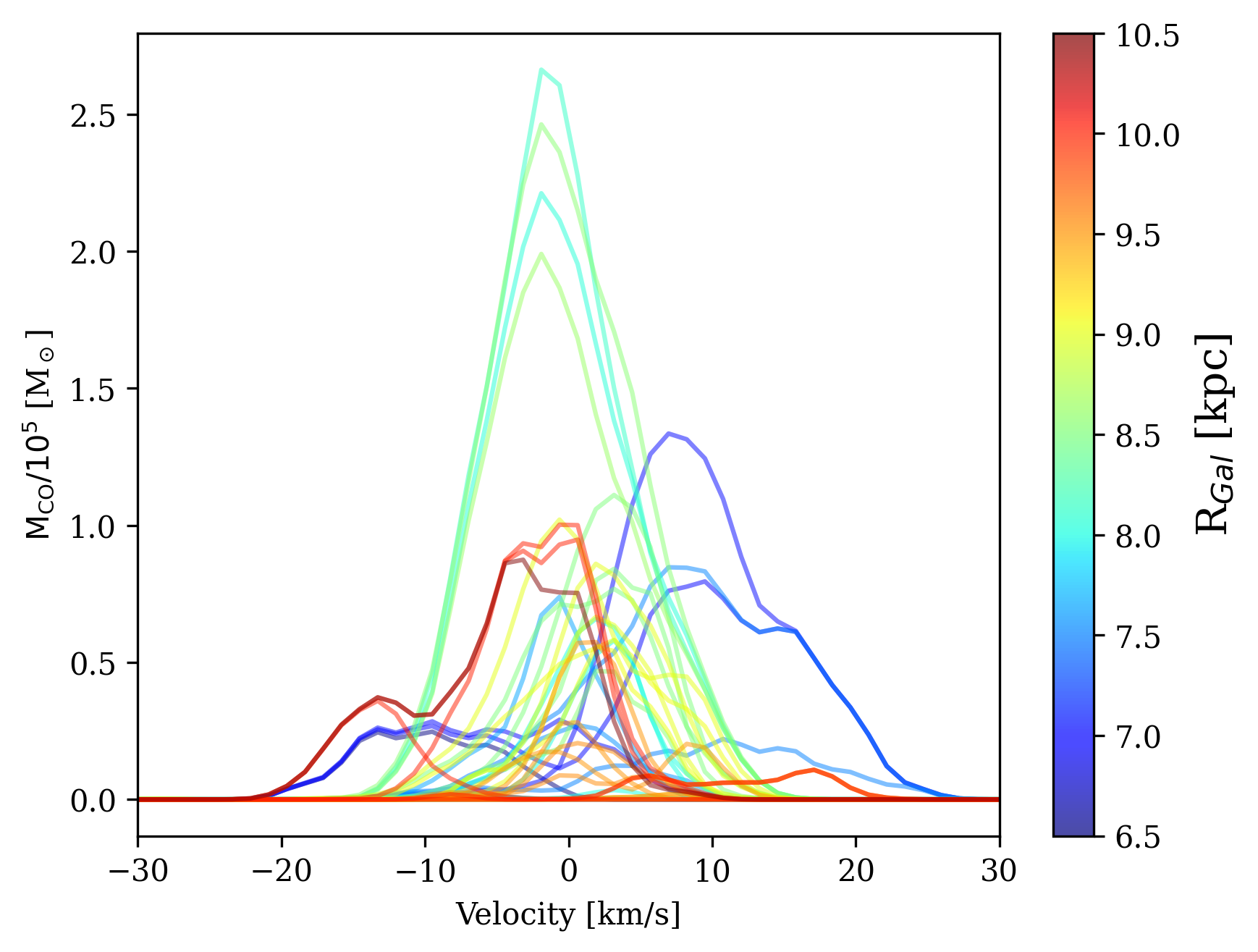}
    \caption{Spectra with apertures of $R=0.5$ kpc covering the whole survey region, showing the variety of shapes between apertures at the same scale. For all apertures of all sizes and methods of smoothing, see Appendix \ref{app:app_diffpos}.}
    \label{fig:diffpos5}
\end{figure}

Table \ref{tab:aps} lists the derived properties for the aperture spectra. As examples of the resulting spectra, Fig. \ref{fig:aps_sun} shows the spectra for Sun-centred apertures. We show the summed spectra expressed in mass units, so that the contribution is largest for the most massive clouds rather than for the most nearby clouds (for temperature units, see Appendix \ref{app:MT}, Fig.\ref{fig:MT_sun}). Fig. \ref{fig:aps_sun} shows that the total spectra clearly widen for larger apertures, and the largest apertures in particular have non-Gaussian wings in the spectrum. To illustrate the diversity of spectra for a given aperture size, Fig. \ref{fig:diffpos5} shows all spectra for the apertures with radius of 0.5 kpc (see Fig. \ref{fig:app_diffpos} for the same figure for all aperture sizes). It shows that the variety of spectra at a given scale is large, both in the centroid velocity and in the exact shape of the spectrum. This variety seems to cause the non-Gaussian wings in the aperture spectra.

% 2: Scale dependencies

\begin{figure}%[h!]
    \centering
    \includegraphics[width=0.49\textwidth]{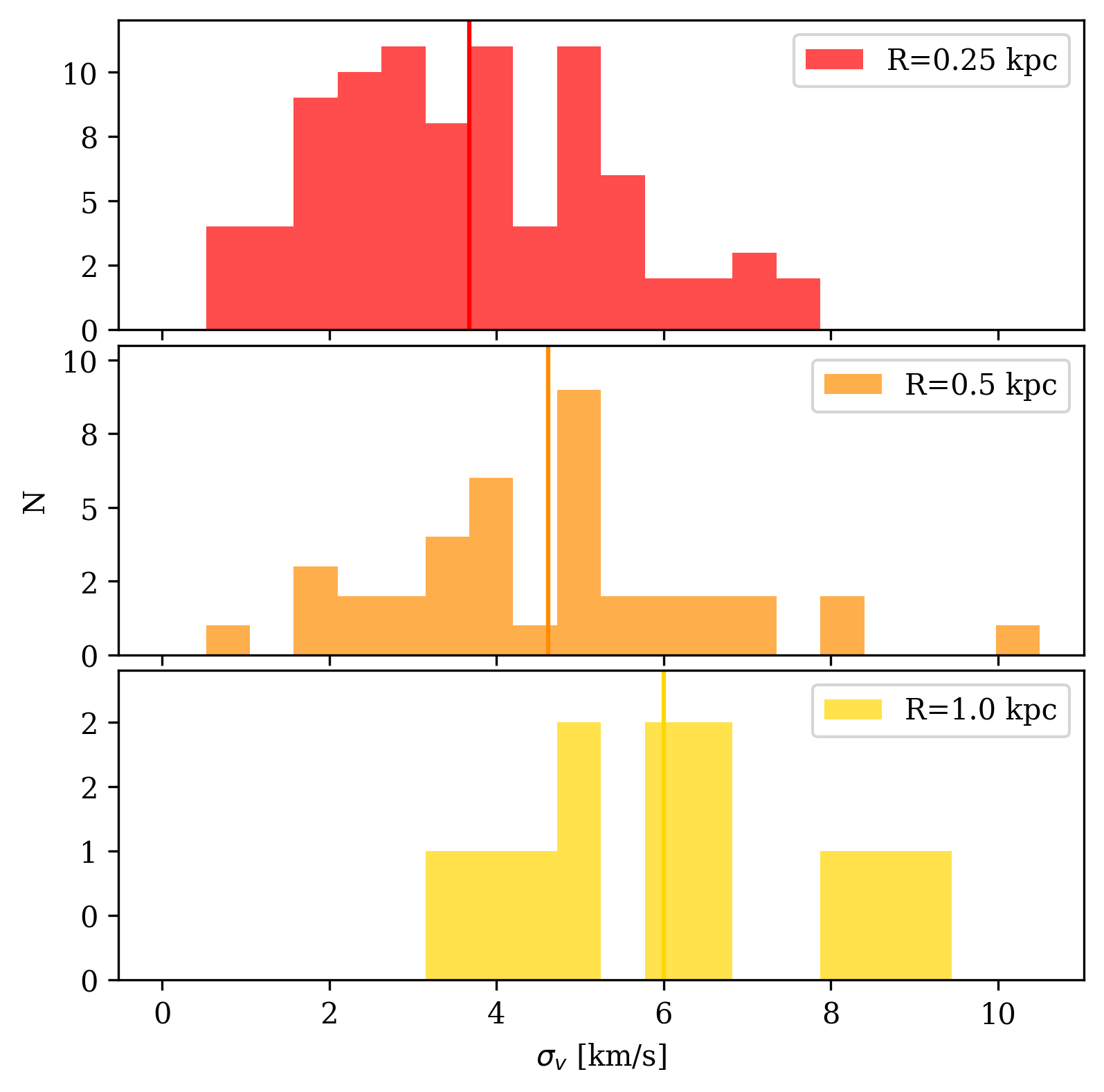}
    \caption{Histograms of the standard deviations of the total CO spectra of the apertures. The means are denoted by the vertical lines (the values are shown in Table \ref{tab:aps}).}
    \label{fig:histaps}
\end{figure}

The results also show that the velocity dispersion, measured by both $\sigma_\mathrm{v}$ and FWHM, depends on the aperture radius. This scale dependence is  illustrated in Fig. \ref{fig:larson} (top panel), which shows the Larson relations, and Fig. \ref{fig:histaps}, which shows the histograms of the velocity standard deviations. A linear fit to the linewidth-size relation for the apertures yields $\sigma_v=0.5\cdot R^{0.35\pm0.01}$. We note that due to the wide range of possible slopes for the clouds (cf. Fig. \ref{fig:chi2_larson}), the slope obtained for the apertures agrees with the slope obtained for the clouds. A linear fit to the mass-size relation for the apertures yields $\mathrm{M_{CO}}= 229 \cdot R^{1.4\pm0.1}$, where the slope is consistent with the fit for the clouds. 

% 3: Environment

\begin{figure*}
    \centering
        \includegraphics[width=0.32\textwidth]{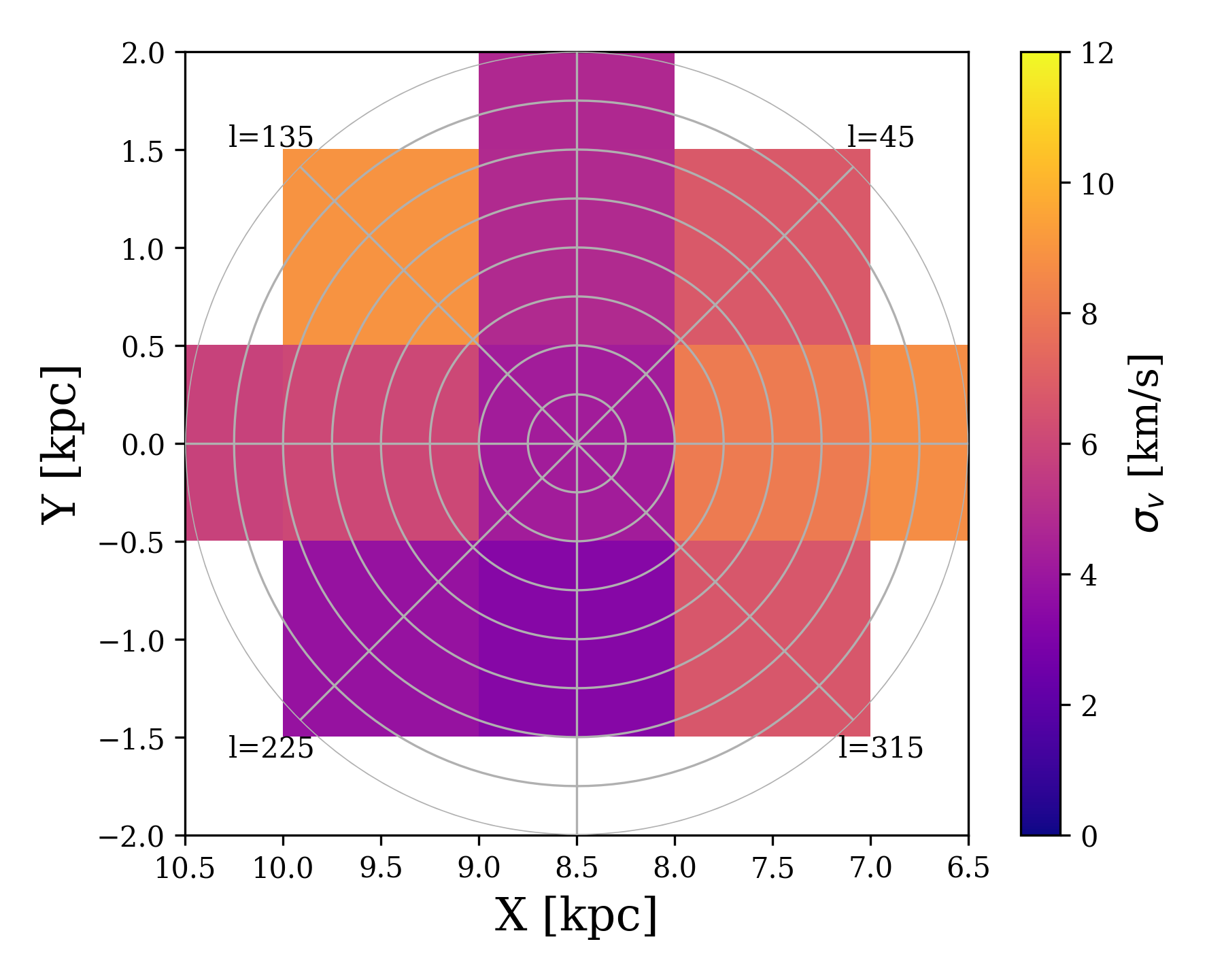}
        \includegraphics[width=0.32\textwidth]{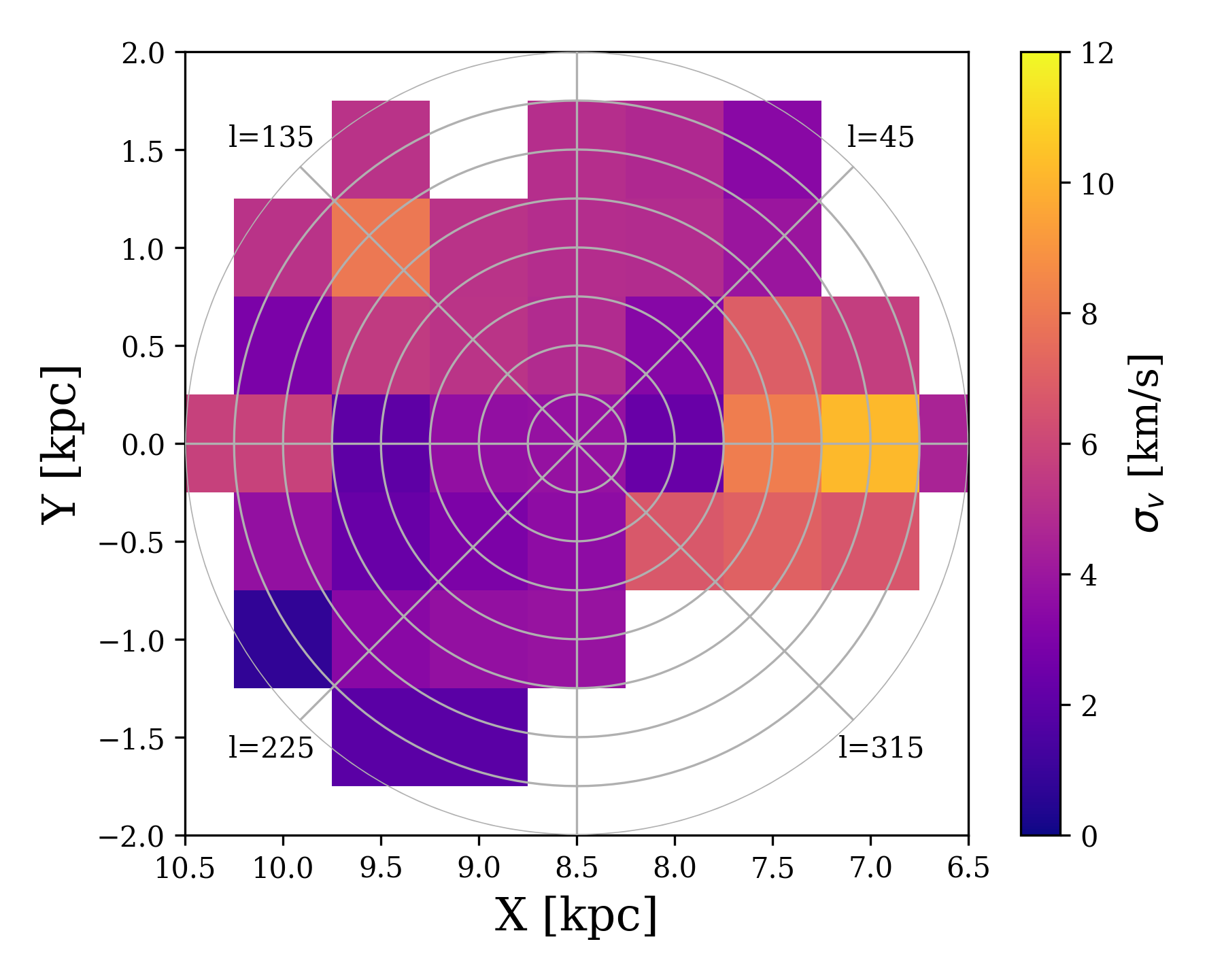}
        \includegraphics[width=0.32\textwidth]{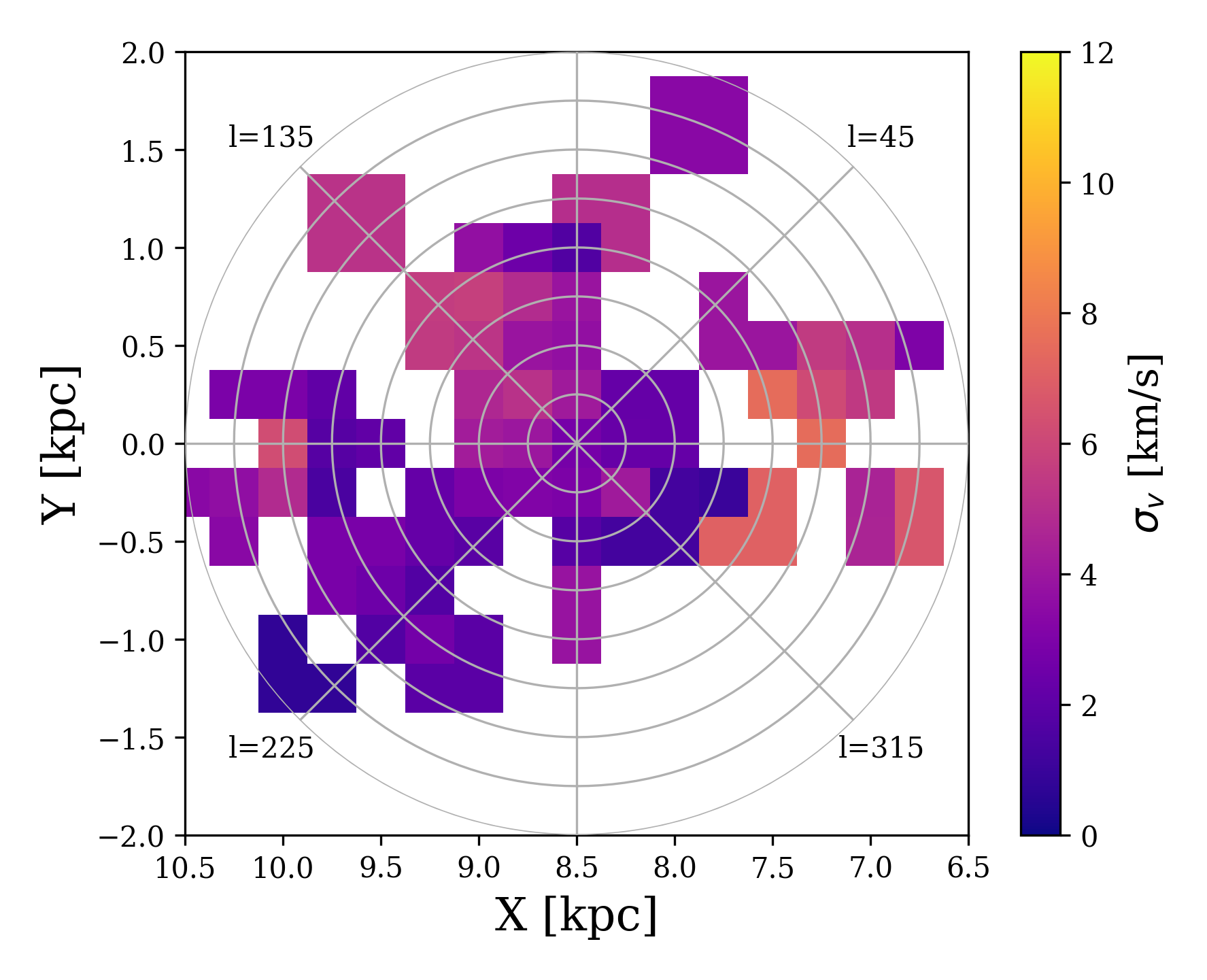}
        
        \includegraphics[width=0.32\textwidth]{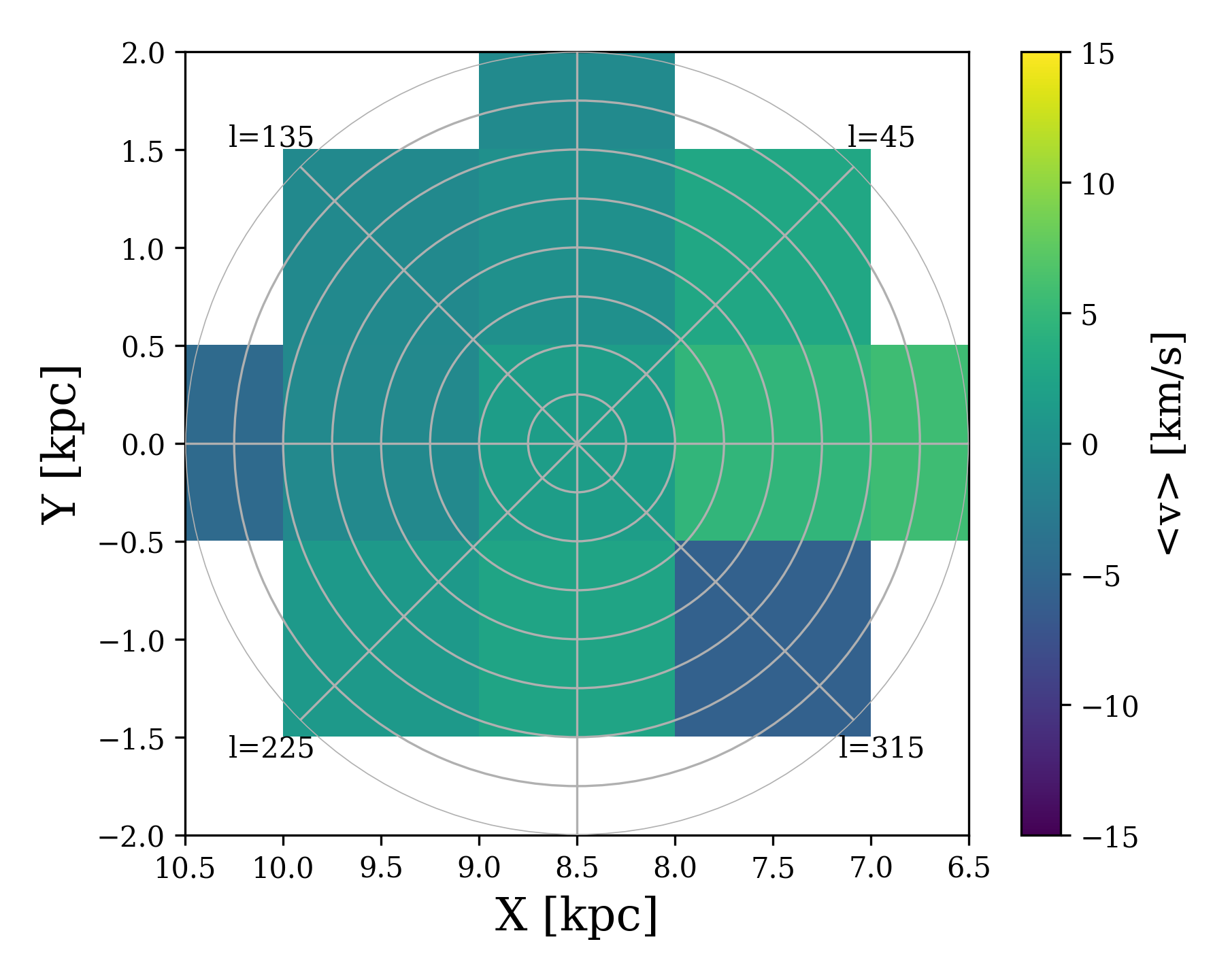}
        \includegraphics[width=0.32\textwidth]{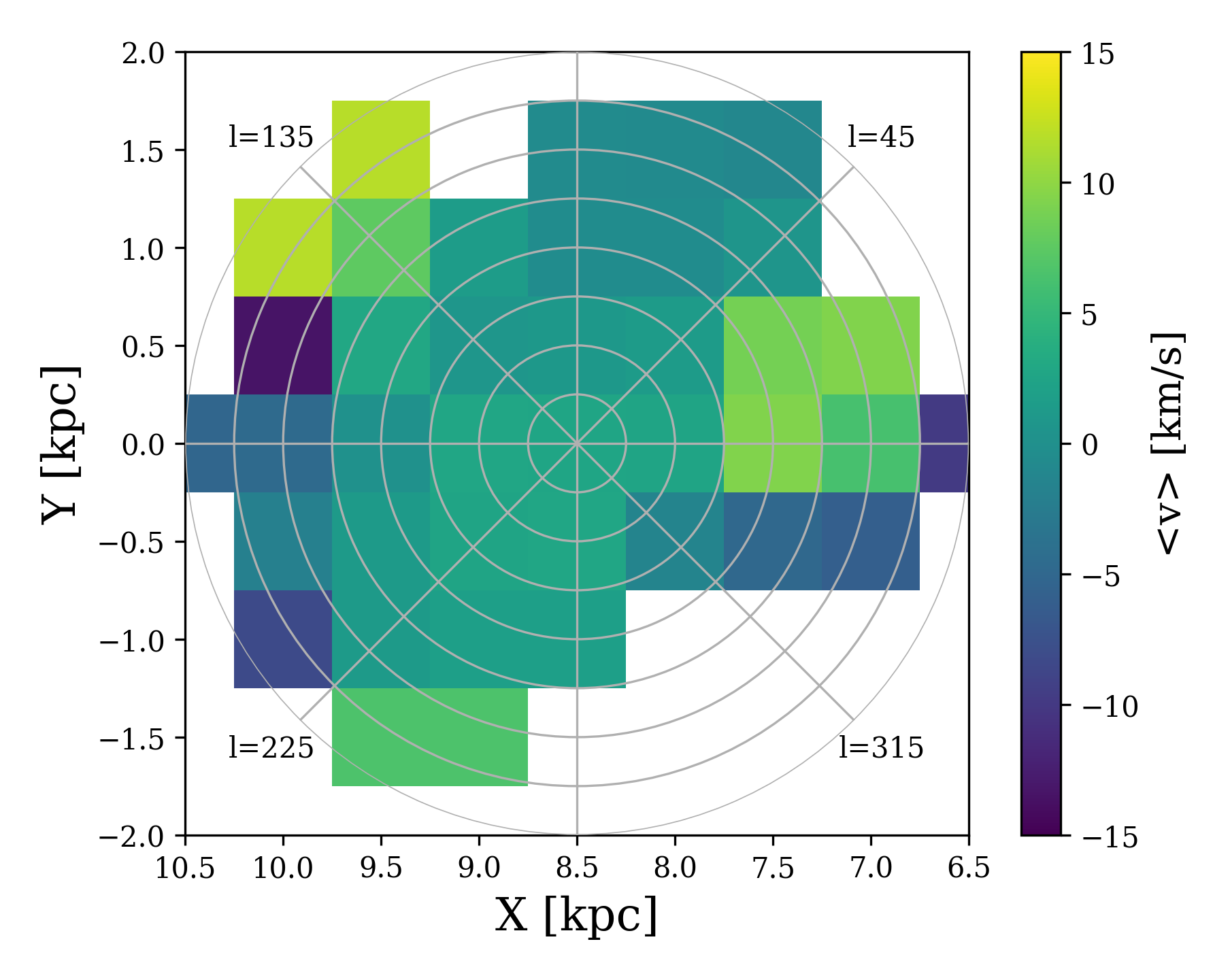}
        \includegraphics[width=0.32\textwidth]{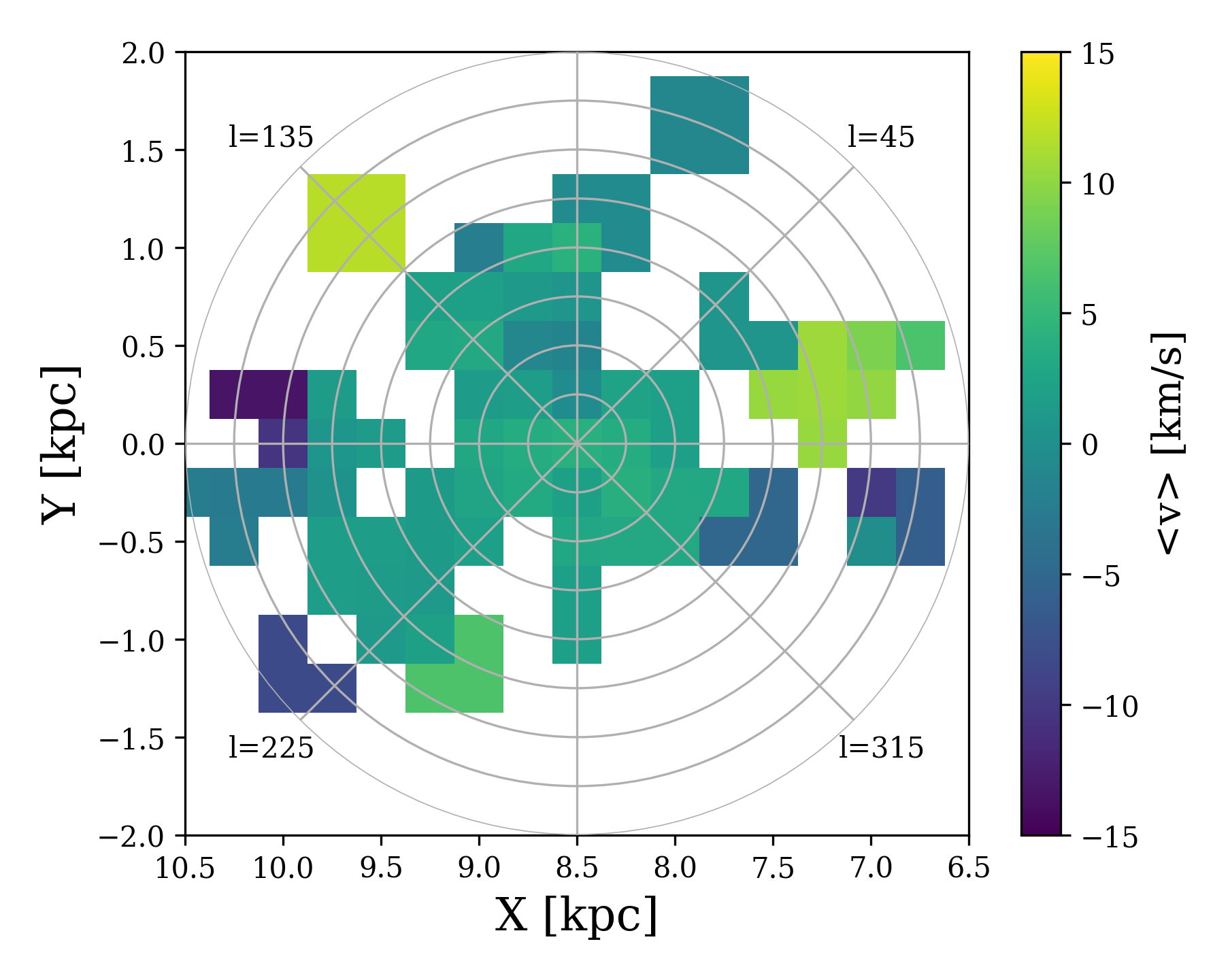}
    \caption{Characteristics of CO velocity spectra within the local Galactic environment ($<2$ kpc distance). The top row shows maps of the standard deviation ($\sigma$), and the bottom row shows the mean velocity of the clouds within the beam. The columns show the data for the apertures with $R=1$ kpc (left), $R=0.5$ kpc (middle), and $R=0.25$ kpc (right). The pixels correspond to Nyquist beams covering the sample.}
    \label{fig:maps}
\end{figure*}

We also see an indication of a trend in the velocity statistics as a function of Galactic environment (Figs. \ref{fig:diffpos5}, \ref{fig:maps}, \ref{fig:app_diffpos}). Apertures closer to the Galactic centre and the Sagittarius spiral arm have slightly broader spectra and higher mean velocities. The larger velocity dispersions originate from the fact that the clouds in this area are larger; a consequence of the linewidth-size relation is that they also have higher velocity dispersion. It is unclear whether the differences in the mean velocity have a physical meaning. It is possible that due to the low number of apertures, the trend is a random fluctuation in mean velocities. It is also possible that the peculiar velocities might exhibit a trend with respect to to the spiral arms, ranging from highest velocities at the spiral arm (lowest $R_\mathrm{Gal}$ values) to the lowest velocities in the interarm region (higher $R_\mathrm{Gal}$ values).

Finally, we note that our aperture spectra, and hence all results in this section, only include the identified clouds in the survey area, and not any possible diffuse gas component that might be located between the clouds. The possible effect of a diffuse component is discussed in Sect. \ref{sec:d_diffuse}. 

%%%%%%%%%%%%%%%%%%%%%%%%%%%%%%%%%
%%%%%%%%%%%%%%%%%%%%%%%%%%%%%%%%%
\section{Discussion}
\label{sec:discussion}
%%%%%%%%%%%%%%%%%%%%%%%%%%%%%%%%%
%%%%%%%%%%%%%%%%%%%%%%%%%%%%%%%%%

%We focus the discussion of this paper on the possible effect of a diffuse component (Sect. \ref{sec:d_diffuse}) and on elaborating on the qualitative connection between this and extragalactic works (Sect. \ref{sec:d_exgal}). %\LEt{a single sentence does not constitute a paragraph. Please either add to this or remove}

%%%%%%%%%%%%%%%%%%%%%%%%%%%%%%%%%%%%%%%%%%%%%%%%%%%%%%%%%%%%%%%
\subsection{Possible effects of a diffuse gas component}
\label{sec:d_diffuse}
%%%%%%%%%%%%%%%%%%%%%%%%%%%%%%%%%%%%%%%%%%%%%%%%%%%%%%%%%%%%%%%

The molecular cloud sample in this work by construction includes the molecular gas that resides in cloud-like objects, that is, in objects that are typically separable in $p$-$p$-$v$ space. However, it has been shown in the Milky Way that a significant amount of molecular gas is also more diffuse and not necessarily distributed in such clouds \citep[e.g.][]{romanduval2016diffusegas}. If a component like this is normally present in galaxies, it would be included in any extragalactic observational apertures. However, it is not included in our bird's eye experiment. In this section, we speculate about the effect that a significant diffuse component would have on our Milky Way data and especially on the linewidth-size relation. This knowledge is necessary to understand how our results could be compared to measurements in other galaxies. 

We created a simple cartoon model that consists of gas that has a constant density and velocity dispersion to mimic a diffuse component. The basic problem is that we have little knowledge of the nature and properties of the diffuse component. Studies of external galaxies suggest that a diffuse component could be more broadly distributed vertically than the thick molecular gas disk and that its velocity dispersion is relatively large, close to that of the neutral HI gas \citep{caldu2013high,caldu2016molecular,mogotsi2016hi}. Guided by these works, we here tested velocity dispersions between 7 and 17 km/s for the diffuse component. The mass fraction of dense gas in the solar neighbourhood is $\sim 50\%$ \citep{romanduval2016diffusegas}. We used these numbers to estimate the mass of the diffuse component in the survey area and simulated the spectra for the diffuse gas. We then summed the original spectra and the diffuse component spectra for each aperture to generate simulated total spectra.

\begin{figure}%[h!]
    \centering
    \includegraphics[width=0.49\textwidth]{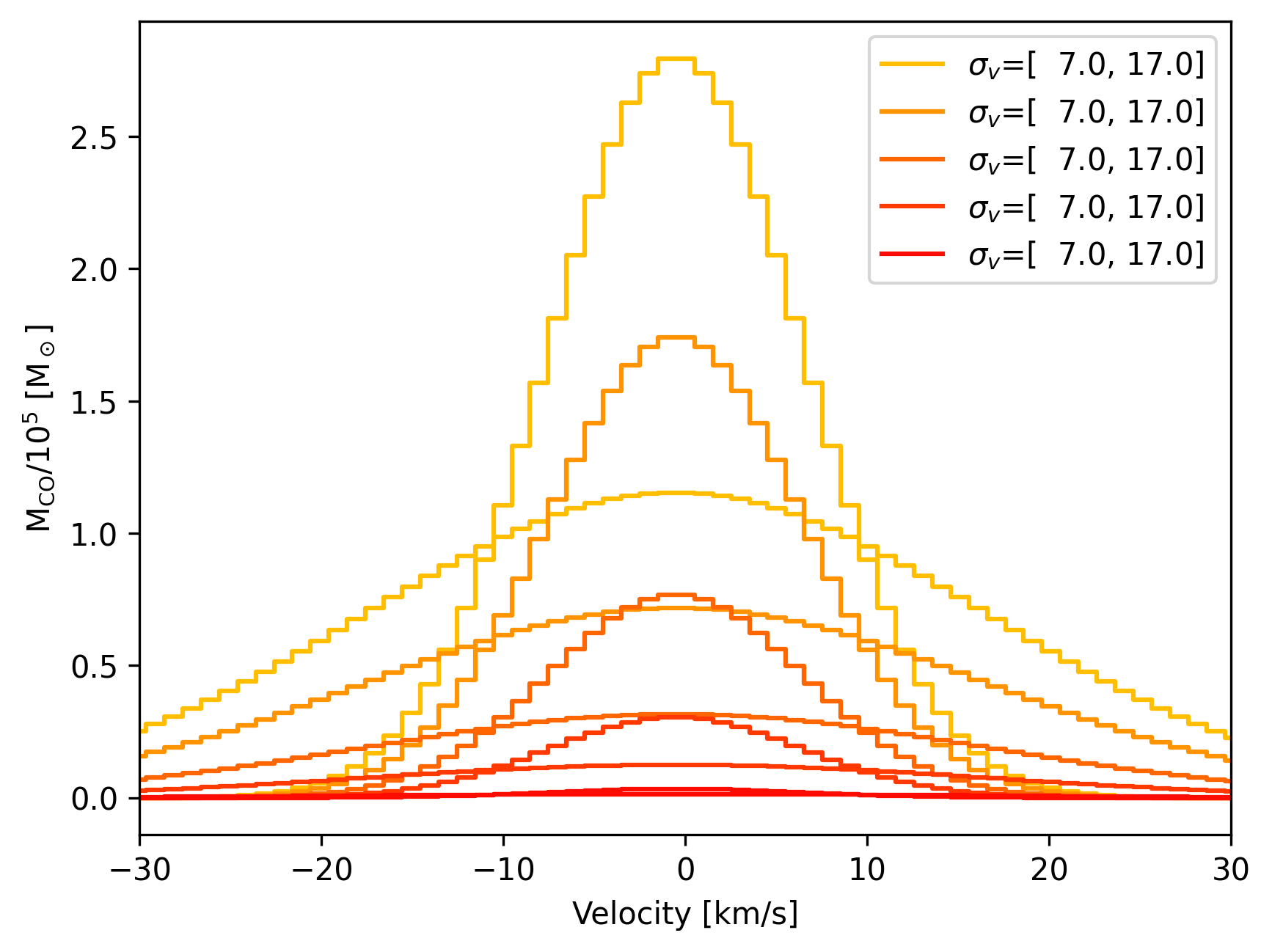}
    \includegraphics[width=0.49\textwidth]{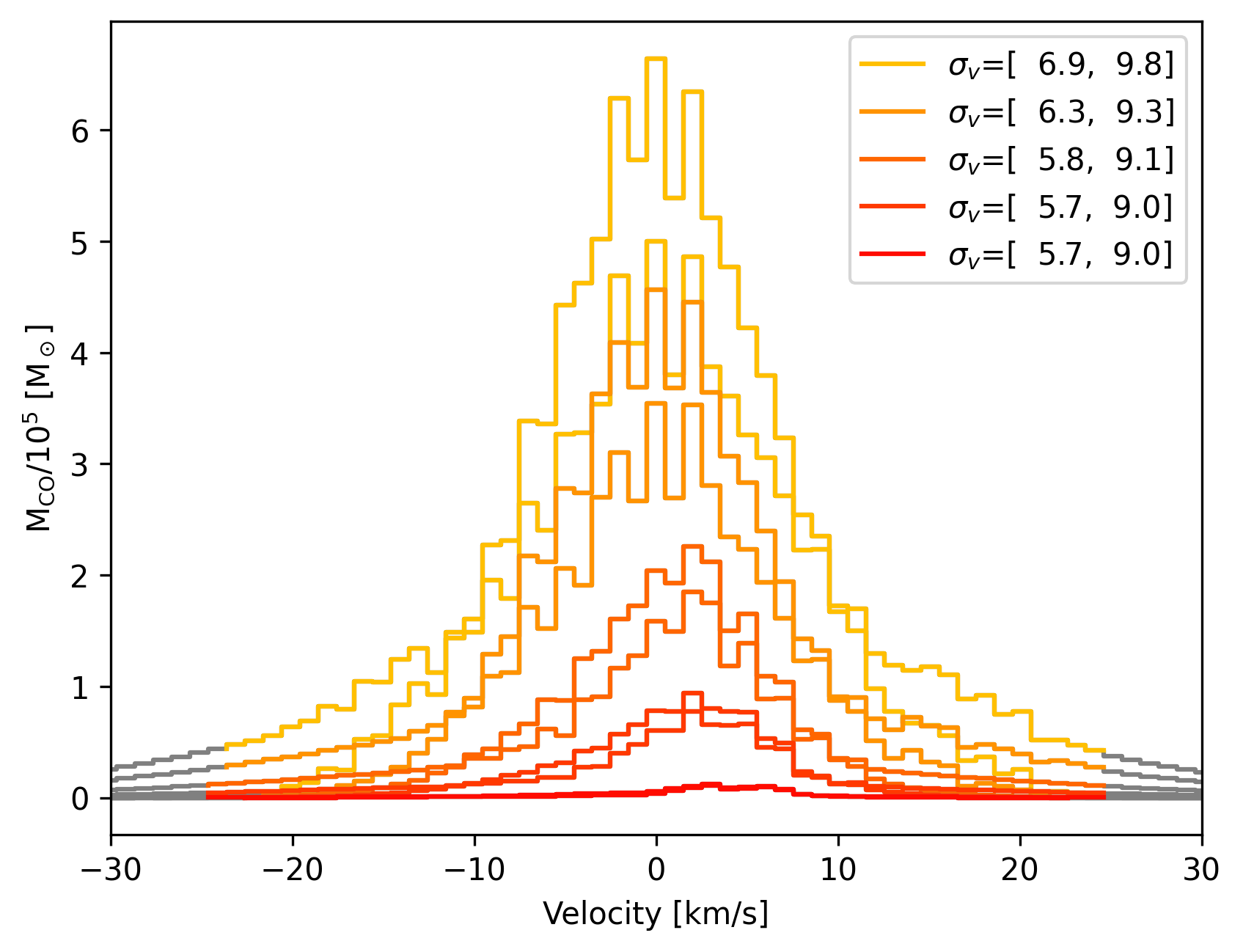}
    \includegraphics[width=0.49\textwidth]{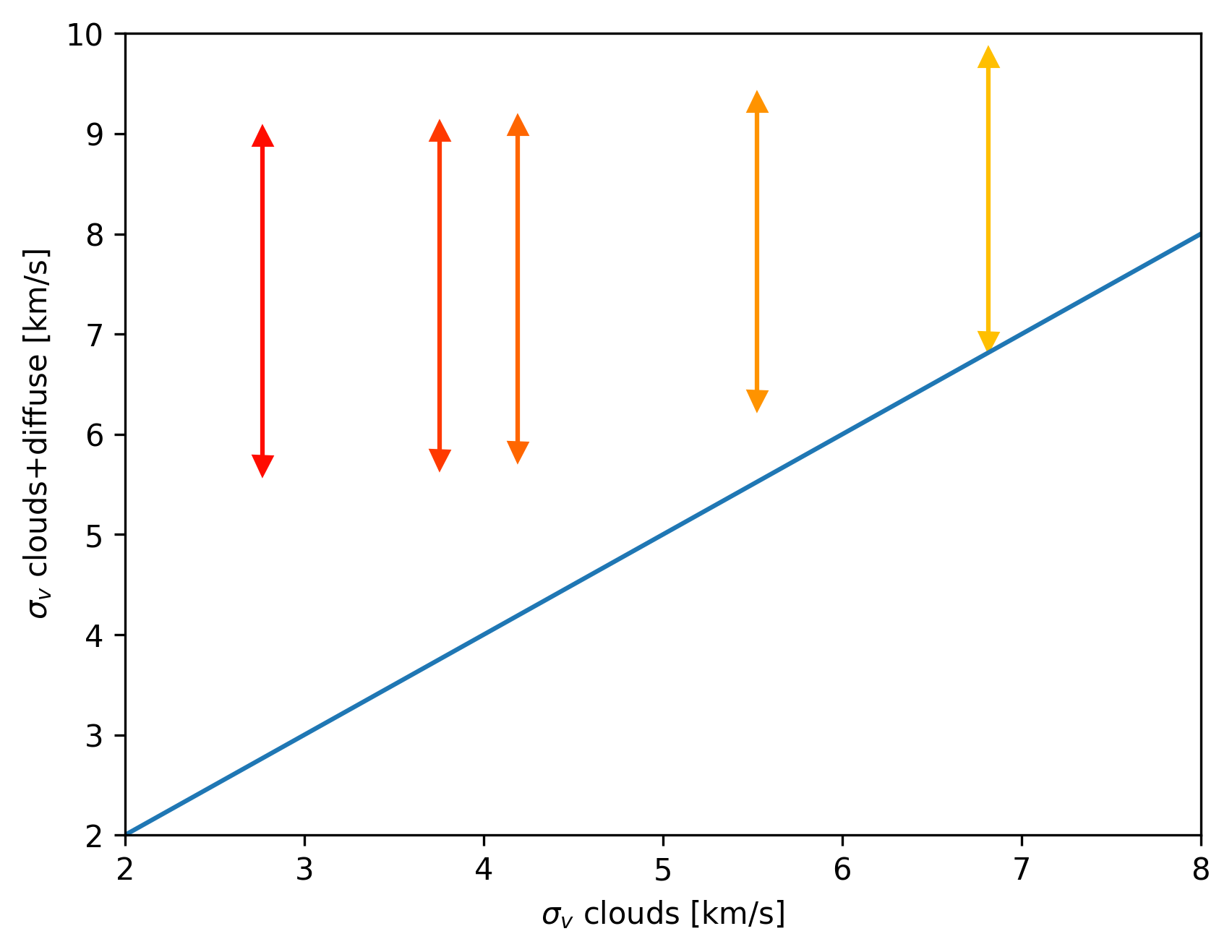}

    \caption{CO velocity distributions for Sun-centred apertures of various sizes, with standard deviations noted in the legends. \emph{Top:} Estimated spectra for the diffuse component in the survey area, with a minimum and maximum velocity dispersion. \emph{Middle:} Combination of the two, how they would look if observed together. The grey lines show what we estimate as the observation threshold (see main text), and we estimate the velocity standard deviations above this baseline. \emph{Bottom:} Standard deviations of the apertures containing the clouds alone vs the estimated total spectra. The blue line corresponds to a one-to-one relation.
    }
    \label{fig:diffuse}
\end{figure}

Figure \ref{fig:diffuse} shows the simulated spectra of diffuse gas in Sun-centred apertures, the resulting total spectra, and the relation between the velocity dispersions of the original spectra and the simulated total spectra. When we computed the velocity standard deviations, we subtracted a baseline in order to be more similar to observational studies. This baseline is defined as the mean mass in bins between $\pm$30 and $\pm$20 km/s, analogously to the mean emission outside of the peak. The part of the spectra below this value was not considered and is marked in grey. 

\begin{figure}%[h!]
    \centering
    \includegraphics[width=0.49\textwidth]{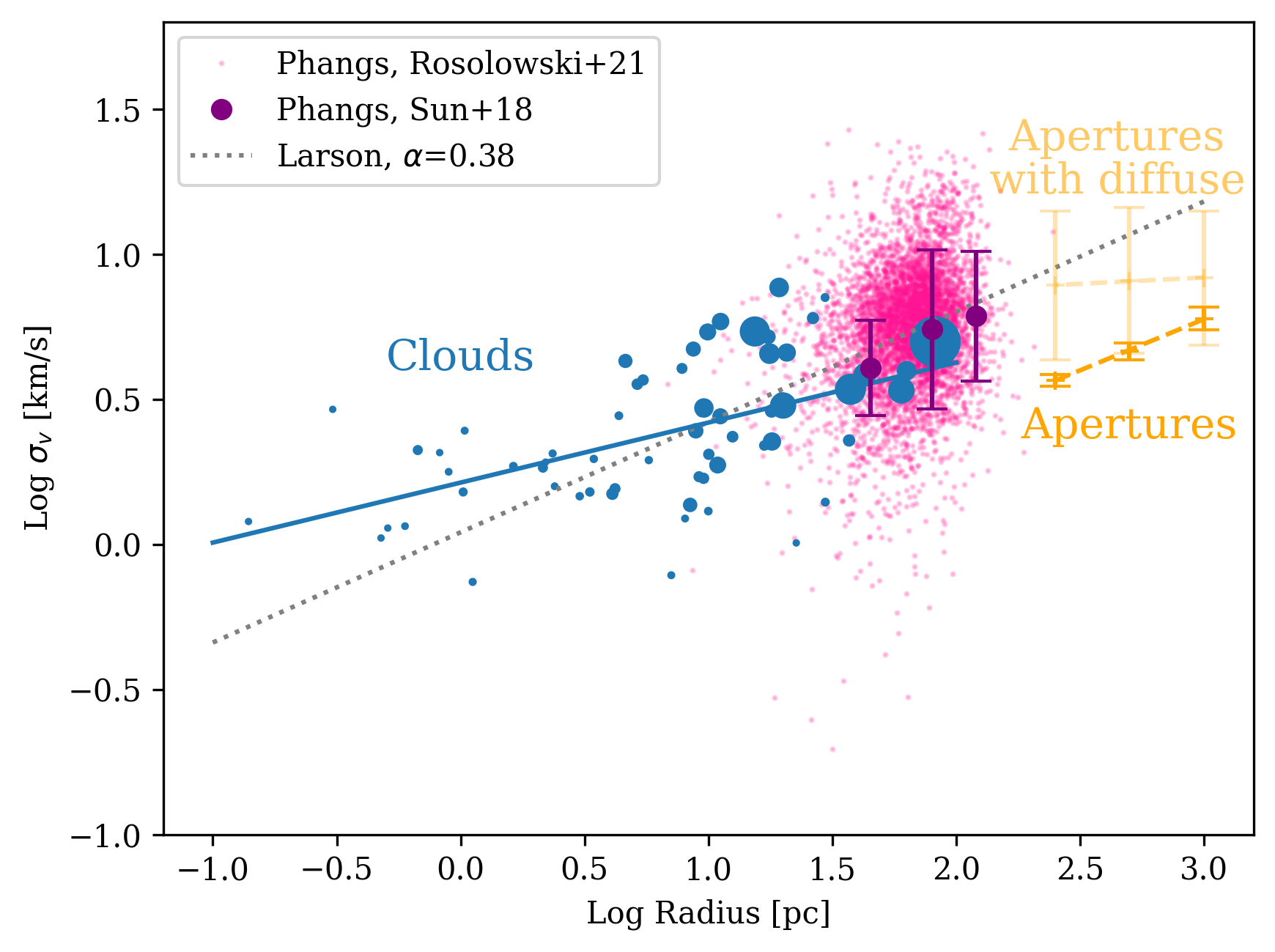}
    \caption{Linewidth-size relation for clouds and apertures compared with apertures with diffuse gas and data from extragalactic works. The estimated relation for the apertures with a diffuse component (Sect. \ref{sec:d_diffuse}) are shown in pale orange. The extragalactic data of molecular clouds from \cite{rosolowsky2021giant} are shown in pink, and the extragalactic data of apertures at various scales from \cite{sun2018cloud} are shown in purple. 
    }
    \label{fig:wphangs}
\end{figure}

The results show that when the aperture spectra are combined with the estimated spectra from the diffuse component, the velocity dispersion increases. This follows naturally from the facts that the diffuse component has a larger linewidth than the denser and more cloud-like gas, and that the diffuse component is significant mass-wise. The effect is especially strong for the broadest version of the diffuse component and for the smaller apertures. This is not only true for the Sun-centred apertures, but for all apertures. The overall effect is that the scale-dependence of the velocity structure is affected: the velocity dispersion increases less with scale than without the diffuse component. In other words, the diffuse component causes the size-linewidth relation to become flatter. This is shown in Fig. \ref{fig:wphangs}, where the slope for the apertures with the simulated diffuse component is $0.04\pm0.01$.

%%%%%%%%%%%%%%%%%%%%%%%%%%%%%%%%%%%%%%%%%%%%%%%%%%%%%%%%%%%%%%%
\subsection{Relation to extragalactic works}
\label{sec:d_exgal}
%%%%%%%%%%%%%%%%%%%%%%%%%%%%%%%%%%%%%%%%%%%%%%%%%%%%%%%%%%%%%%%

In this section we briefly place our results in the context of recent extragalactic works. The aperture sizes used in our bird's eye view correspond to $25-6\arcsec$ at the distance of the nearby spiral galaxy M51 (8.4 Mpc). This is comparable to aperture sizes that are used to study the ISM properties in some of the most recent survey works such as PAWS \citep{schinnerer2013pdbi,leroy2017cloud} and PHANGS \citep[e.g.][]{leroy2021phangs, schinnerer2019}. The spatial resolution of these works is from a few dozen pc to $\sim$140 pc, similar to the sizes of our clouds. Thus, the scales probed by our work generally overlap with those of studies in the most recent extragalactic ISM works.

However, our data, and hence the apertures, only include the CO gas that is organised in clearly discernible clouds. A potentially present diffuse gas between the clouds that would be included within the beam of extragalactic observations is not part of our data set. It is currently unclear what the role of this component in extragalactic data is. For example, \cite{rosolowsky2021giant} argued that their cloud-finding algorithm is insensitive to isolated diffuse molecular gas, and that diffuse molecular gas in dense regions would be assigned to nearby clouds. \cite{sun2018cloud} included all detected molecular emission at the selected spatial scales. Ultimately, understanding the extent and nature of this component is clearly beyond our paper, but the simple experiment we present in section \ref{sec:d_diffuse} demonstrates that it is essential for properly linking the Galactic and extragalactic studies.

Acknowledging the differences between our data and extragalactic studies, we set in Fig. \ref{fig:wphangs} the size-linewidth relation from our work in the context of PHANGS data from \cite{sun2018cloud} and \cite{rosolowsky2021giant}. \cite{sun2018cloud} studied the velocity dispersion of CO spectral cubes in 15 nearby galaxies at resolutions between 45 and 120 pc, while \cite{rosolowsky2021giant} decomposed the CO data of 10 nearby galaxies into almost 5000 clouds and measured the velocity dispersion of those clouds at 90 pc resolution. The comparison illustrates, to the zeroth degree, the similarity of the data and the potential of further work to study the build-up of the relation over almost four orders of magnitudes in size scales.
More detailed comparisons are needed to fully understand the differences between extragalactic and Galactic studies of molecular clouds, but it is encouraging to see these first similarities between the data.

%%%%%%%%%%%%%%%%%%%%%%%%%%%%%%%%%%%%% 
\section{Conclusions}
\label{sec:conclusion}
%%%%%%%%%%%%%%%%%%%%%%%%%%%%%%%%%%%%%
%%%%%%%%%%%%%%%%%%%%%%%%%%%%%%%%%%%%%

We have analysed the kinematics of the most complete molecular cloud sample to date within a distance of 2 kpc using CO$(J=1-0)$ data. We analysed the kinematics of the individual clouds and revisited Larson's relations. We also performed an experiment in which we mimicked a face-on view of the Milky Way and analysed the kinematics of the clouds within apertures of 0.25--2 kpc in size. This Bird's Eye view enabled us to study the scale dependence of gas kinematics from cloud scales to kpc scales. Our main results and conclusions are summarised below.

\begin{enumerate}
    \item The molecular clouds within 2 kpc from the Sun show a linewidth-size relation with large scatter. We obtain the best-fit relation of $\sigma_v=1.5\cdot R^{0.3\pm0.1}$. The mass-size relation is stronger and has the best-fit relation of $\mathrm{M_{CO}}= 794 \cdot R^{1.5\pm0.5}$. The slopes of both relations are consistent with earlier works within 2$\sigma$ uncertainty.
    \item With our Bird's Eye view experiment, we describe the CO spectra in circular apertures within 2 kpc from the Sun. These apertures describe the total CO spectra in the solar neighbourhood conditions of the Milky Way disk (Fig. \ref{fig:aps_sun} shows Sun-centred aperture spectra of different radii). The aperture spectra are relatively symmetric, but exhibit non-Gaussian wings at extreme velocities. The origin of the wings is unclear. They might be caused by peculiar motions of the clouds (i.e. motions that do not follow the assumed rotation curve).
    \item We describe the scale dependence of the velocity dispersion and mass in our survey area. The best-fit size-linewidth relation for the apertures is $\sigma_v=0.5\cdot R^{0.35\pm0.01}$, and the best-fit mass-size relation is $\mathrm{M_{CO}}= 229 \cdot R^{1.4\pm0.1}$. Both slopes are consistent with the cloud fits within 2$\sigma$.
    \item We demonstrate that if the Milky Way were viewed from outside, a diffuse, widespread CO gas component could significantly flatten the size-linewidth relation from what we derive for apertures between 0.25-2 kpc. If a diffuse component like this is present in galaxies in general, the relations derived at these scales may also be flattened due to the diffuse component. Understanding the nature of the potential diffuse component is crucial for connecting Galactic and extragalactic works. 
\end{enumerate}

This work continues our new approach from Paper I to work towards connecting Galactic and extragalactic studies of ISM and key star formation relations. Galactic and extragalactic data have important differences that need to be further studied and taken into account when we continue to reconcile the observations across the scales. In future works, we plan to investigate the relations between the kinematics described in this paper and the density distributions and star formation results from Paper I, and study how the results fit the current theoretical models for star formation.

\iffalse
\begin{acknowledgements}
This project has received funding from the European Union's Horizon 2020 research and innovation programme under grant agreement No 639459 (PROMISE). Jan Orkisz acknowledges funding from the Swedish Research Council, grant No. 2017-03864.
\end{acknowledgements}
\fi

%\newpage
%\tableofcontents
%{\setlength\LTcapwidth{0.85\linewidth}
\bibliographystyle{aa}%apalike}%unsrt}{plain}
\bibliography{X_bib}
%\newpage

%%%%%%%%%%%%%%%%%%%%%%%%%
\begin{appendix}
\onecolumn
\let\part=\chapter\appendix
%\input{Z_appendix}
%\appendix
%\appendixpage
%\addappheadtotoc
%\section{Appendix}

\section{List of clouds in the sample}

This appendix includes the list of the 64 clouds in the sample we used in this paper; see Table \ref{tab:veltable}. They are ordered according to distance, and we list their size, kinematic properties, and mass. For locations in the Galaxy, see \cite{spilker2021bird}.

%\clearpage
%\onecolumn
%\setcounter{table}{0}
% THE LONG TABLES OF CLOUD PARAMETERS
\begin{longtable}{lllllll}
\label{tab:veltable}
\\
\caption{Clouds in our sample, with some of their properties listed.}\\
\hline\hline%\hline
Name    & Dist\tablefootmark{a} &  Radius\tablefootmark{b} & $\sigma_\mathrm{v}$   &  <v>$_{corrected}$\tablefootmark{c}    & <v>$_{original}$ &   $\mathrm{M_{CO}}$    \\
\hline
 & [pc] & [pc] & [km/s] & [km/s] & [km/s] & [M$_\odot$] \\
\hline      

\hline
\endfirsthead
\caption{Continued.} \\
\hline
Name    & Dist\tablefootmark{a} &  Radius\tablefootmark{b} &  $\sigma_\mathrm{v}$   &  <v>$_{corrected}$\tablefootmark{c}    & <v>$_{original}$ &   $\mathrm{M_{CO}}$    \\
\hline
 & [pc] & [pc] & [km/s] & [km/s] & [km/s] & [M$_\odot$] \\
\hline      
\endhead
\hline
\endfoot
\hline
\endlastfoot
Taurus  &       130     &       3.32    &       1.51    &        6.76   &        6.33    &       1.24$\cdot 10^4$         \\ 
Aquila-South    &       135     &       0.00    &       0.86    &        1.24    &        3.23   &       3.70$\cdot 10^2$         \\ 
Ophiuchus       &       139     &       4.20    &       1.55    &        2.88    &        2.94   &       2.48$\cdot 10^4$         \\ 
RCrA    &       155     &       1.12    &       0.74    &        5.44   &        5.56    &       4.13$\cdot 10^3$         \\ 
Chamaeleon      &       161     &       2.40    &       1.58    &        4.95    &        3.27   &       2.31$\cdot 10^3$         \\ 
Pipe    &       180     &       8.05    &       1.23    &        4.18   &        4.12    &       3.16$\cdot 10^3$         \\ 
Coalsack        &       187     &       2.35    &       2.06    &       -1.37   &       -3.53   &       4.13$\cdot 10^3$    \\ 
Lupus   &       197     &       3.02    &       1.46    &        7.09   &        5.16    &       6.00$\cdot 10^3$         \\ 
Hercules        &       223     &       0.60    &       1.15    &        2.67    &        6.42   &       2.09$\cdot 10^3$         \\ 
Camelopardalis  &       235     &       0.48    &       1.05    &        2.17    &       -0.45   &       3.70$\cdot 10^2$         \\ 
Aquila  &       254     &       9.16    &       1.71    &        3.77   &        7.36    &       2.51$\cdot 10^4$         \\ 
Pegasus-West    &       258     &       0.30    &       2.92    &       -7.77   &       -3.68   &       9.42$\cdot 10^1$    \\ 
Perseus &       276     &       5.16    &       3.56    &        7.55   &        5.05    &       2.88$\cdot 10^4$         \\ 
L1333   &       283     &       1.04    &       2.46    &        4.74   &        1.16    &       3.17$\cdot 10^3$         \\ 
Pegasus-East    &       292     &       0.82    &       2.07    &       -5.18   &       -7.08   &       2.65$\cdot 10^2$    \\ 
UrsaMaj &       330     &       2.20    &       1.92    &        6.32   &        2.38    &       3.37$\cdot 10^2$         \\ 
Polaris &       343     &       1.02    &       1.51    &        0.71   &       -3.36   &       9.87$\cdot 10^3$    \\ 
L1265   &       344     &       0.51    &       1.14    &        1.43   &       -2.13   &       3.47$\cdot 10^2$    \\ 
CephFlare       &       346     &       4.62    &       4.28    &       -1.04   &       -3.70   &       5.71$\cdot 10^4$    \\ 
GumNeb  &       349     &       1.63    &       1.86    &        3.00   &        3.14    &       7.22$\cdot 10^3$         \\ 
LamOri  &       399     &       5.75    &       1.95    &       -3.53   &        4.75    &       4.97$\cdot 10^3$         \\ 
Gemini  &       400     &       0.14    &       1.20    &       -3.69   &       -0.48   &       1.12$\cdot 10^2$    \\ 
OrionB  &       433     &       8.89    &       2.46    &        4.93   &        9.10    &       7.18$\cdot 10^4$         \\ 
OrionA  &       438     &       11.17   &       2.76    &        2.42   &        7.35    &       8.17$\cdot 10^4$         \\ 
California      &       466     &       8.69    &       4.71    &        1.52    &       -2.18   &       6.83$\cdot 10^4$         \\ 
Serpens &       490     &       18.04   &       2.26    &        1.93   &        7.59    &       1.15$\cdot 10^5$         \\ 
Lacerta &       504     &       0.89    &       1.78    &        4.48   &        0.67    &       2.26$\cdot 10^3$         \\ 
CygOB7  &       561     &       19.78   &       3.12    &       -1.80   &       -2.32   &       9.80$\cdot 10^4$    \\ 
Circinus        &       675     &       9.99    &       1.30    &        3.10    &       -6.01   &       6.88$\cdot 10^3$         \\ 
CepheusOB3b     &       700     &       12.51   &       2.35    &       -3.68   &       -7.87   &       3.03$\cdot 10^4$    \\ 
Norma   &       721     &       22.61   &       1.01    &        2.97   &       -15.09  &       3.46$\cdot 10^2$    \\ 
L1307-35        &       741     &       7.82    &       4.04    &        3.19    &       -5.92   &       2.41$\cdot 10^4$         \\ 
MonocerosR2     &       767     &       10.89   &       1.88    &        1.97    &       11.16   &       9.39$\cdot 10^4$         \\ 
MonOB1  &       771     &       10.03   &       2.04    &       -0.87   &        5.63    &       2.74$\cdot 10^4$         \\ 
IC5146  &       792     &       4.35    &       2.77    &        5.44   &        3.07    &       7.69$\cdot 10^3$         \\ 
CephOB4 &       850     &       11.19   &       5.85    &        0.60   &       -9.09   &       1.02$\cdot 10^5$    \\ 
Vela    &       866     &       42.85   &       3.85    &        1.98   &        3.89    &       2.00$\cdot 10^5$         \\ 
AFGL490 &       900     &       16.80   &       2.19    &       -2.06   &       -12.46  &       2.05$\cdot 10^4$    \\ 
L1340-55        &       903     &       9.93    &       5.39    &        4.11    &       -7.10   &       9.57$\cdot 10^4$         \\ 
S140    &       910     &       9.36    &       2.86    &       -0.06   &       -7.44   &       1.81$\cdot 10^4$    \\ 
IC1396  &       941     &       8.44    &       1.37    &        3.95   &       -0.62   &       6.01$\cdot 10^4$    \\ 
L1293-1306      &       977     &       5.45    &       3.68    &       -2.18   &       -11.86  &       2.77$\cdot 10^4$    \\ 
Split   &       1000    &       62.98   &       3.96    &        0.65   &       14.14   &       1.35$\cdot 10^5$    \\ 
Ara     &       1064    &       29.56   &       7.08    &       -5.13   &       -32.48  &       9.16$\cdot 10^3$    \\ 
Cygnus  &       1214    &       82.33   &       4.99    &       -0.58   &        2.65    &       1.22$\cdot 10^6$         \\ 
Lagoon  &       1220    &       26.41   &       6.02    &       13.54   &       18.05   &       3.48$\cdot 10^4$    \\ 
S147    &       1220    &       0.67    &       2.11    &        1.50   &        1.67    &       1.68$\cdot 10^4$         \\ 
M20     &       1234    &       19.28   &       7.67    &        9.64   &       14.13   &       1.38$\cdot 10^5$    \\ 
Rosette &       1261    &       17.98   &       2.88    &        1.66   &       13.37   &       4.79$\cdot 10^4$    \\ 
CMaOB1  &       1262    &       9.56    &       1.69    &        1.10   &       16.48   &       2.72$\cdot 10^4$    \\ 
NGC2362 &       1317    &       3.45    &       1.97    &        6.58   &       21.74   &       5.47$\cdot 10^3$    \\ 
L291    &       1348    &       29.62   &       1.40    &        2.78   &       11.21   &       8.73$\cdot 10^3$    \\ 
GGD4    &       1349    &       4.09    &       1.49    &        0.46   &        2.41    &       3.56$\cdot 10^4$         \\ 
NGC6604 &       1352    &       36.94   &       2.28    &       15.60   &       22.23   &       3.68$\cdot 10^4$    \\ 
M17     &       1509    &       15.39   &       5.41    &       10.35   &       31.67   &       3.85$\cdot 10^5$    \\ 
S235    &       1560    &       9.59    &       2.95    &       -13.45  &       -17.63  &       1.34$\cdot 10^5$    \\ 
IC443   &       1593    &       2.15    &       1.84    &       -9.43   &       -3.53   &       1.63$\cdot 10^4$    \\ 
W3-W4-W5        &       1647    &       17.43   &       5.19    &       11.80   &       -29.72  &       6.47$\cdot 10^4$    \\ 
NGC6334 &       1700    &       17.64   &       4.54    &       -9.95   &       -19.26  &       1.60$\cdot 10^5$    \\ 
M16     &       1731    &       20.01   &       3.01    &        6.56   &       21.90   &       2.82$\cdot 10^5$    \\ 
Cyg-West        &       1800    &       60.07   &       3.39    &       -1.08   &        9.80    &       2.82$\cdot 10^5$         \\ 
Cartwheel       &       1800    &       20.71   &       4.58    &       -0.22   &       -14.66  &       1.14$\cdot 10^5$    \\ 
GemOB1  &       1865    &       37.42   &       3.42    &       -2.28   &        5.42    &       4.13$\cdot 10^5$         \\ 
Maddalena       &       1888    &       7.08    &       0.78    &       -8.27   &       12.78   &       3.40$\cdot 10^3$    \\ 

\hline
\end{longtable}
\tablefoot{\\
\tablefoottext{a}{See references in \citet{spilker2021bird}.}\\
\tablefoottext{b}{Radius is from the area of the cloud with extinction $> 3$ mag, assumed spherical.}\\
\tablefoottext{c}{Calculated after clouds were cropped and disk rotation was subtracted.}\\
}

%\twocolumn
\clearpage

\section{PPV overview of clouds}
\label{app:overview}

\begin{figure}[h!]
    \centering
    \includegraphics[width=0.49\textwidth]{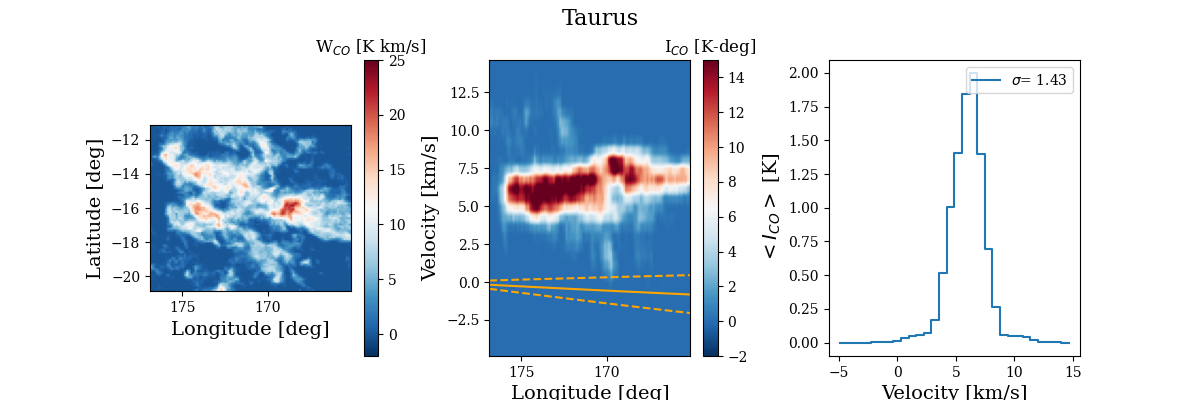}
    \includegraphics[width=0.49\textwidth]{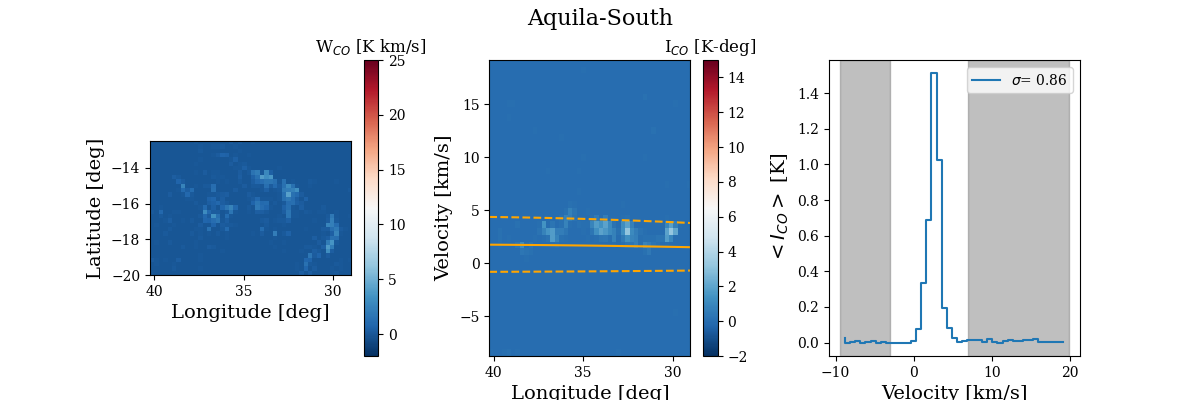}
    \includegraphics[width=0.49\textwidth]{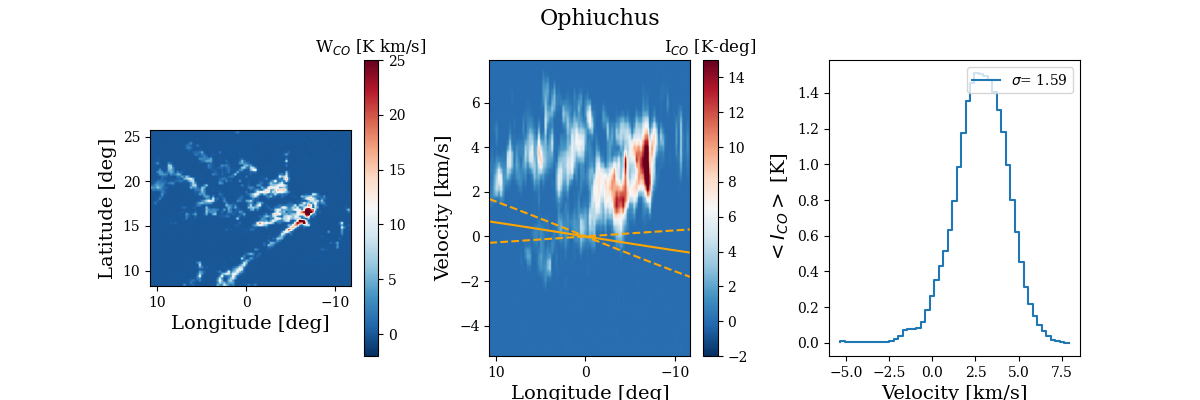}
    \includegraphics[width=0.49\textwidth]{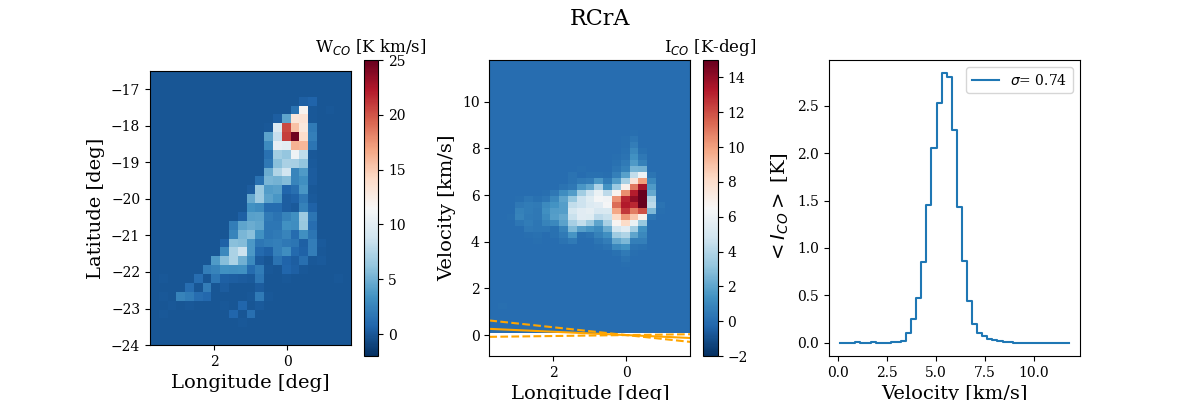} %png}

    \includegraphics[width=0.49\textwidth]{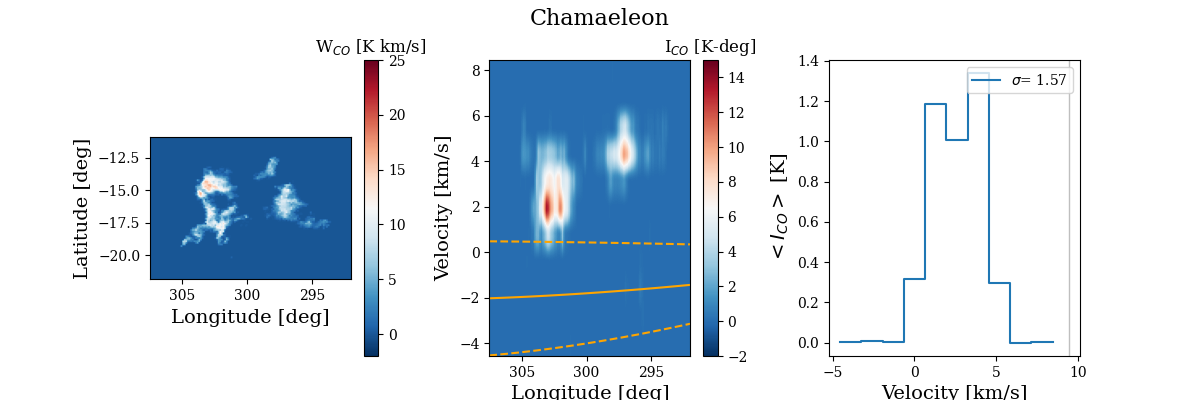}
    \includegraphics[width=0.49\textwidth]{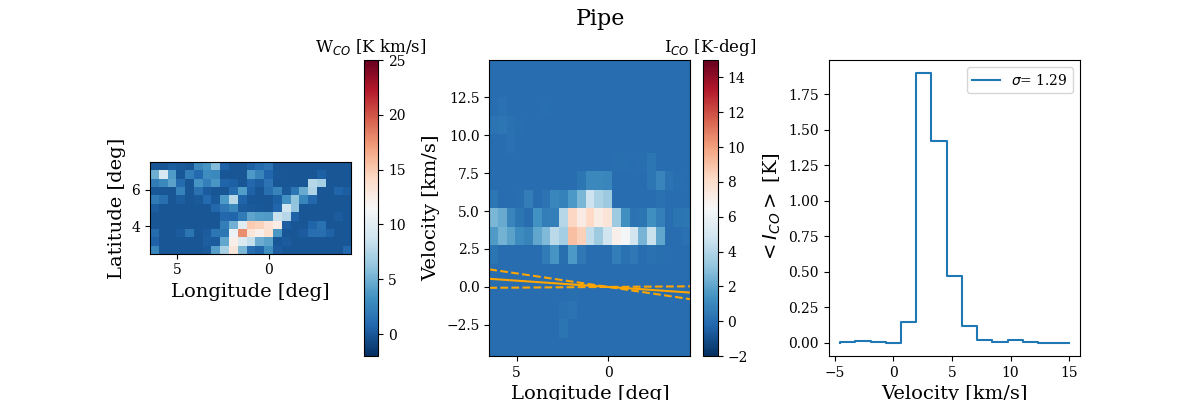}
    \includegraphics[width=0.49\textwidth]{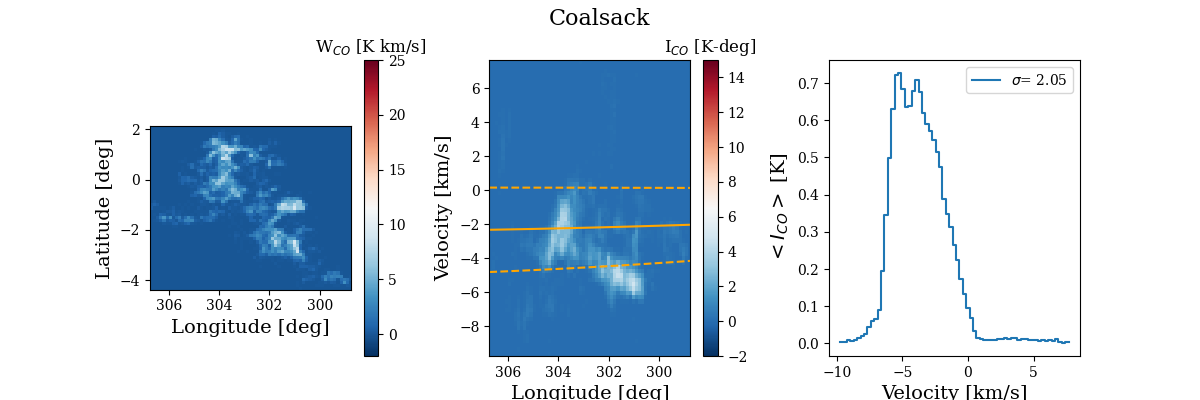}
    \includegraphics[width=0.49\textwidth]{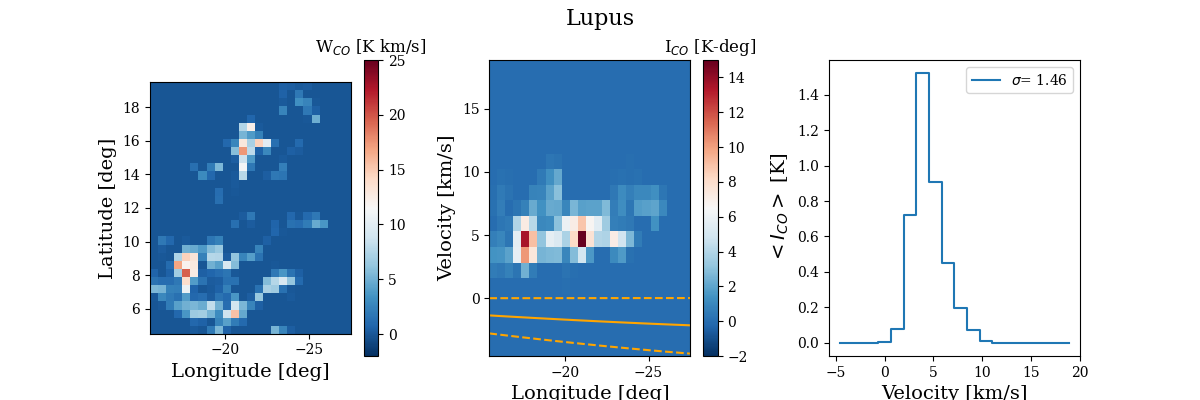}
    
    \includegraphics[width=0.49\textwidth]{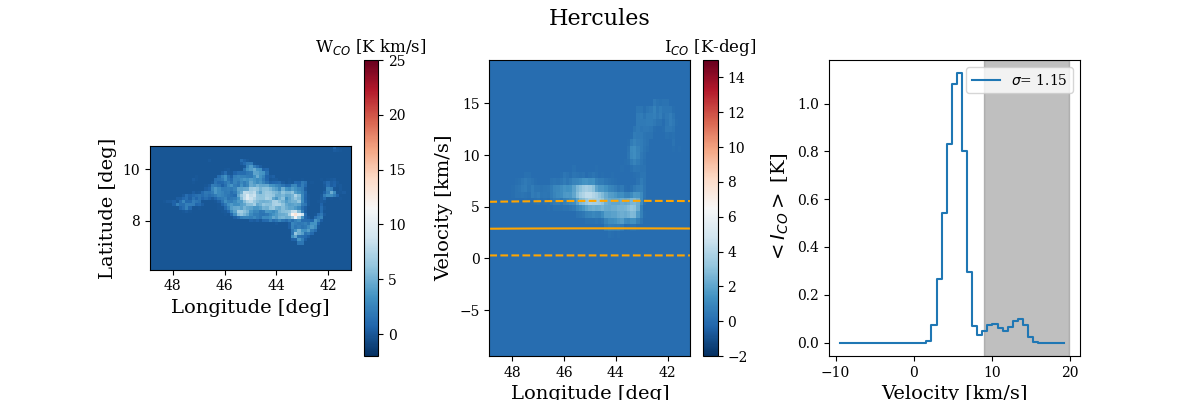}
    \includegraphics[width=0.49\textwidth]{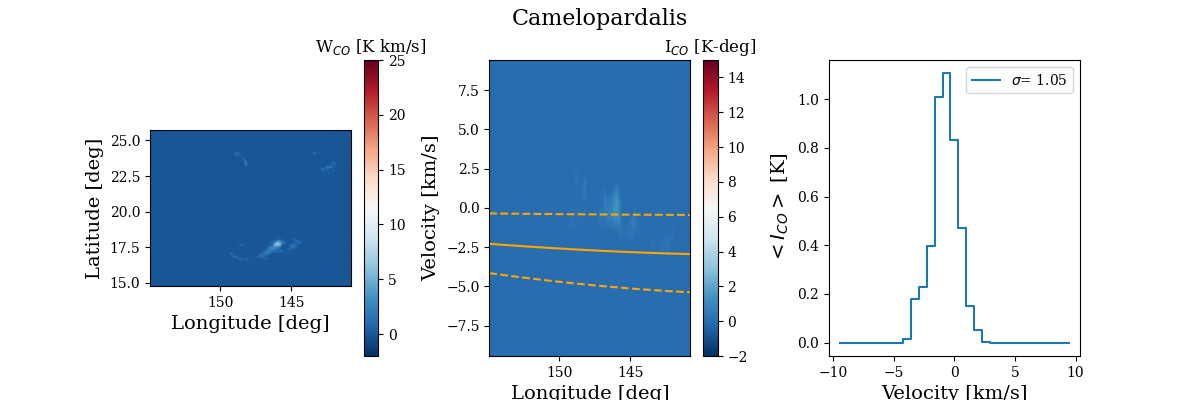}
    \includegraphics[width=0.49\textwidth]{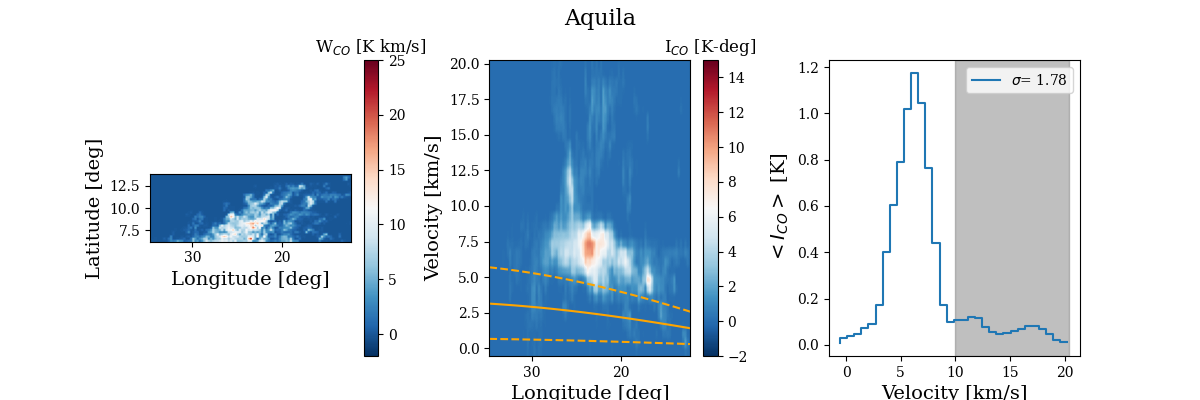}
    \includegraphics[width=0.49\textwidth]{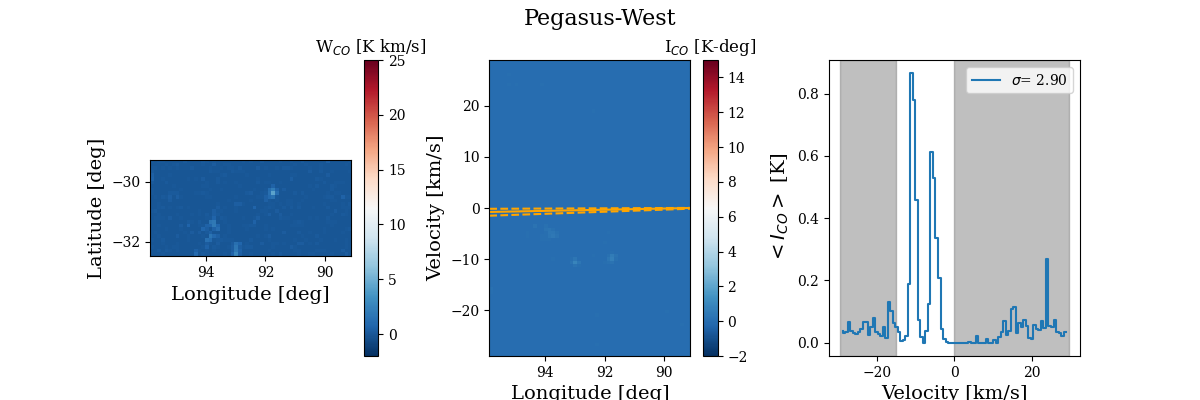}
    \caption{Overview of the PPV data of the molecular clouds in our sample, seen in CO position and velocity from \citet{dame2001milky}. The grey regions of the spectra show the regions that were masked out of the analysis, as they do not seem to be associated with the main cloud. The solid orange line in velocity space shows the estimated disk velocity at the cloud distance, which was subtracted from the original spectra.}
    \label{fig:3pan_1}
\end{figure}

\begin{figure}[h!]
    \centering
    \includegraphics[width=0.49\textwidth]{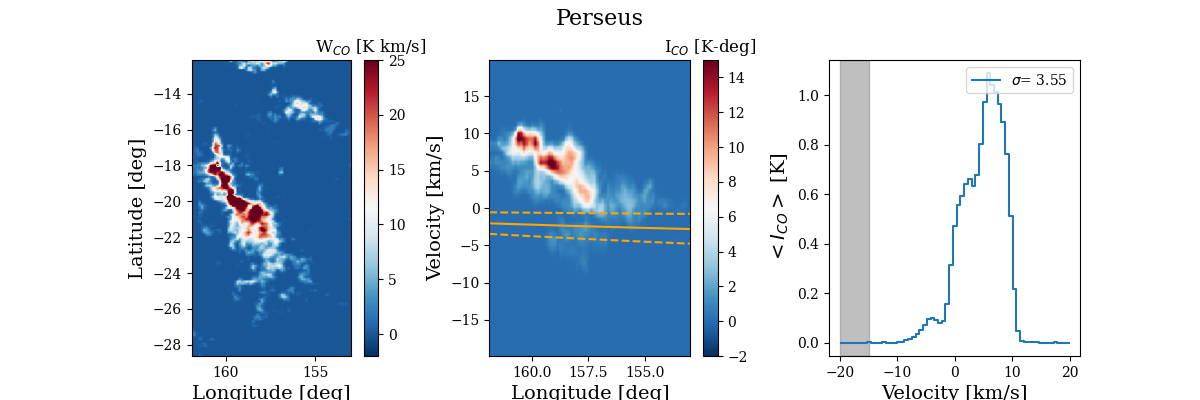}
    \includegraphics[width=0.49\textwidth]{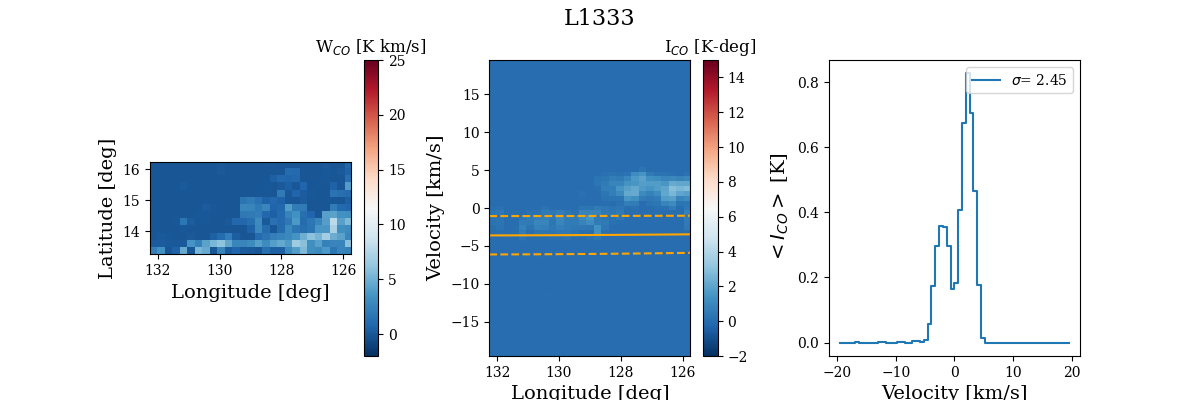}
    \includegraphics[width=0.49\textwidth]{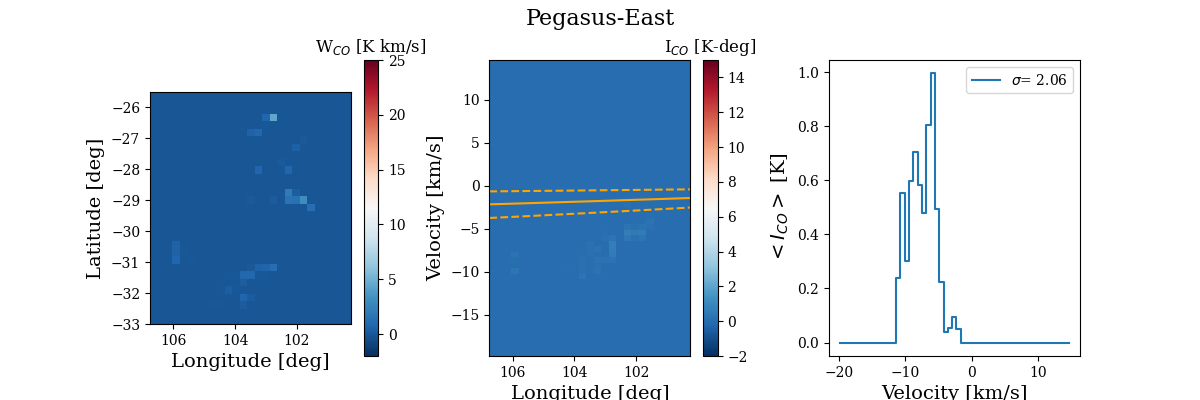}
    \includegraphics[width=0.49\textwidth]{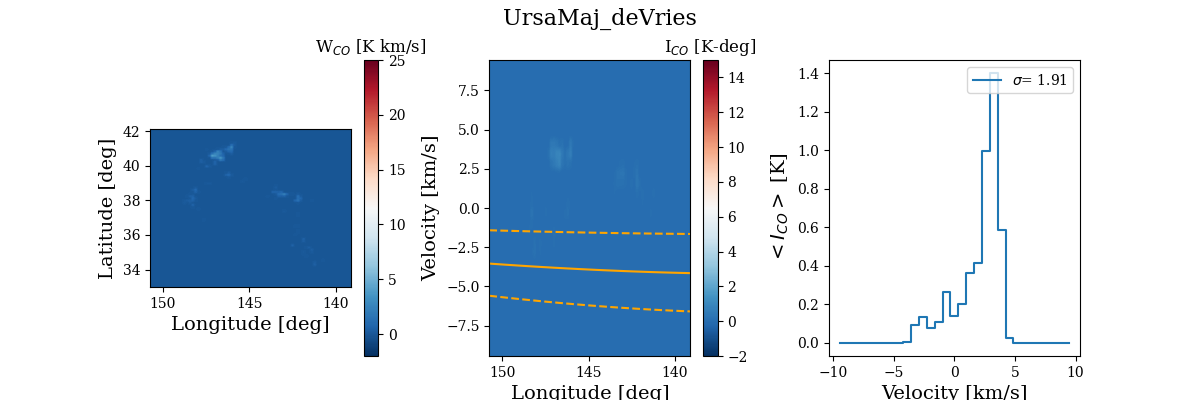}

    \includegraphics[width=0.49\textwidth]{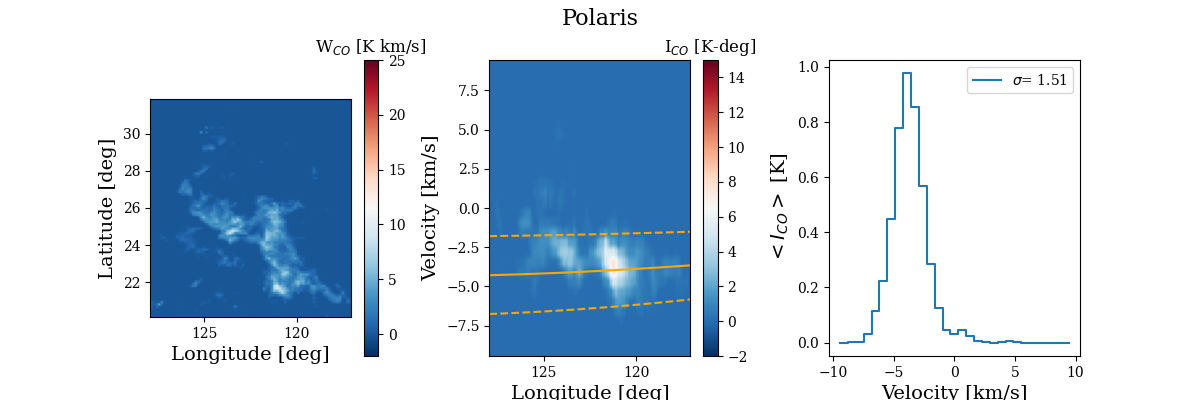}
    \includegraphics[width=0.49\textwidth]{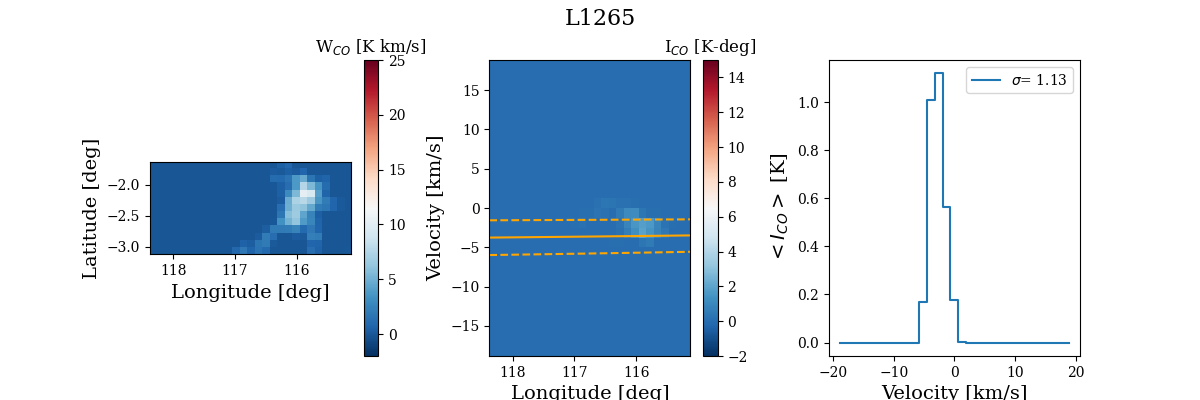}
    \includegraphics[width=0.49\textwidth]{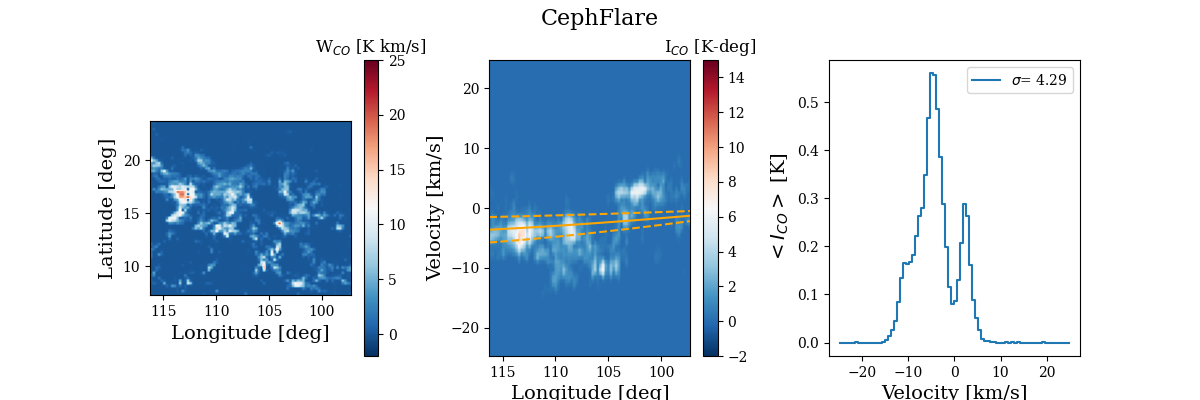}
    \includegraphics[width=0.49\textwidth]{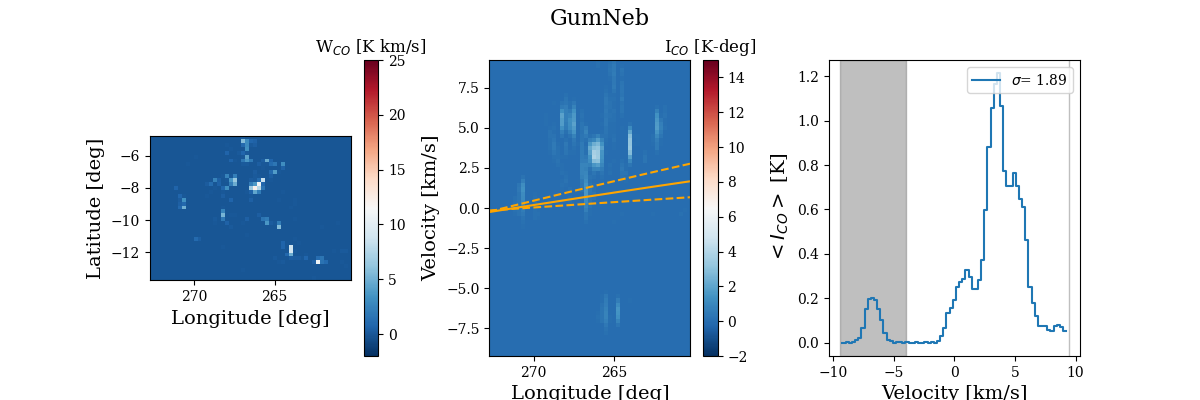}
    
    \includegraphics[width=0.49\textwidth]{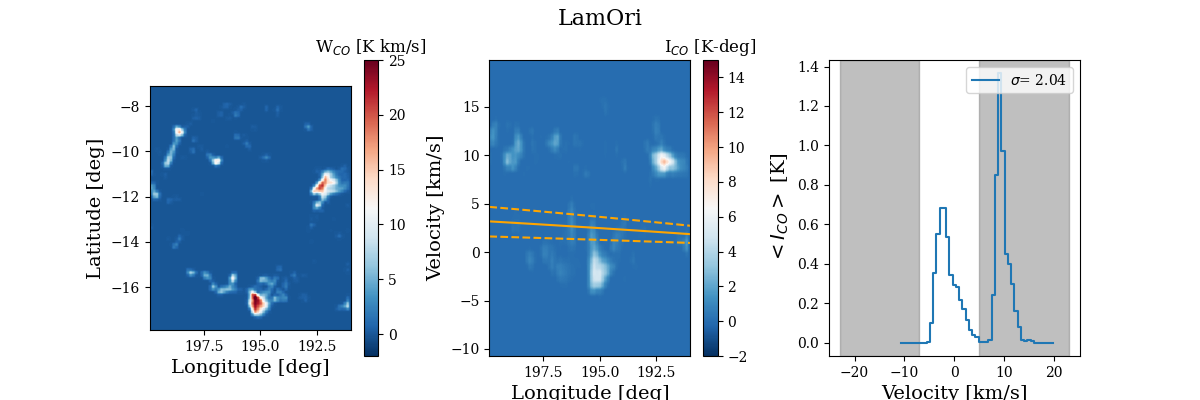}
    \includegraphics[width=0.49\textwidth]{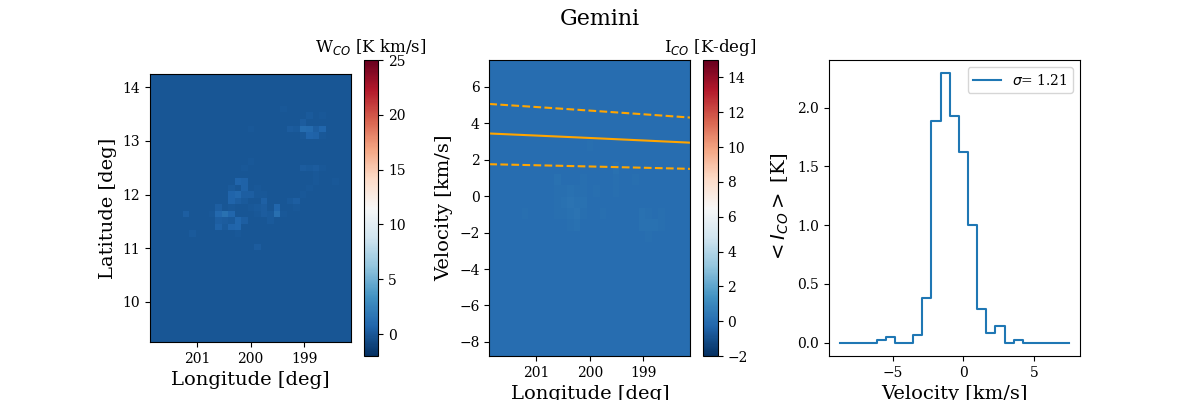}
    \includegraphics[width=0.49\textwidth]{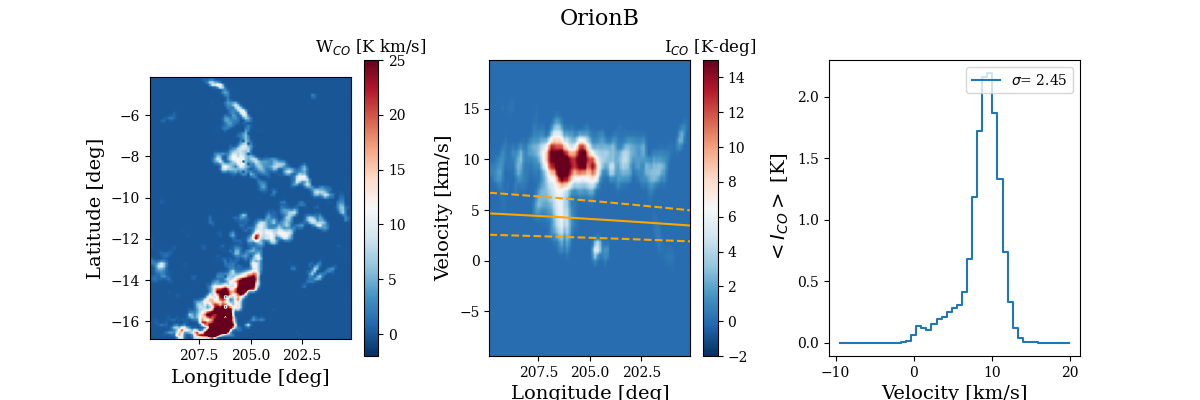}
    \includegraphics[width=0.49\textwidth]{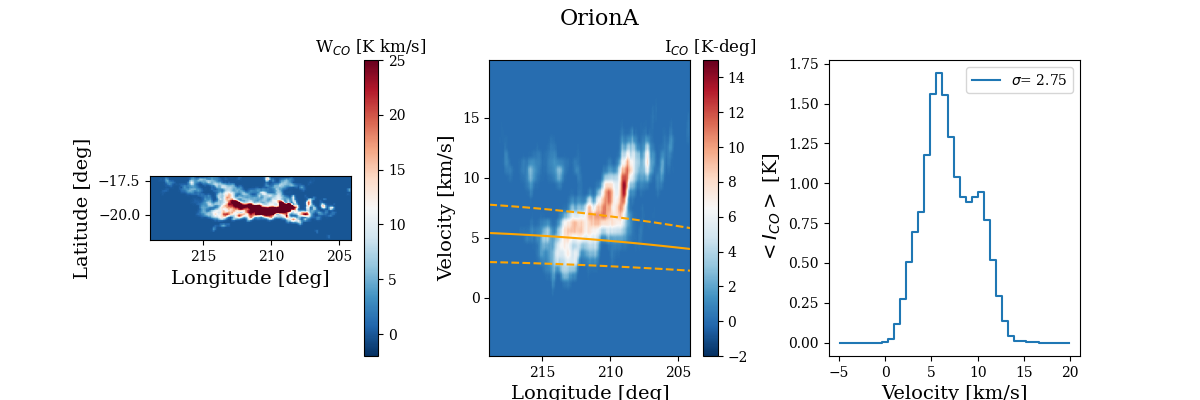}

    \includegraphics[width=0.49\textwidth]{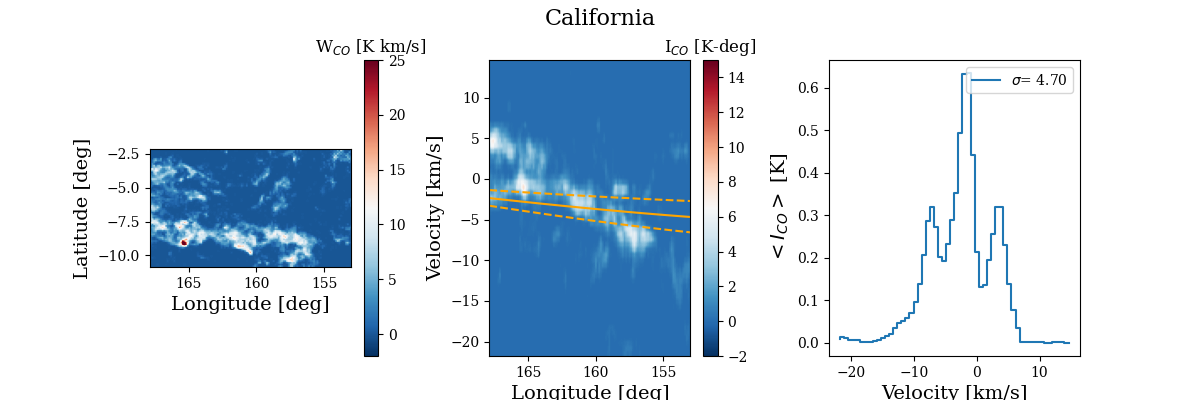}
    \includegraphics[width=0.49\textwidth]{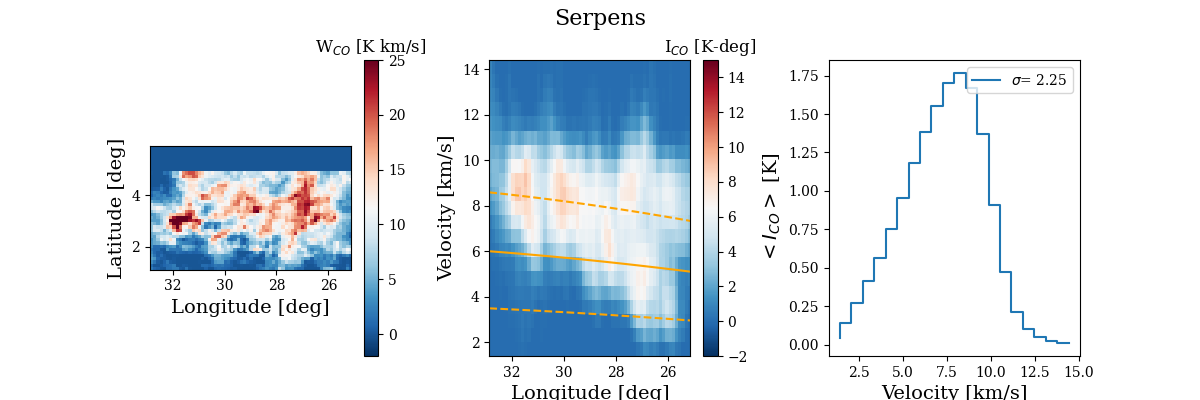}

    \caption{Overview of the PPV data of the molecular clouds in our sample, seen in CO position and velocity. CO from the \citet{dame2001milky} survey.}
    \label{fig:3pan_2}
\end{figure}

\begin{figure}[h!]
    \centering
    \includegraphics[width=0.49\textwidth]{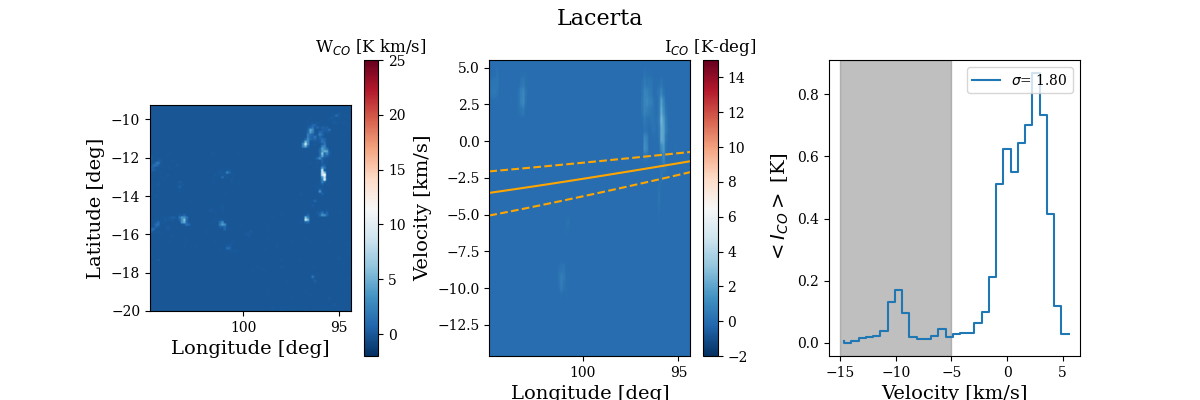}
    \includegraphics[width=0.49\textwidth]{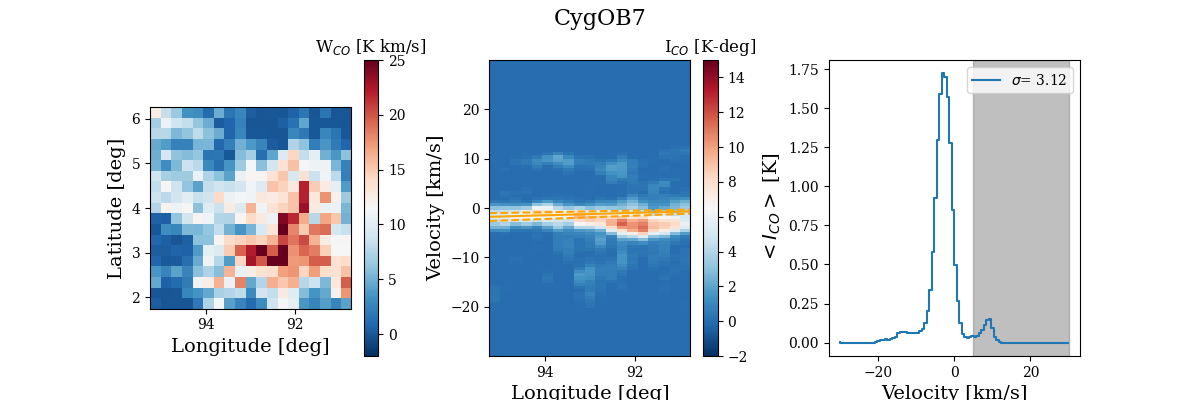}

    \includegraphics[width=0.49\textwidth]{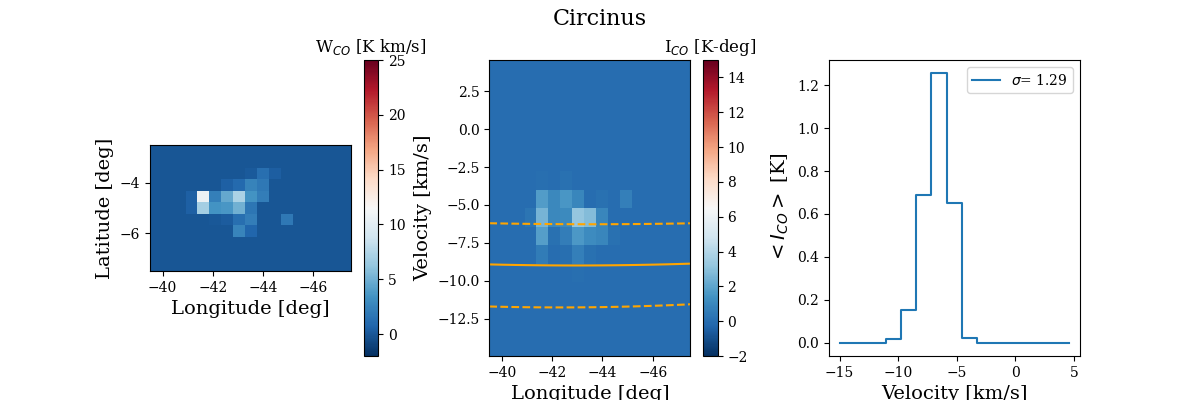}
    \includegraphics[width=0.49\textwidth]{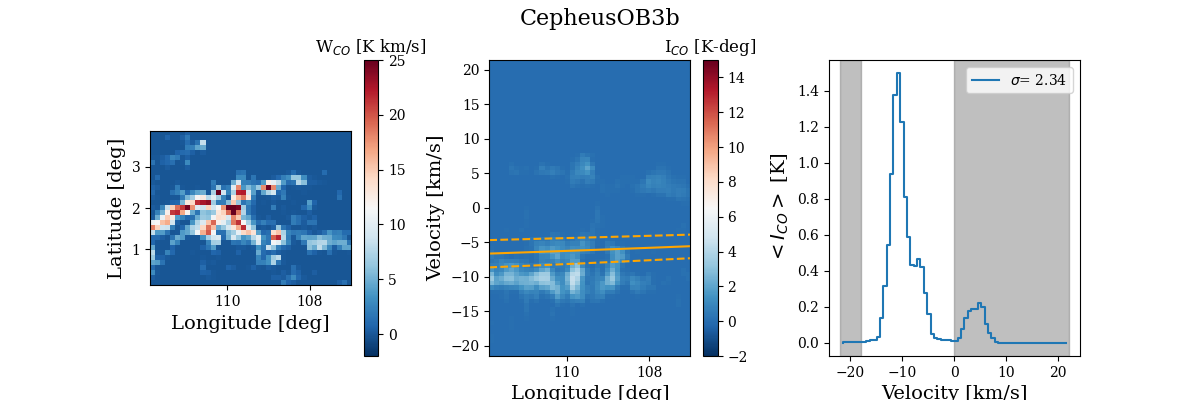}
    \includegraphics[width=0.49\textwidth]{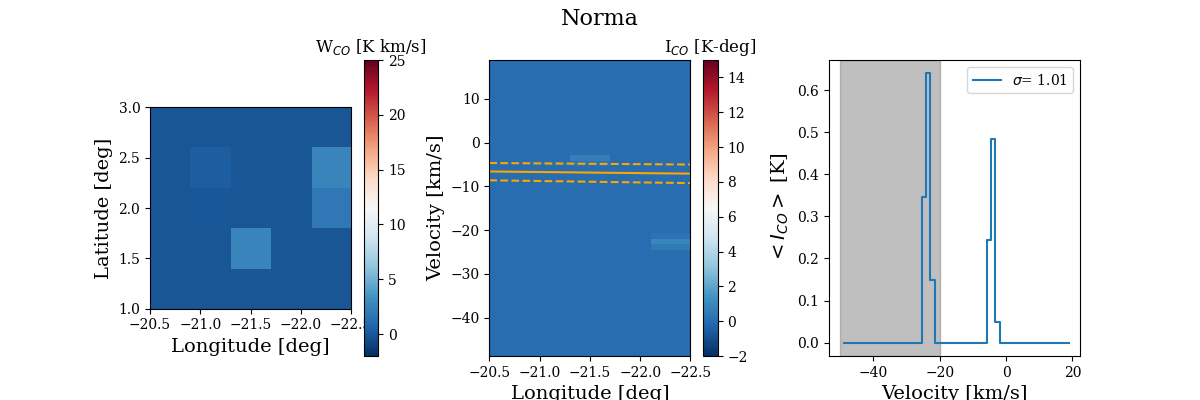}
    \includegraphics[width=0.49\textwidth]{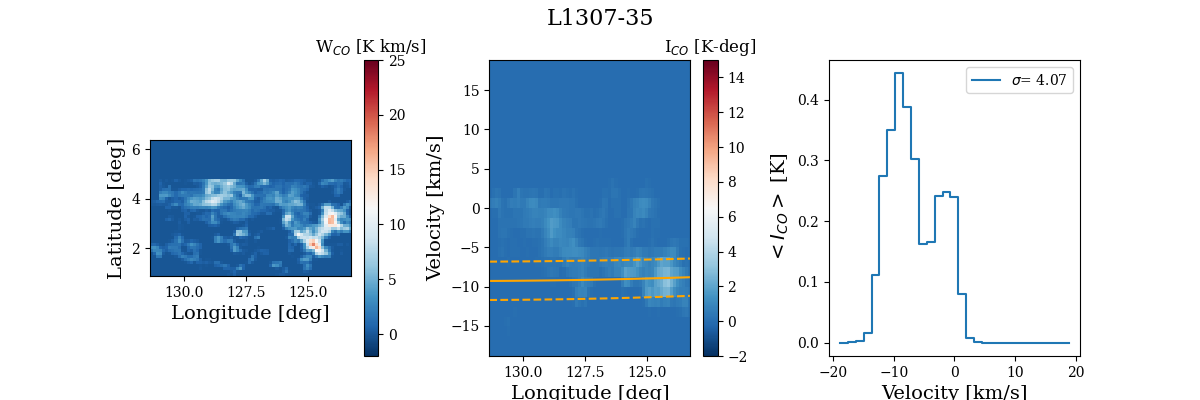}
    
    \includegraphics[width=0.49\textwidth]{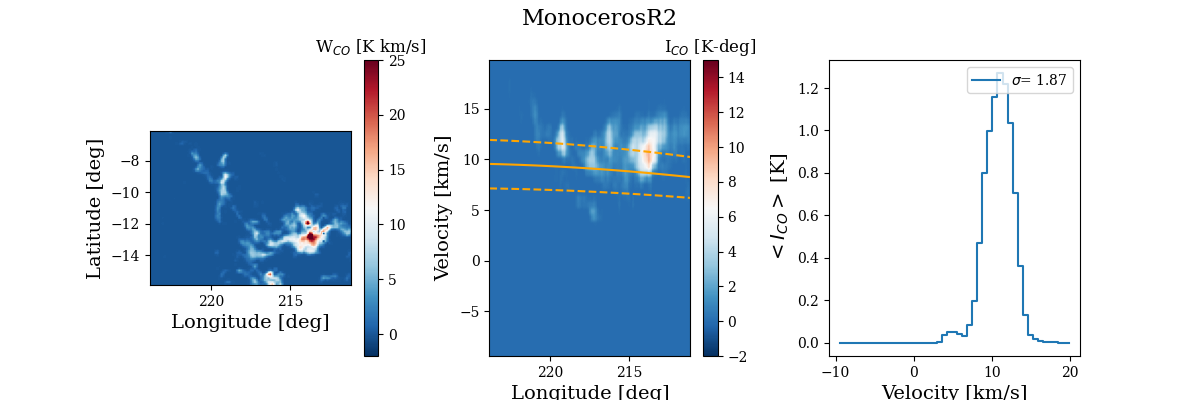}
    \includegraphics[width=0.49\textwidth]{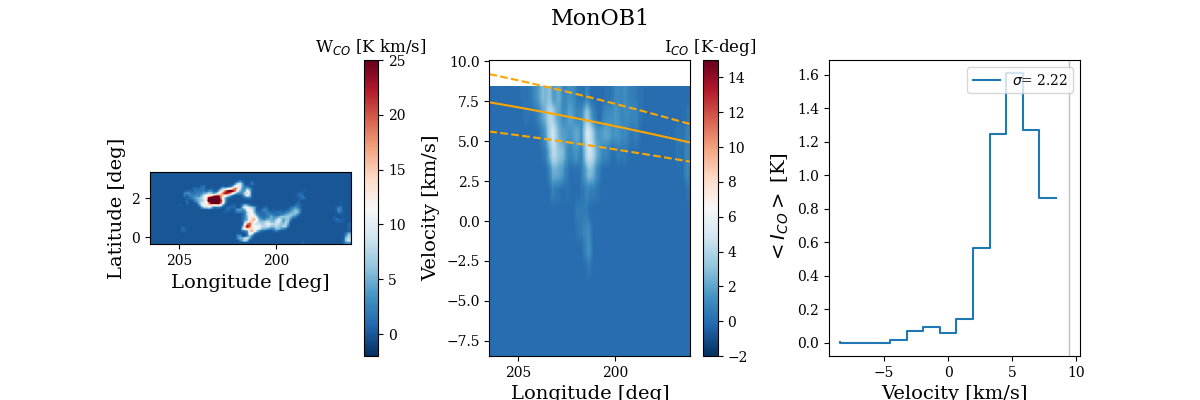}
    \includegraphics[width=0.49\textwidth]{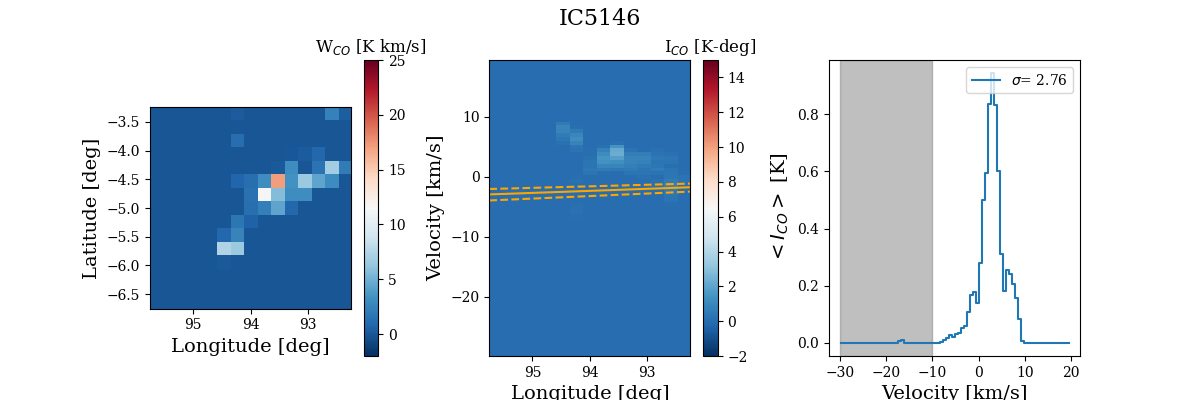}
    \includegraphics[width=0.49\textwidth]{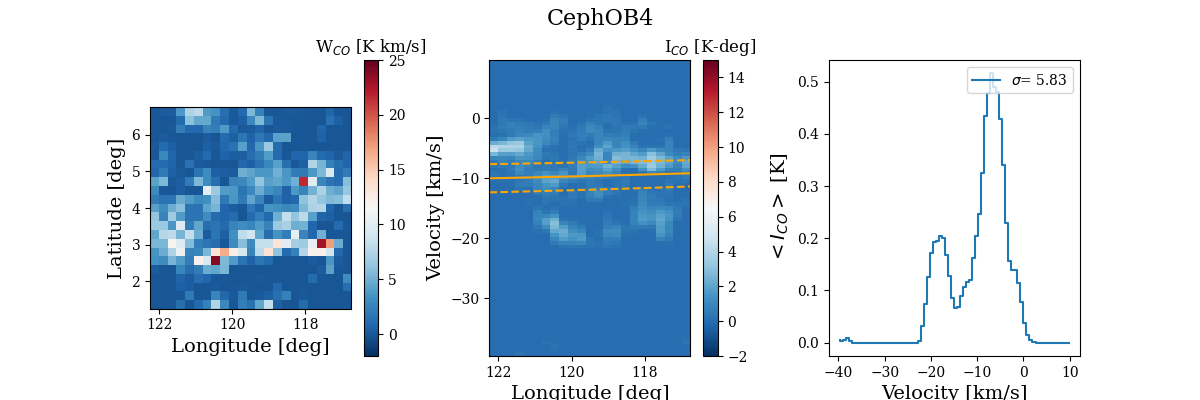}
    
    \includegraphics[width=0.49\textwidth]{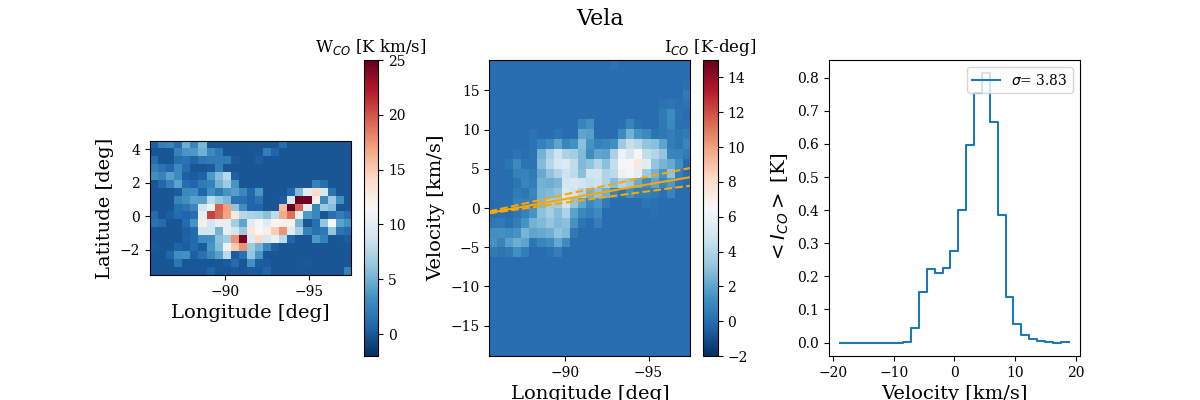}
    \includegraphics[width=0.49\textwidth]{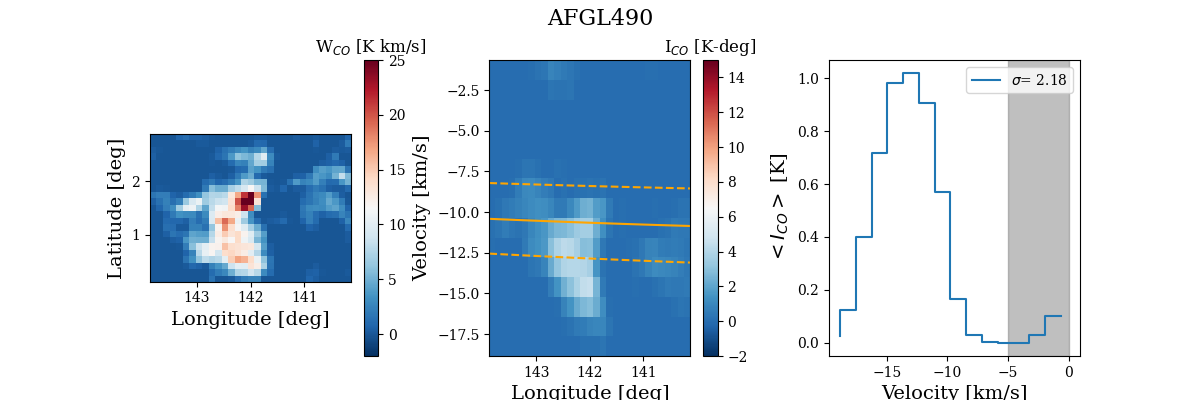}
    \includegraphics[width=0.49\textwidth]{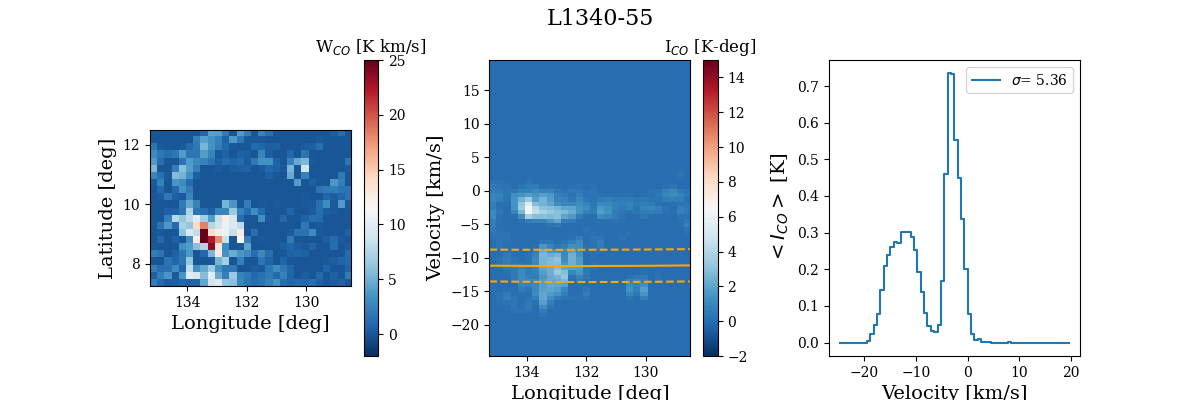}
    \includegraphics[width=0.49\textwidth]{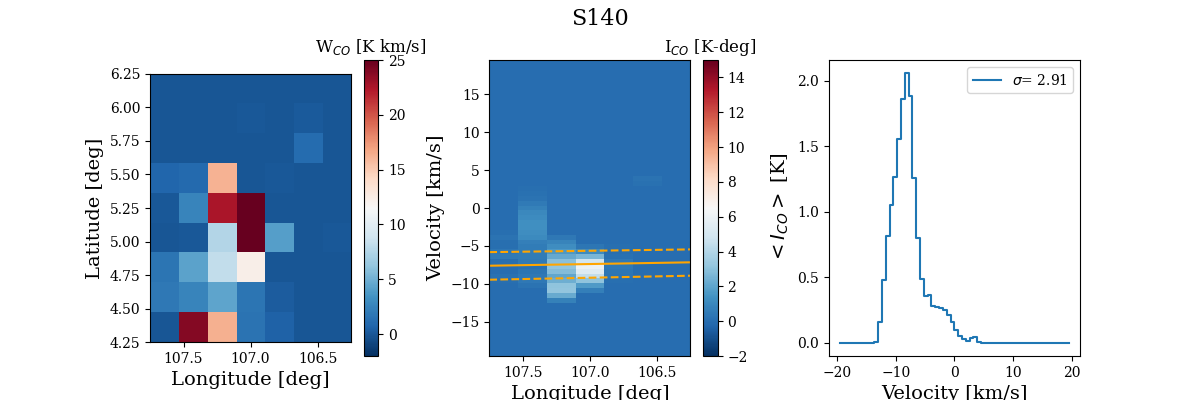}
    \caption{Overview of the PPV data of the molecular clouds in our sample, seen in CO position and velocity. CO from the \citet{dame2001milky} survey.}
    \label{fig:3pan_3}
\end{figure}

\begin{figure}[h!]
    \centering
    \includegraphics[width=0.49\textwidth]{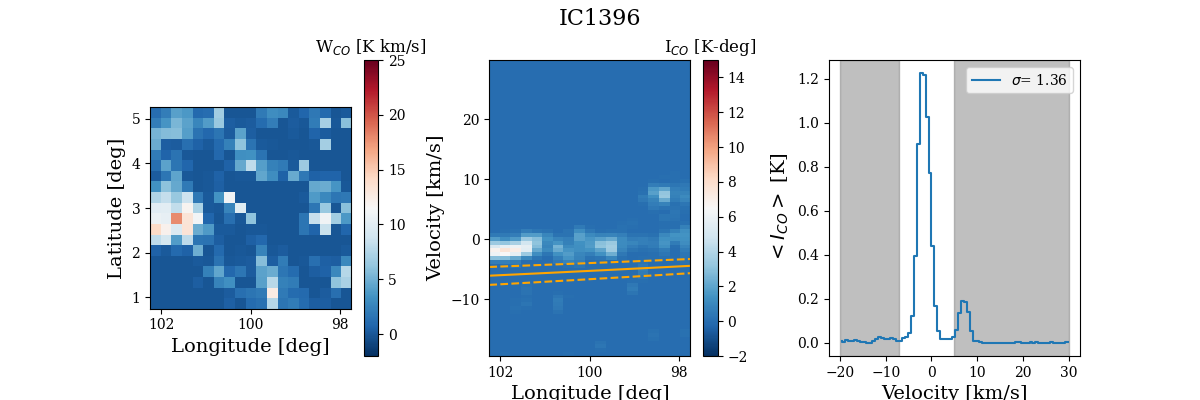}
    \includegraphics[width=0.49\textwidth]{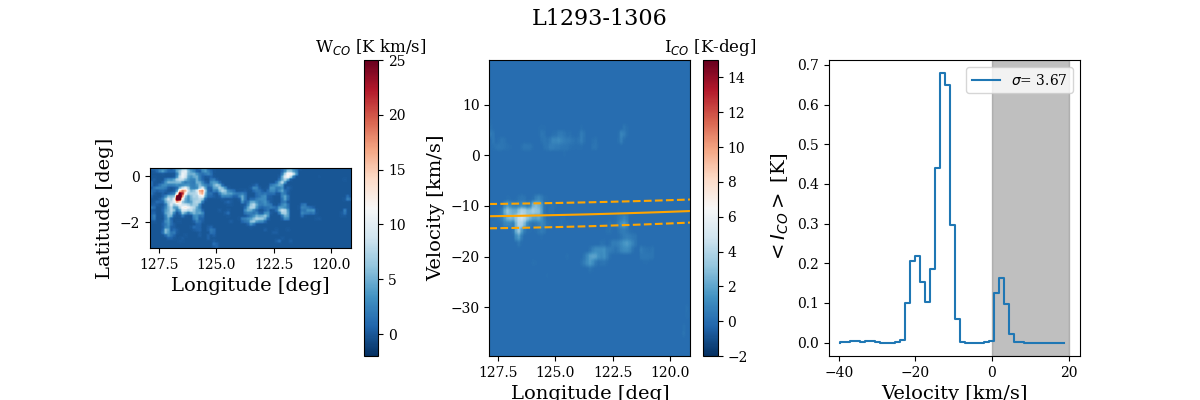}
    \includegraphics[width=0.49\textwidth]{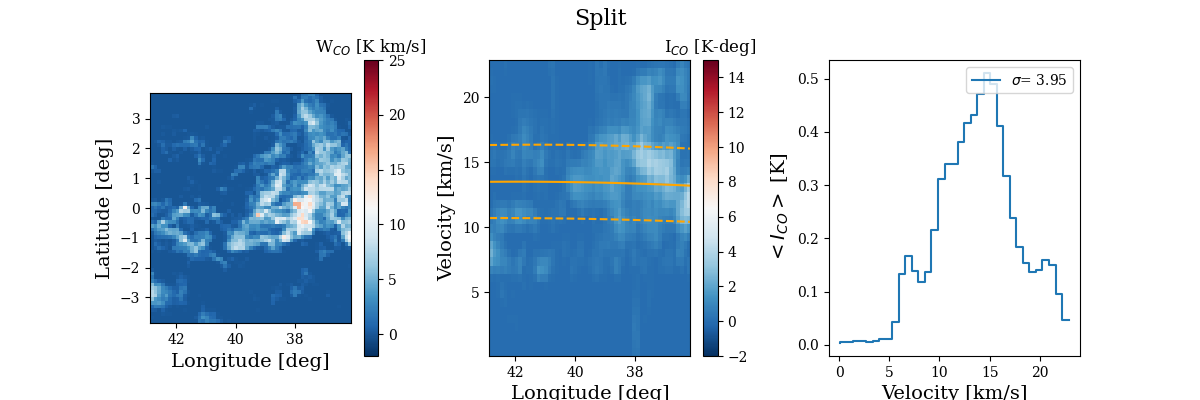}
    \includegraphics[width=0.49\textwidth]{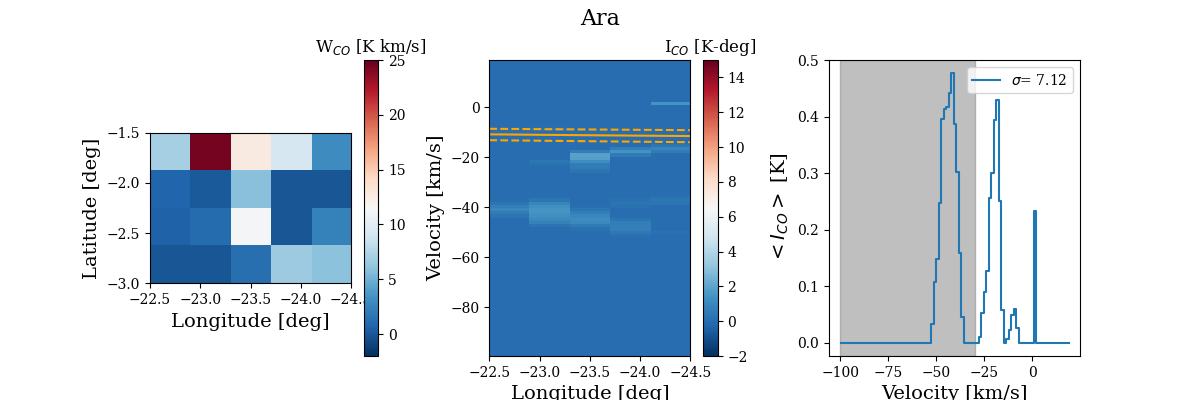}
    
    \includegraphics[width=0.49\textwidth]{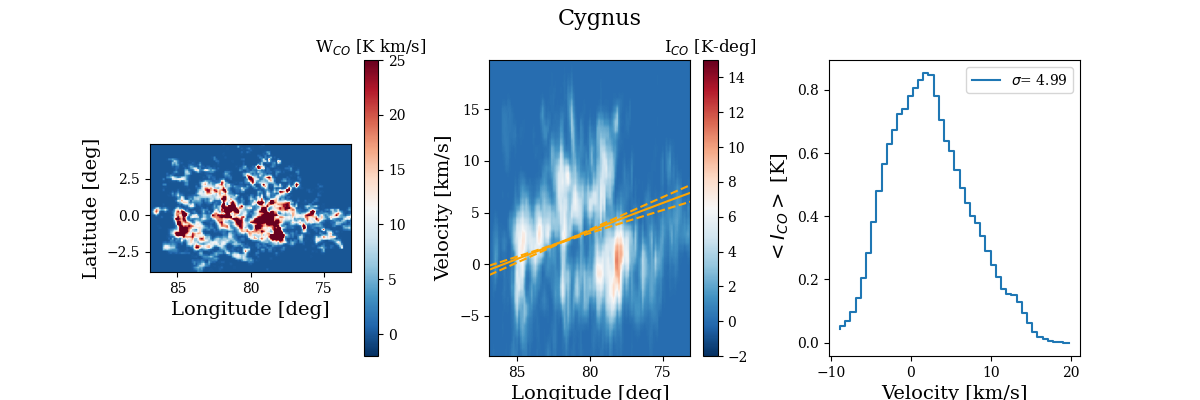}
    \includegraphics[width=0.49\textwidth]{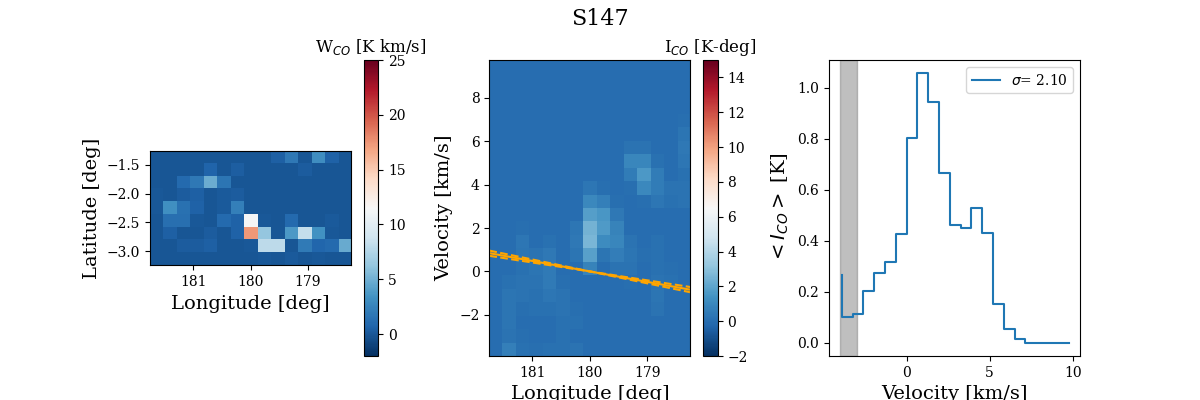}
    \includegraphics[width=0.49\textwidth]{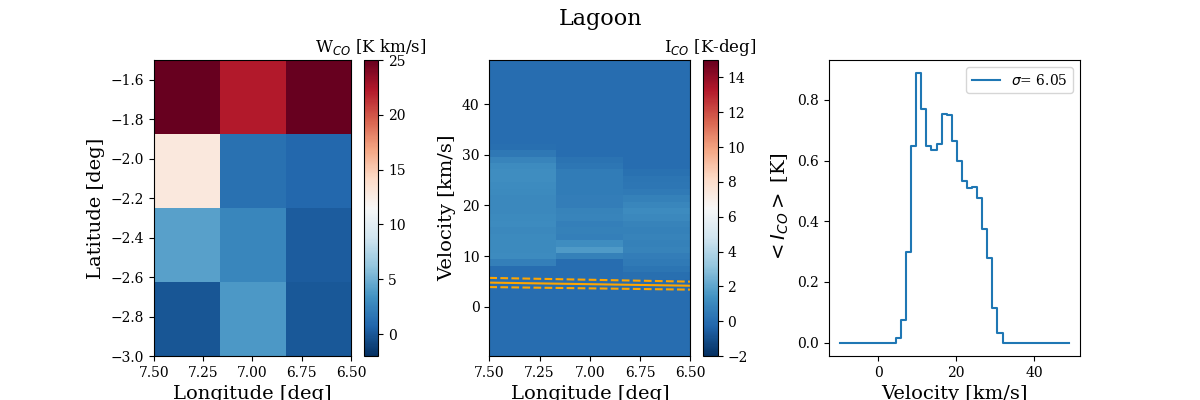}
    \includegraphics[width=0.49\textwidth]{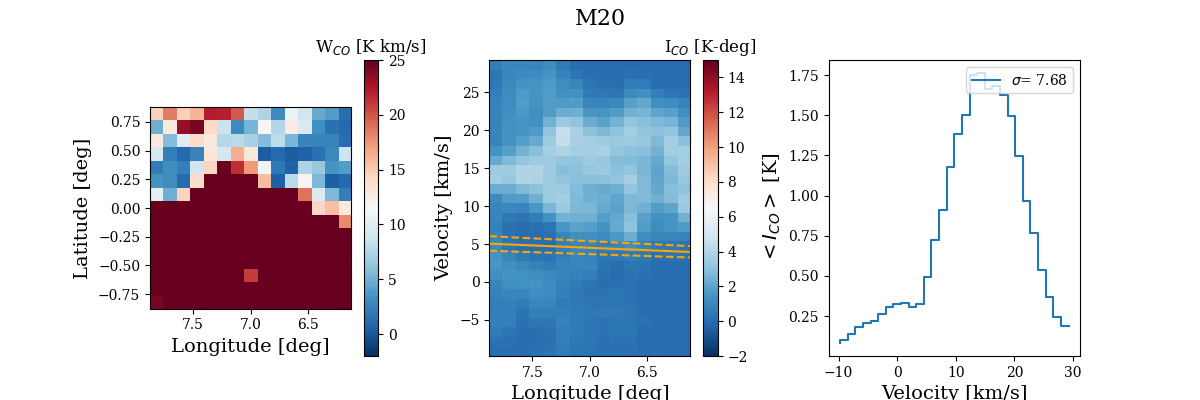}
    
    \includegraphics[width=0.49\textwidth]{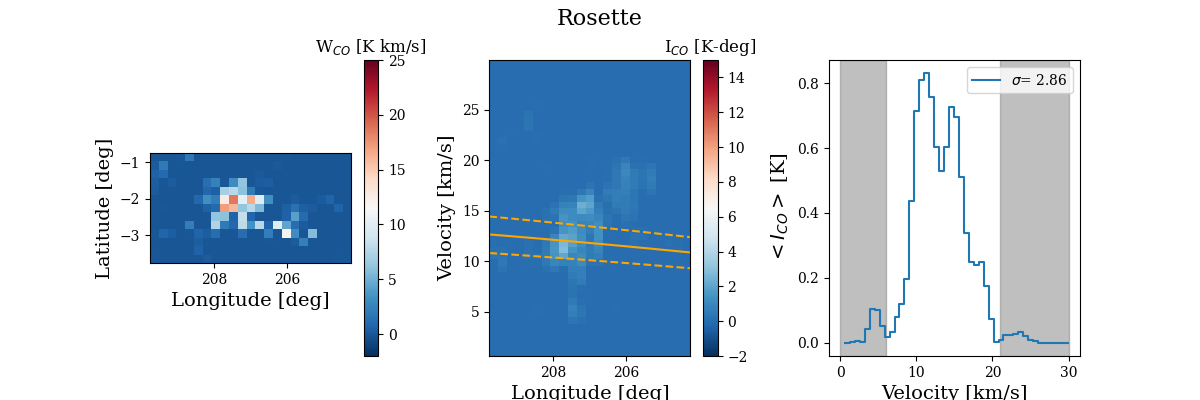}
    \includegraphics[width=0.49\textwidth]{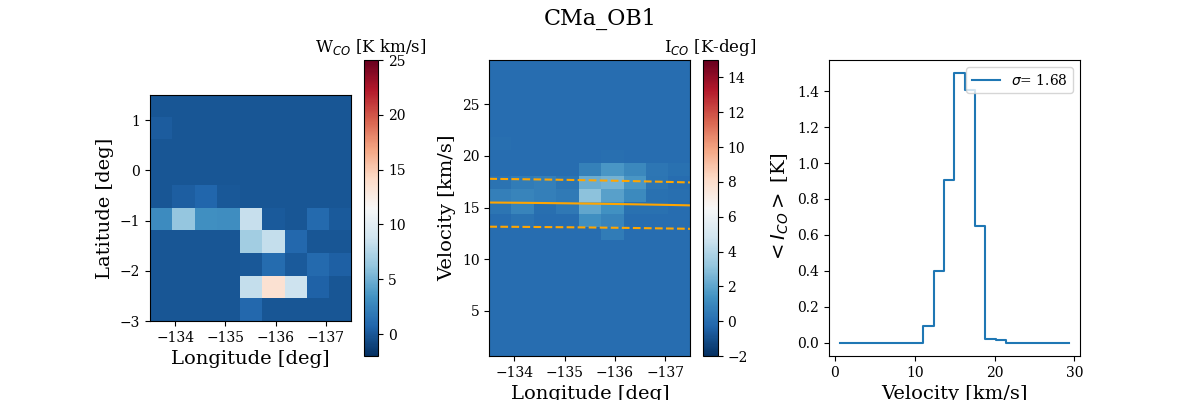}
    \includegraphics[width=0.49\textwidth]{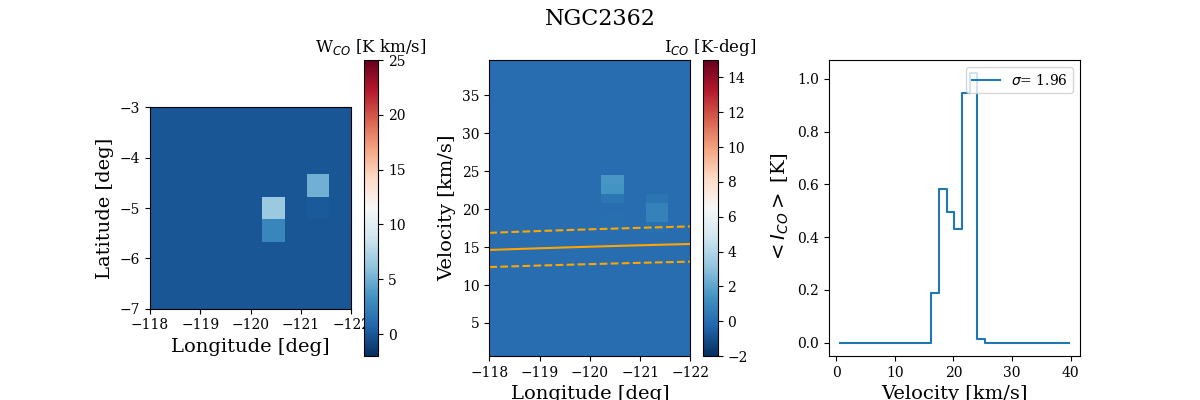}
    \includegraphics[width=0.49\textwidth]{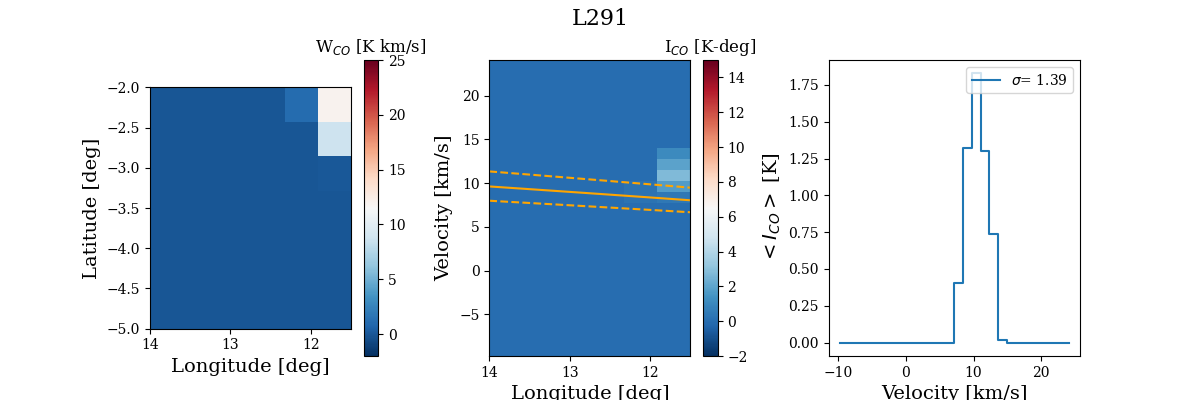}

    \includegraphics[width=0.49\textwidth]{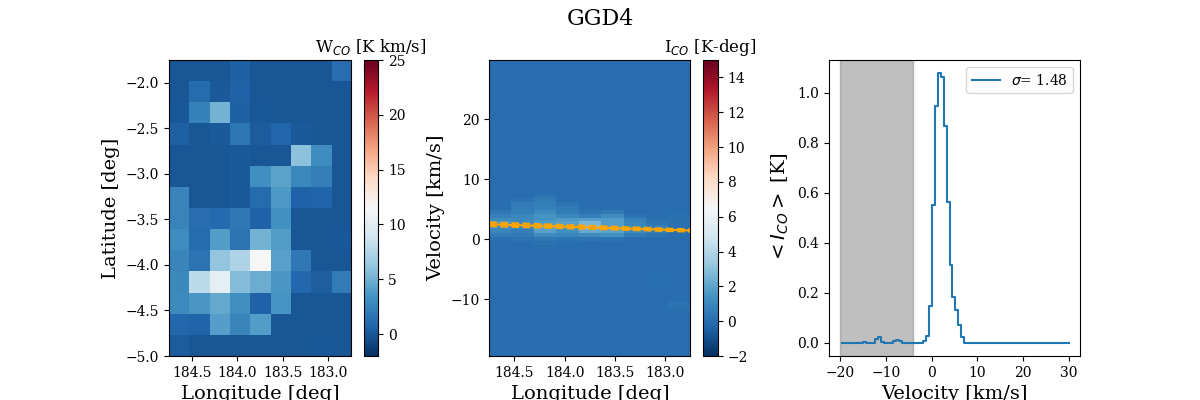}
    \includegraphics[width=0.49\textwidth]{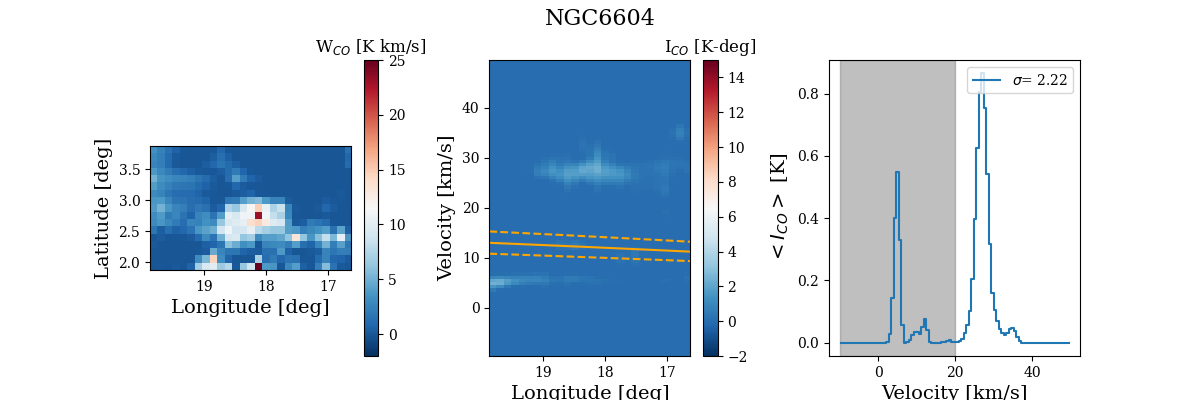}
    \caption{Overview of the PPV data of the molecular clouds in our sample, seen in CO position and velocity. CO from the \citet{dame2001milky} survey.}
    \label{fig:3pan_4}
\end{figure}

\begin{figure}[h!]
    \centering
    \includegraphics[width=0.49\textwidth]{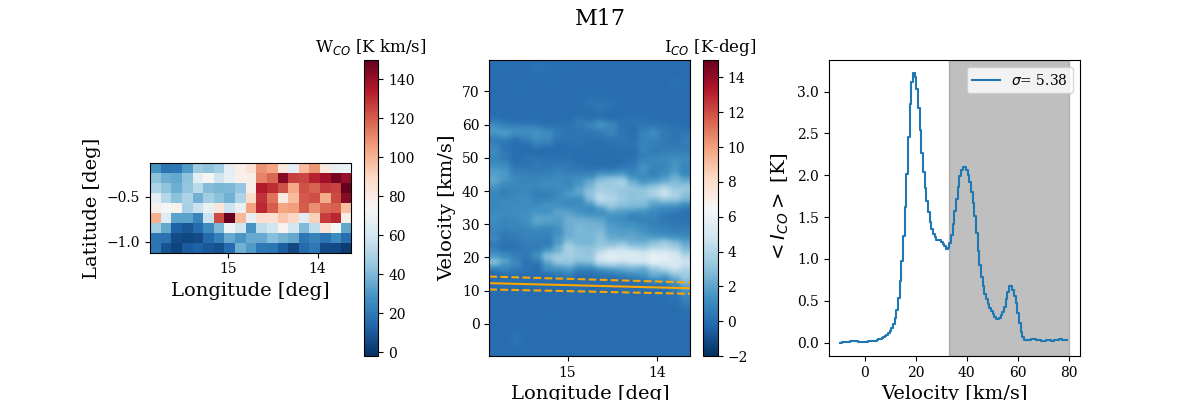}
    \includegraphics[width=0.49\textwidth]{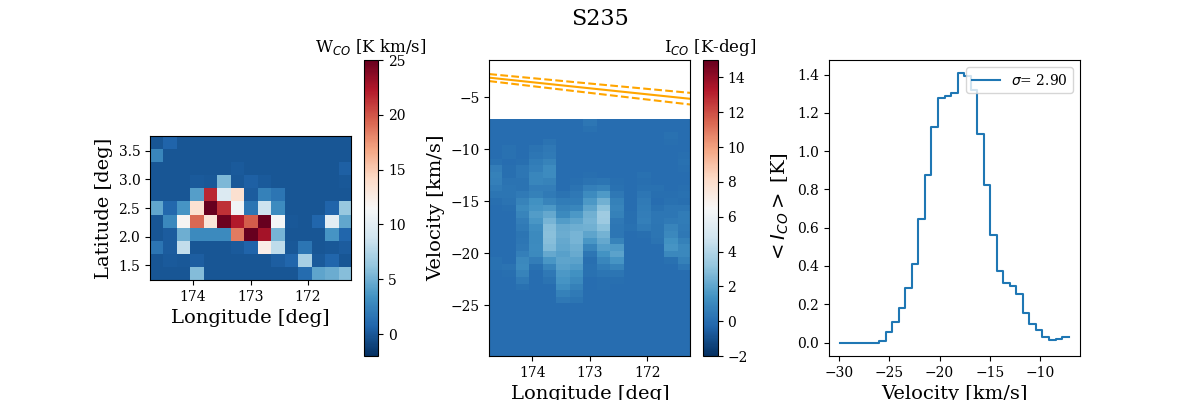}
    
    \includegraphics[width=0.49\textwidth]{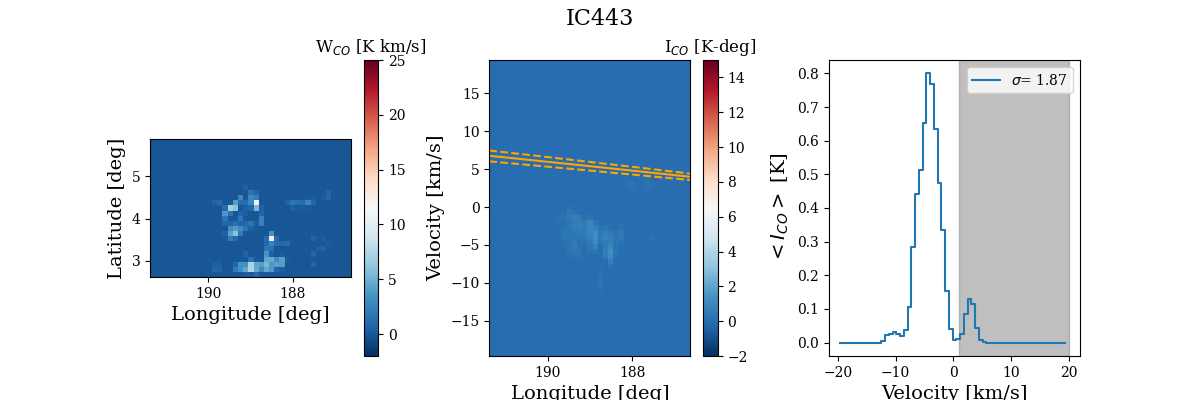}
    \includegraphics[width=0.49\textwidth]{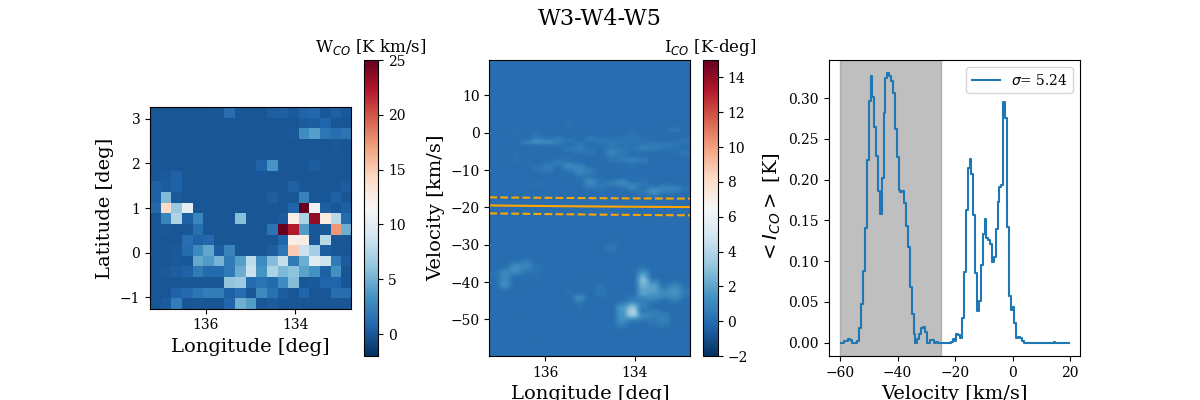}
    \includegraphics[width=0.49\textwidth]{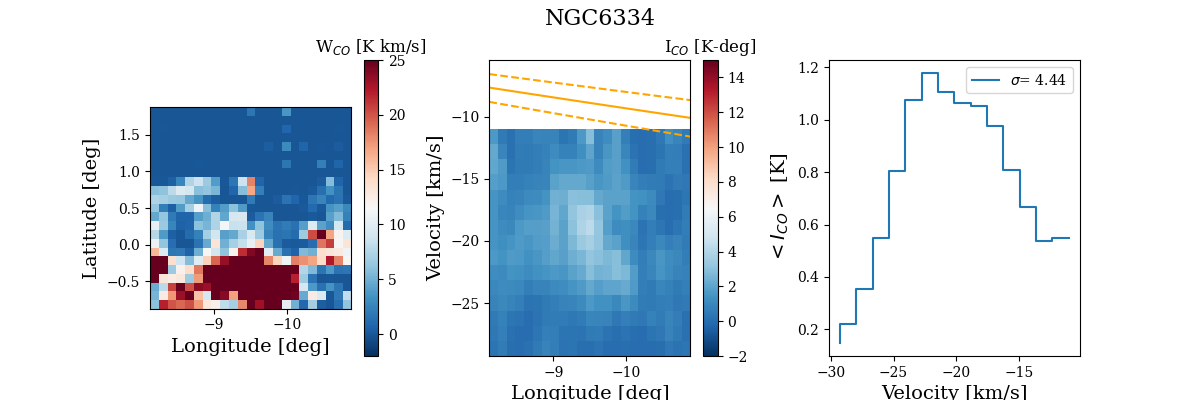}
    \includegraphics[width=0.49\textwidth]{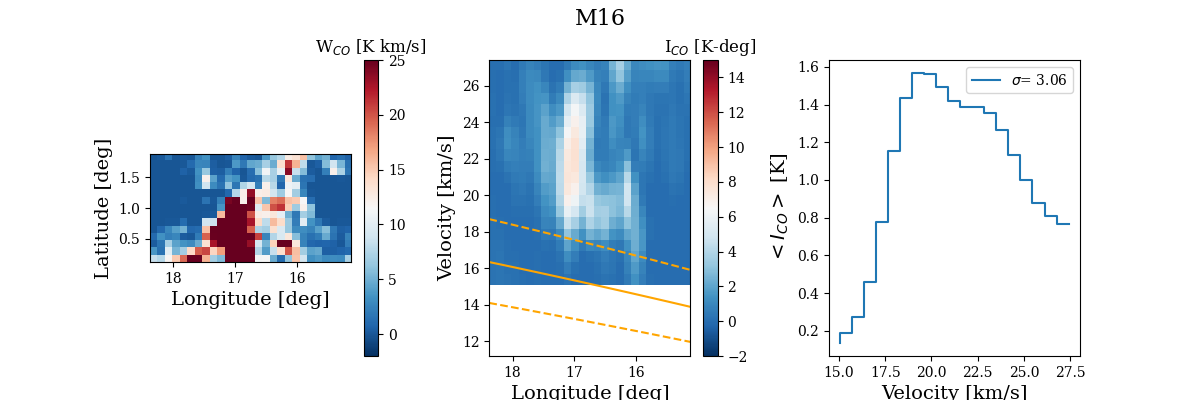}
    
    \includegraphics[width=0.49\textwidth]{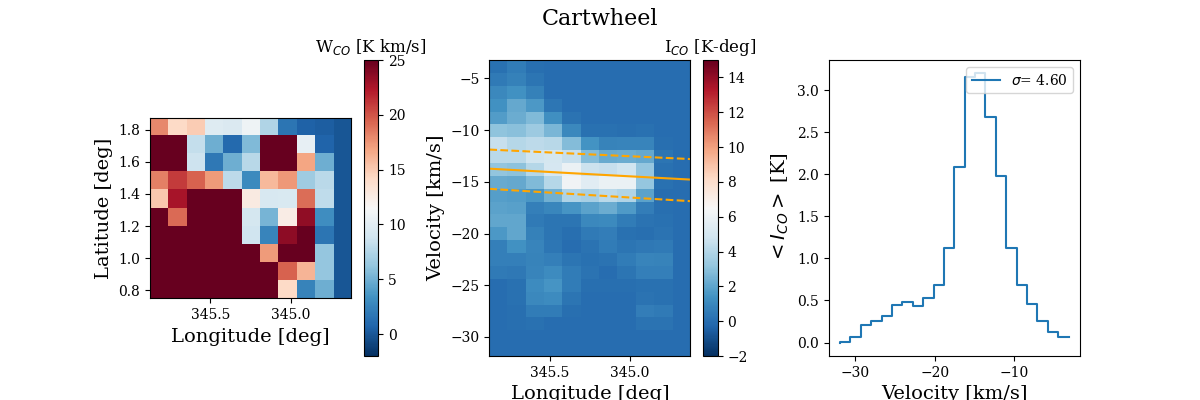}
    \includegraphics[width=0.49\textwidth]{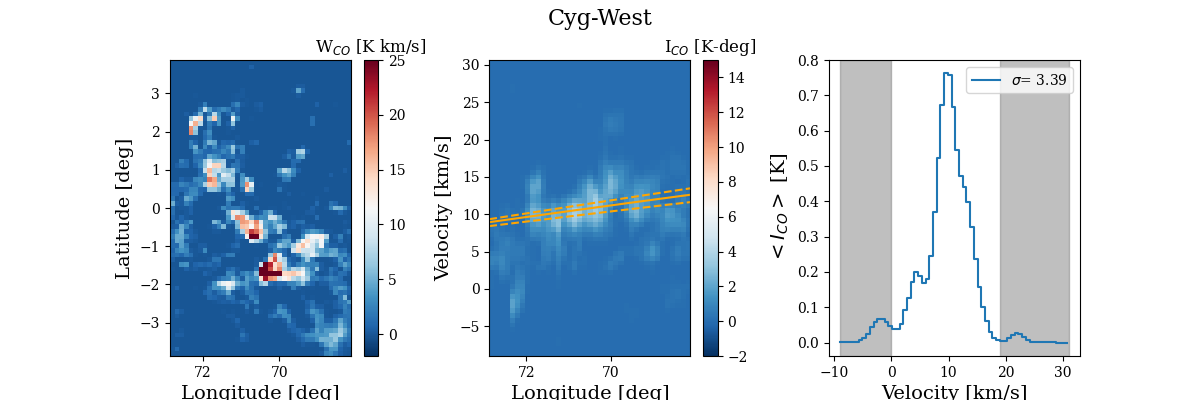}
    \includegraphics[width=0.49\textwidth]{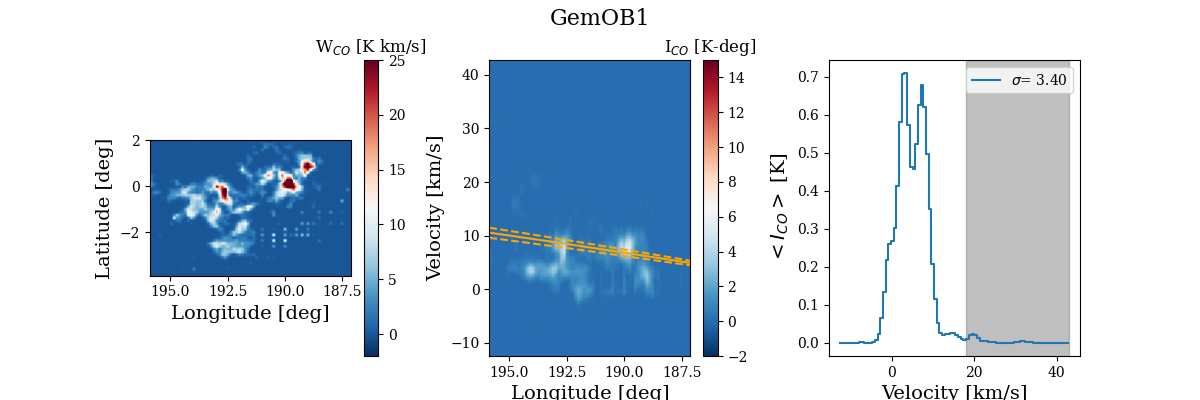}
    \includegraphics[width=0.49\textwidth]{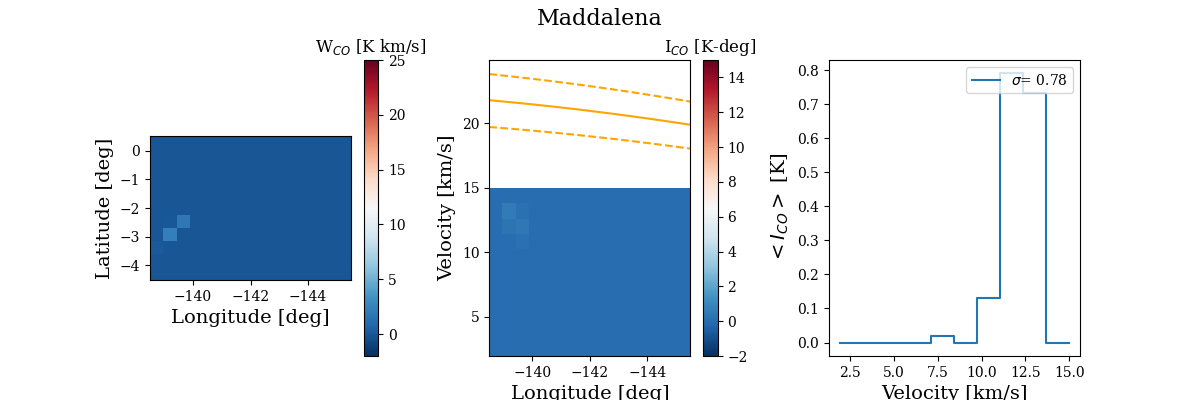}
    \caption{Overview of the PPV data of the molecular clouds in our sample, seen in CO position and velocity. CO from the \citet{dame2001milky} survey.}
    \label{fig:3pan_5}
\end{figure}

This appendix includes an overview of the cloud sample in $p$-$p$-$v$ space; see Figs \ref{fig:3pan_1}-\ref{fig:3pan_5}. For each of the 64 clouds, we show the plane of the sky CO intensity (left), the $p$-$v$ map (centre) and the CO velocity spectrum (right). The orange line in the velocity map shows the velocity expected from the rotation of the galaxy at the position of the cloud. This velocity is subtracted from the data when we correct for the rotation of the Galaxy. The stapled lines show the expected velocity with variations in distance of $\pm$100 pc. The grey fields in the spectra show the regions we masked out due to assumed foreground or background gas.

\clearpage

%\FloatBarrier
\section{Mass and temperature units of spectra}
\label{app:MT}

\begin{figure}[H] %!htb]
%\begin{figure}[h!]
    \centering
    \includegraphics[width=0.35\textwidth]{figs/test2/total_spectrum_avg_cut_new3.png}
    \includegraphics[width=0.35\textwidth]{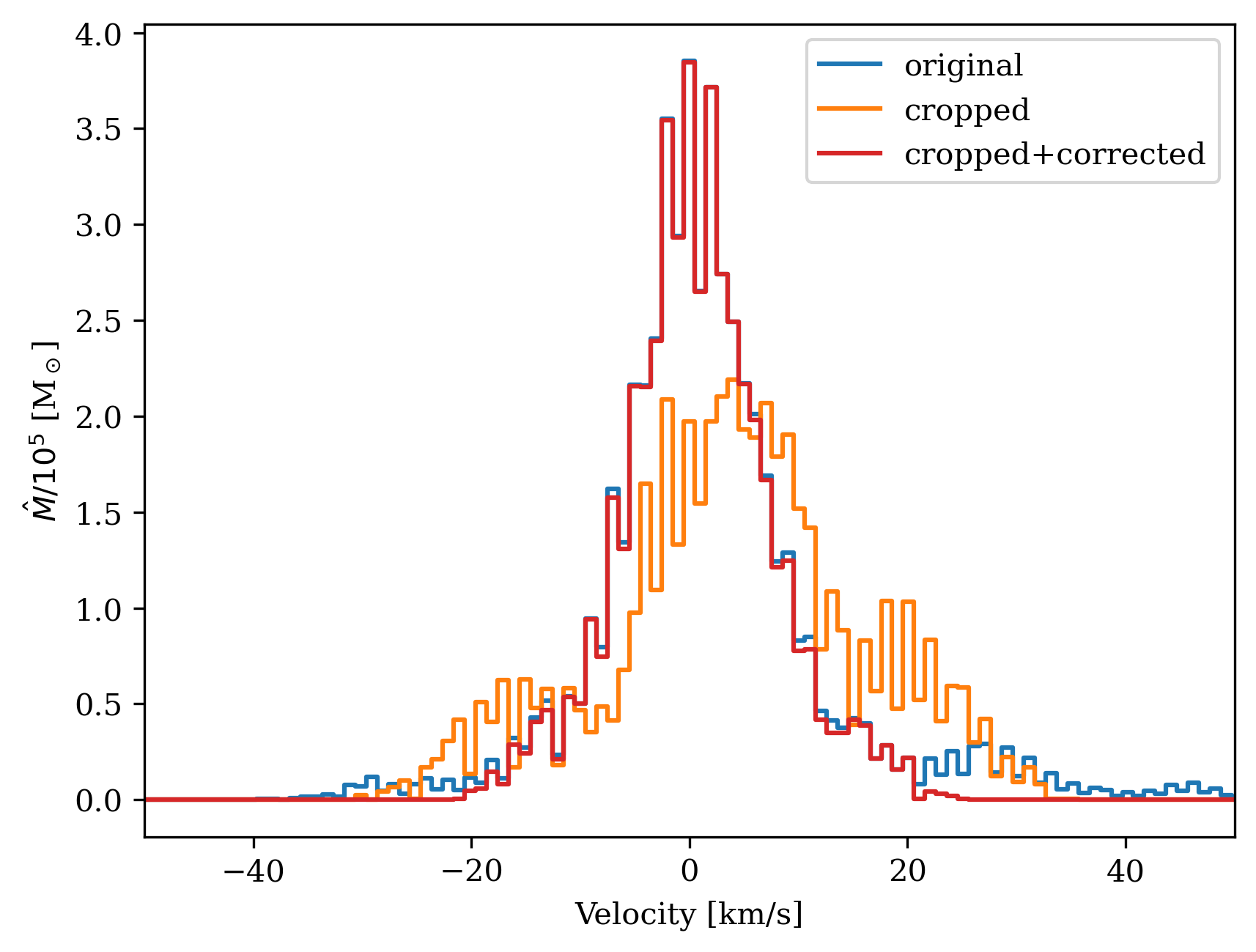}
    \caption{Spectra for the full survey area from Fig.\ref{fig:spect_tot} (left), compared to the same spectra expressed in mass units (right). The average temperature in the survey area is calculated per pixel per cloud.}
    \label{fig:MT_tot}
\end{figure}

\begin{figure}[H] %!htb]
    \centering
    \includegraphics[width=0.35\textwidth]{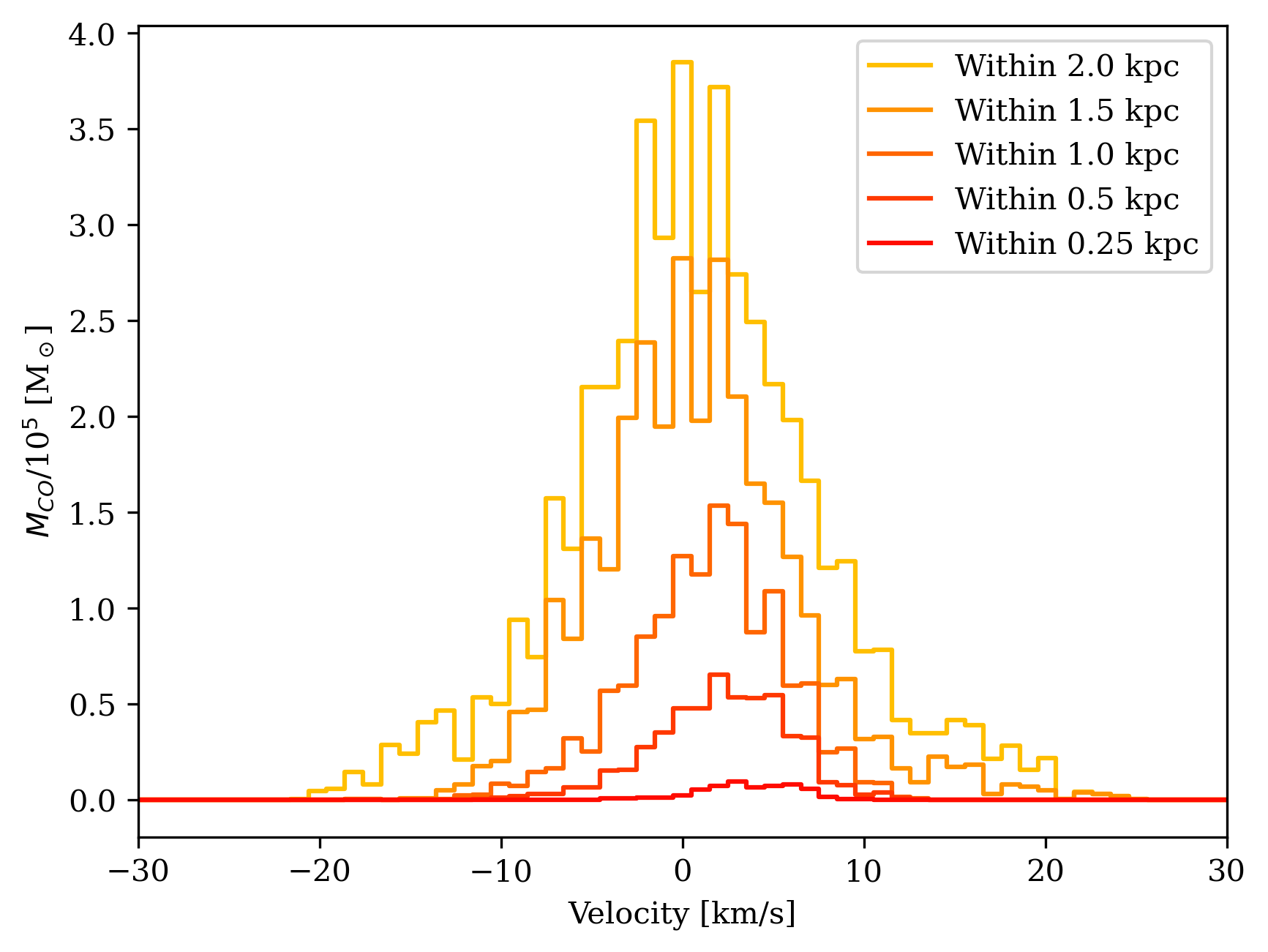}
    \includegraphics[width=0.35\textwidth]{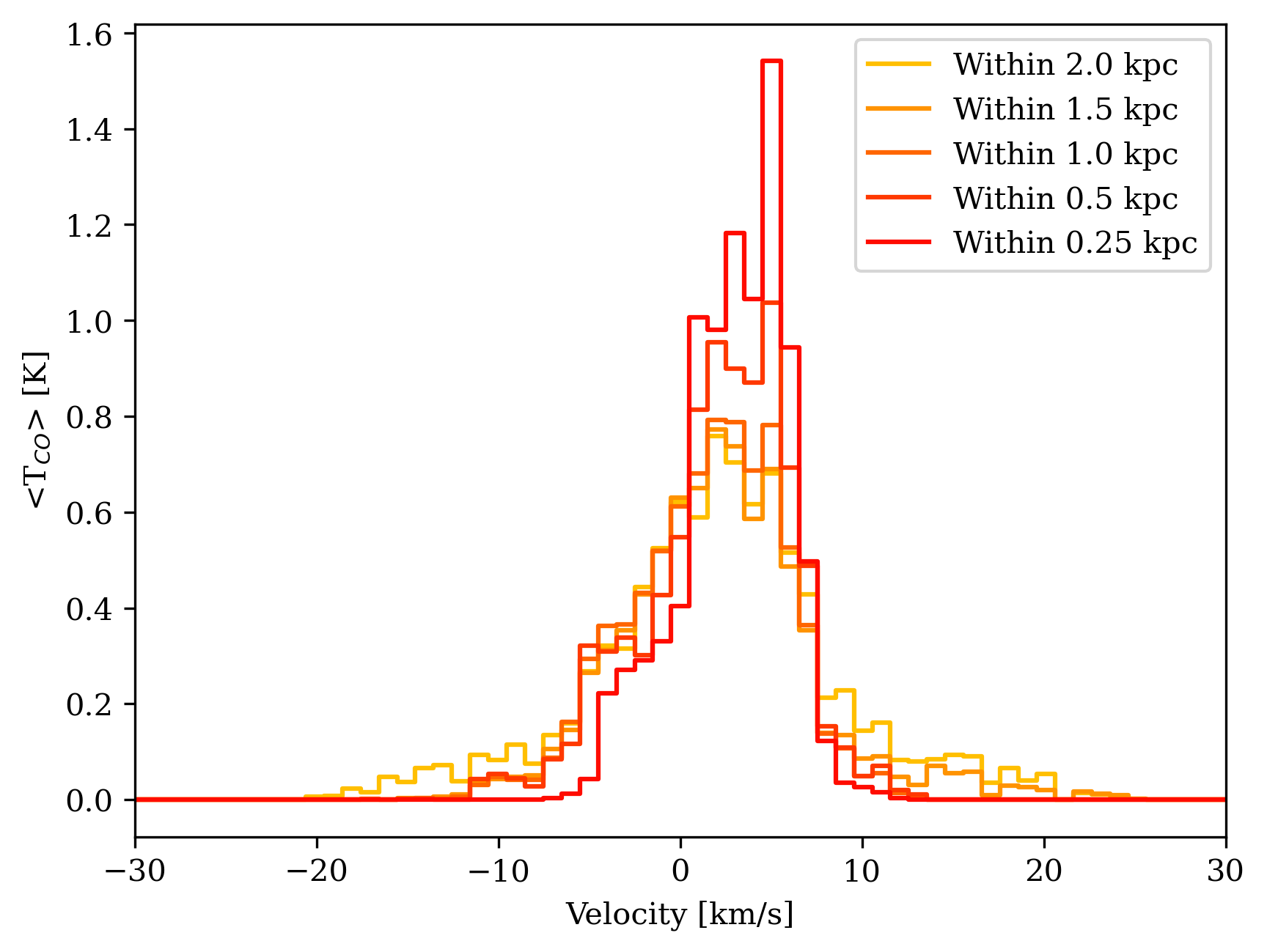}
    \caption{Sun-centred apertures of various sizes from Fig. \ref{fig:aps_sun} (left), compared to the same spectra expressed in average temperature per pixel per cloud (right).}
    \label{fig:MT_sun}
\end{figure}

In this appendix we compare spectra in mass and temperature units to facilitate comparison with extragalactic works. Figure \ref{fig:MT_tot} shows the spectra from Fig.\ref{fig:spect_tot}, compared to the same spectra expressed in mass units. There is some difference between the shape of the spectra between the two panels. This difference is also shown in Fig. \ref{fig:MT_sun}, where we show the spectra from Fig. \ref{fig:aps_sun} compared to the same spectra expressed in average temperature. The spectra in temperature units look different than the mass spectra because the clouds are at various distances, and this affects the calculated mass (see eq.\ref{eq:mass}), but not the measured temperature. We are also more complete at nearer distances, leading to higher mean temperatures in these apertures.

\section{Variation of aperture spectra}
\label{app:app_diffpos}

\begin{figure}[H] %!htb]
%\begin{figure}[h!]
    \centering
    \includegraphics[width=0.32\textwidth]{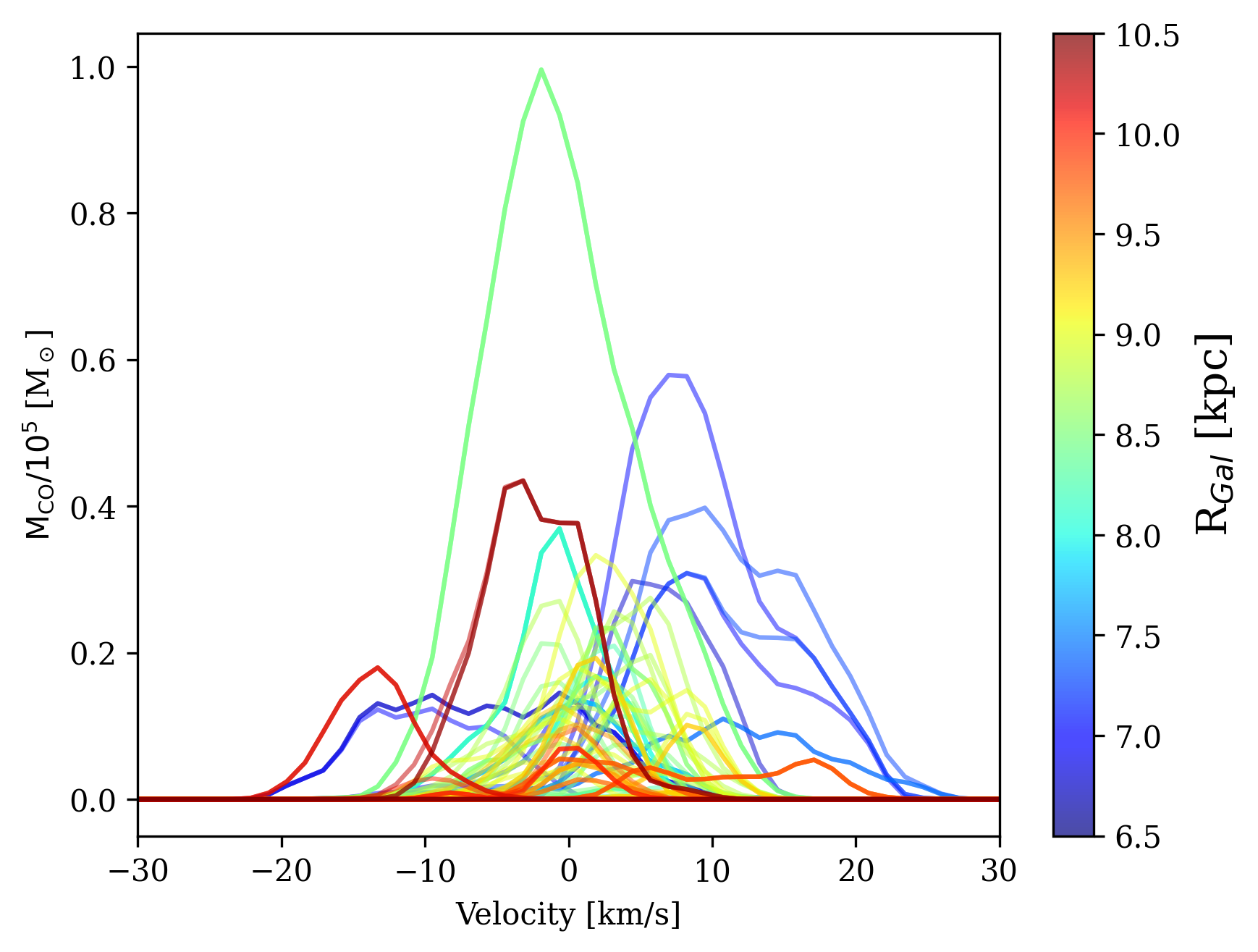}
    \includegraphics[width=0.32\textwidth]{figs/app_diffpos/SPECTRA_diffpos_nostep_interp_1kpc_d2.png}
    \includegraphics[width=0.32\textwidth]{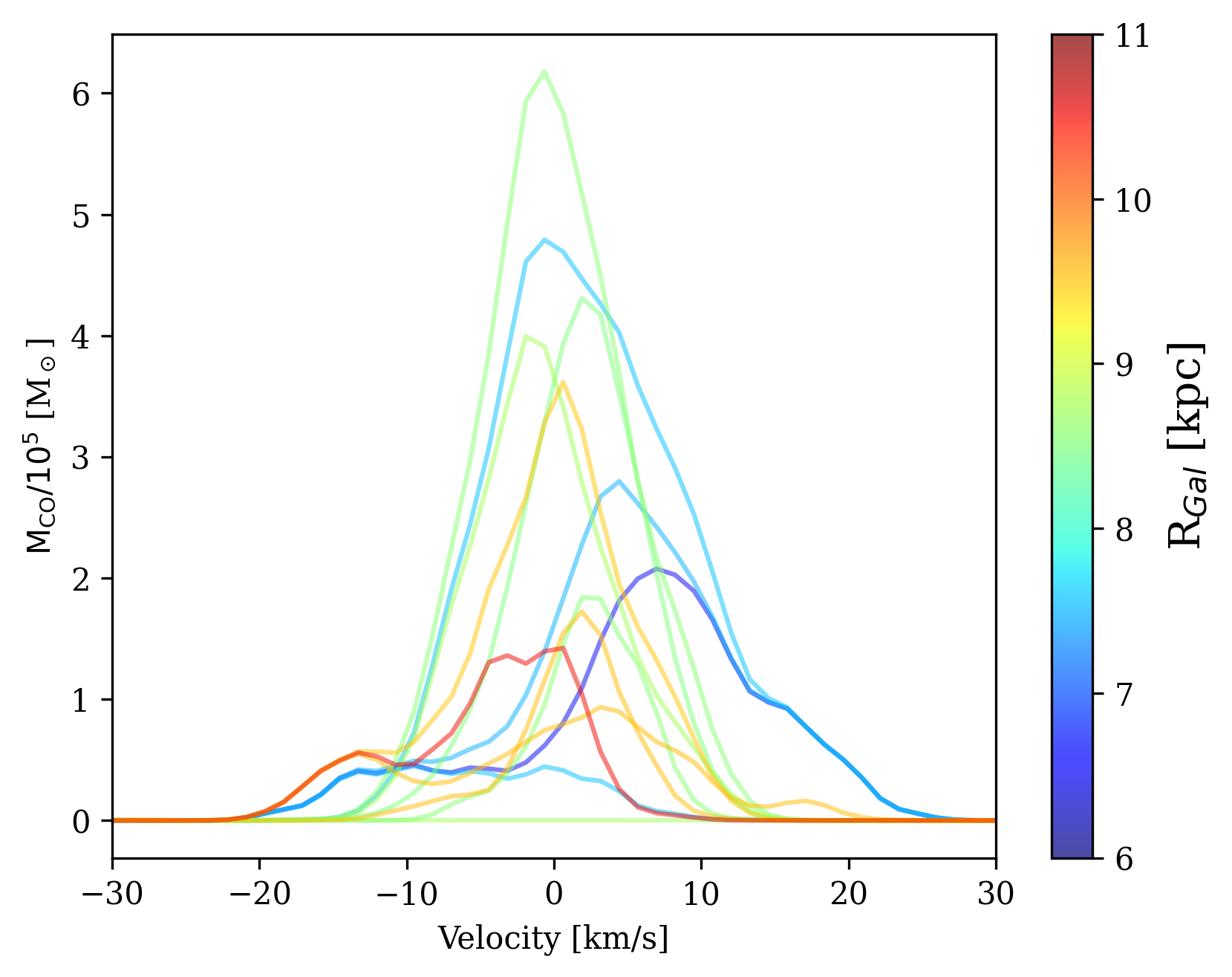}
    \caption{Spectra for all apertures with $R=0.25$ kpc (left), $R=0.5$ kpc (middle), and $R=1.0$ kpc (right).} % \jk{[From these versions, I would only keep the third row. ]}}
    \label{fig:app_diffpos}
\end{figure}

In this appendix we show the spectra of all the apertures covering our survey area; see Fig. \ref{fig:app_diffpos}. There is significant variation between the spectra of apertures of the same size, especially for the smaller apertures. The apertures closer to the Galactic centre and the Sagittarius spiral arm (blue) also tend to have higher mean velocities and possibly more mass than the apertures towards the outer galaxy (red). The spectra were interpolated between points (\texttt{scipy.interpolate.interp1d}) and smoothed with a Gaussian function (\texttt{gaussian$_-$filter1d}) for clarity.

\end{appendix}

\end{document}